\documentclass[final,authoryear,3p,times]{elsarticle}

\usepackage{float}    % control graph position, before "hyperref"
\usepackage{subfig}   % can have multiple captoin of figures
\usepackage{array}    % can help in vertical alignment with text and figures
\usepackage{multirow} 
\usepackage{lineno,hyperref}
\usepackage{url}
\usepackage{amsmath}
\usepackage{amsthm}
\usepackage{amssymb}
\usepackage[ruled,linesnumbered,commentsnumbered]{algorithm2e}  % for amazing algorithm format
\usepackage[titletoc,title]{appendix}                           % can have appendix
\usepackage{verbatim}                                           % can use multi-line comment

\usepackage{enumitem}                                      % make list have paragraph
\setlist{parsep=0pt,listparindent=\parindent}              % make list have paragraph

\modulolinenumbers[5]

\newcommand{\real}{\ensuremath{\mathbb{R}}}

\newcommand{\ltwo}{\ensuremath{\mathbb{L}^2}}

\journal{Computational Statistics and Data Analysis}

\theoremstyle{plain}

\newtheorem{remark}{Remark}

\usepackage{numcompress}\biboptions{authoryear}

\begin{document}

\begin{frontmatter}

\title{Trend and Variable-Phase Seasonality Estimation from Functional Data}

\author[math]{Liang-Hsuan Tai\corref{cor1}}
\ead{ltai@math.fsu.edu}
\author[stat]{Anuj Srivastava}
\ead{anuj@stat.fsu.edu}
\author[math]{Kyle A. Gallivan}
\ead{gallivan@math.fsu.edu}

\cortext[cor1]{Corresponding author at 208 Love building, 1017 Academic Way Tallahassee, FL 32306-4510}
%\fntext[fn1]
\address[math]{Department of Mathematics, Florida State University, Tallahassee, FL 32306, United States}
\address[stat]{Department of Statistics, Florida State University, Tallahassee, FL 32306, United States}

\begin{abstract}
The problem of estimating {\it trend} and {\it seasonal variation} in time-series data
has been studied over several decades, although mostly using single time series. 
This paper studies the problem 
of estimating these components from functional data, i.e. multiple time series, 
in situations where seasonal effects exhibit arbitrary time warpings or phase variability across
different observations. Rather than ignoring the phase variability, or using an off-the-shelf
alignment method to remove phase,  
we take a model-based approach and seek MLEs of the trend and the seasonal effects, while performing 
alignments over the seasonal effects at the same time. The MLEs of trend, 
seasonality, and phase are computed using a coordinate-descent based optimization method. 
We use bootstrap replication for computing confidence
bands and for testing hypothesis about the estimated components. 
We also utilize log-likelihood for selecting  the trend subspace, and for comparisons with other candidate models. 
This framework is demonstrated using 
experiments involving synthetic data and three real data 
(Berkeley Growth Velocity, U.S. electricity price, and USD exchange fluctuation).  

\end{abstract}


\end{frontmatter}

%\tableofcontents
%\linenumbers

%%%%%%%%%%%%%%%%%%%%%%%%%%%%%%%%%%%%%%%%%%%%%%%%%%%%%%%%%%%%%%%%%%%%%%%%%%%%%%%%%%%%
%%%%%%%%%%%%%%%%%%%%%%%%%%%%%%%%%%%%%%%%%%%%%%%%%%%%%%%%%%%%%%%%%%%%%%%%%%%%%%%%%%%%
\section{Introduction} \label{sec: Introduction}
We investigate the classical problem of separating trend and seasonal components 
in a time-series data, but with a few differences. 
Firstly, we assume the availability of multiple observations, i.e. multiple time series, 
as opposed to the classical formulation that mostly uses a single time series to perform such estimation. 
Secondly, we tackle a difficult problem where the seasonal effects exhibit arbitrary time warping, or 
phase variability (see \cite{Marron2015functional} for the notion of phase variation), in each observation. 
This situation arises often in practical situations where the seasonal effect displays cyclostationary behavior,
but are seldom aligned perfectly in the observed data. 

To make the discussion concrete, let us assume that each individual observation, $f_i:[0,1] \to \real$, is  
made up of two main components, in addition to the observation noise, according to the superposition model: 
\begin{equation}
f_i(t) = h(t) + (g, \gamma_i)(t) + \epsilon_i(t),\ i= 1,2, \dots, n\ .
\label{eqn:general-model}
\end{equation}
On the right, the different terms are: 
\begin{itemize}
\item {\bf Trend}, denoted by 
$h:[0,1] \to \real$, captures the long-term evolution of the data. Generally, 
we are interested in $h$ being either a lower 
order polynomial representing a null, constant, linear, quadratic shapes, or slowly varying sinusoid. 
\item {\bf Seasonal effect}, denoted by  $g:[0,1] \to \real$, captures seasonal or period effects in the data, 
Instead of assuming the seasonal effect to be fixed, or perfectly aligned across observations,
we make the model more general by including a temporally-misaligned version of $g$. 
That is, we utilize the term $(g, \gamma_i)(t)$, instead of $g(t)$, 
which represents a time warping of $g$ by a function $\gamma_i$, where
$\gamma_i:[0,1] \to [0,1]$ is a boundary-preserving diffeomorphism.  
There are several ways to express this warping. The most common 
form  is simply $g(\gamma_i(t))$ but, as discussed later, there are other possibilities. 

\item {\bf Observation noise}, denoted by $\epsilon_i: [0,1] \to \real$ with the assumption that $\epsilon_i$ 
are {\it i. i. d.} with $E[\epsilon_i(t)] = 0$ for all $i$ and $t$. 
\end{itemize}

\begin{figure}[!htb] x
\begin{center}
   \subfloat[observed functions]
	    {\includegraphics[scale=0.26]{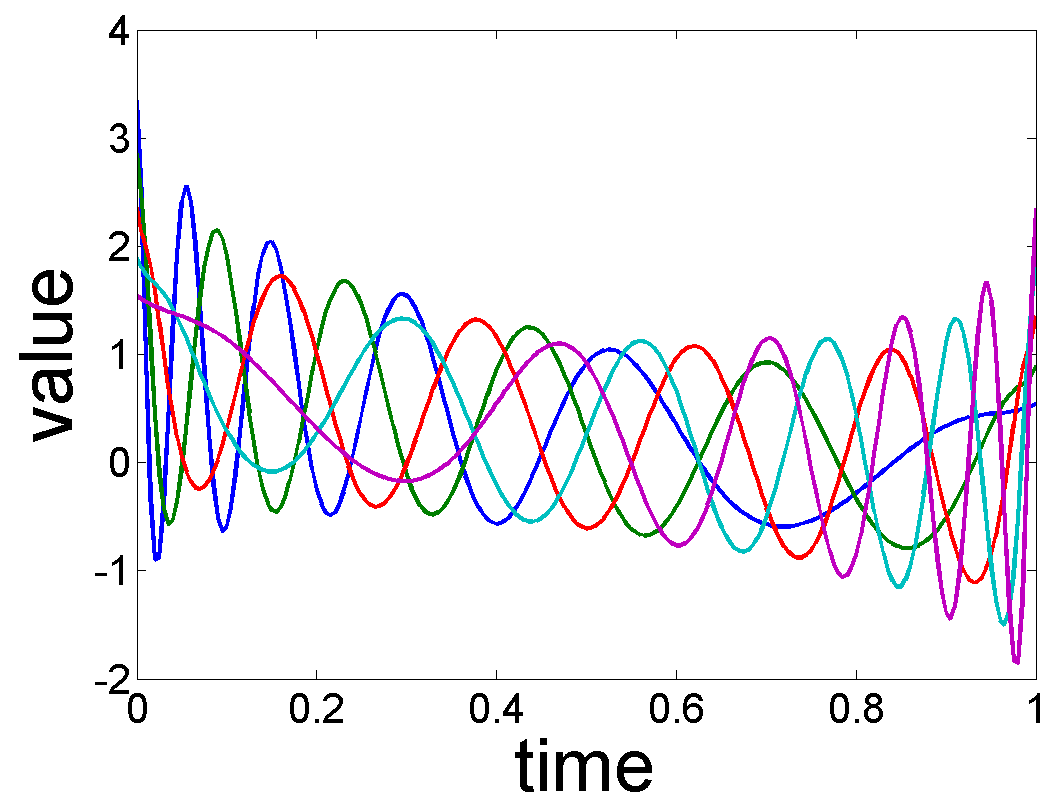}}
   \subfloat[cross-sectional mean]
	    {\includegraphics[scale=0.26]{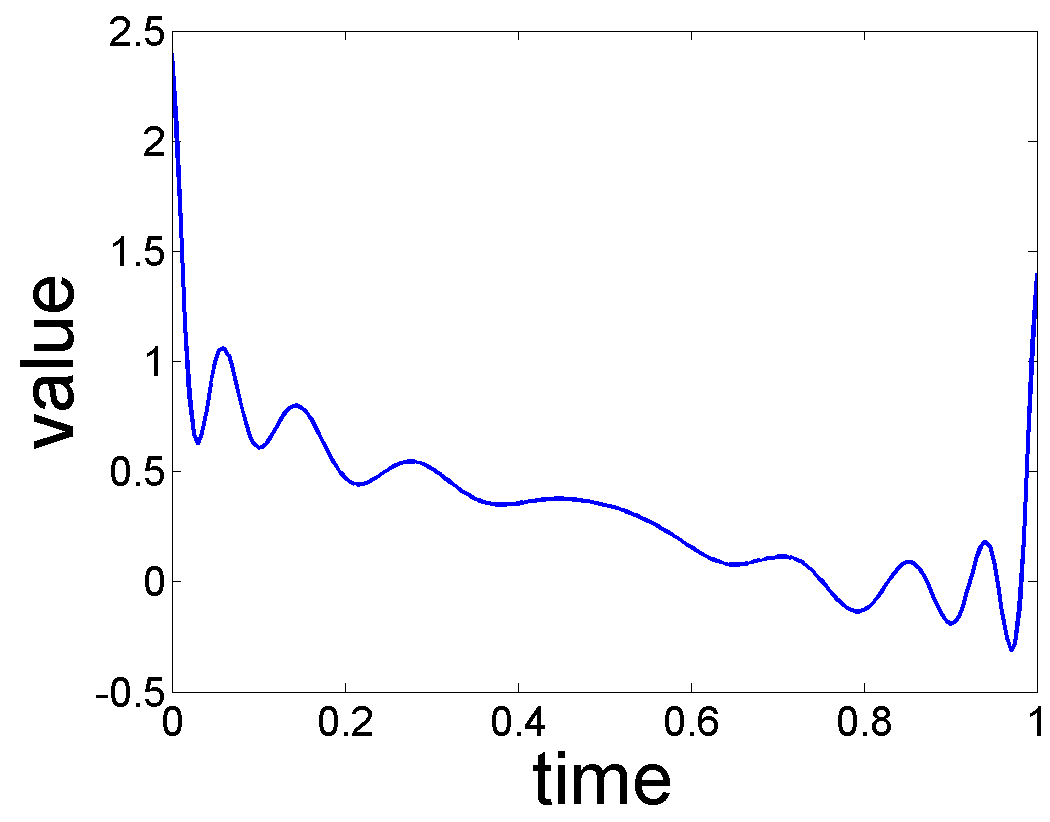}}\\[-2ex]
   \subfloat[trend]
	    {\includegraphics[scale=0.23]{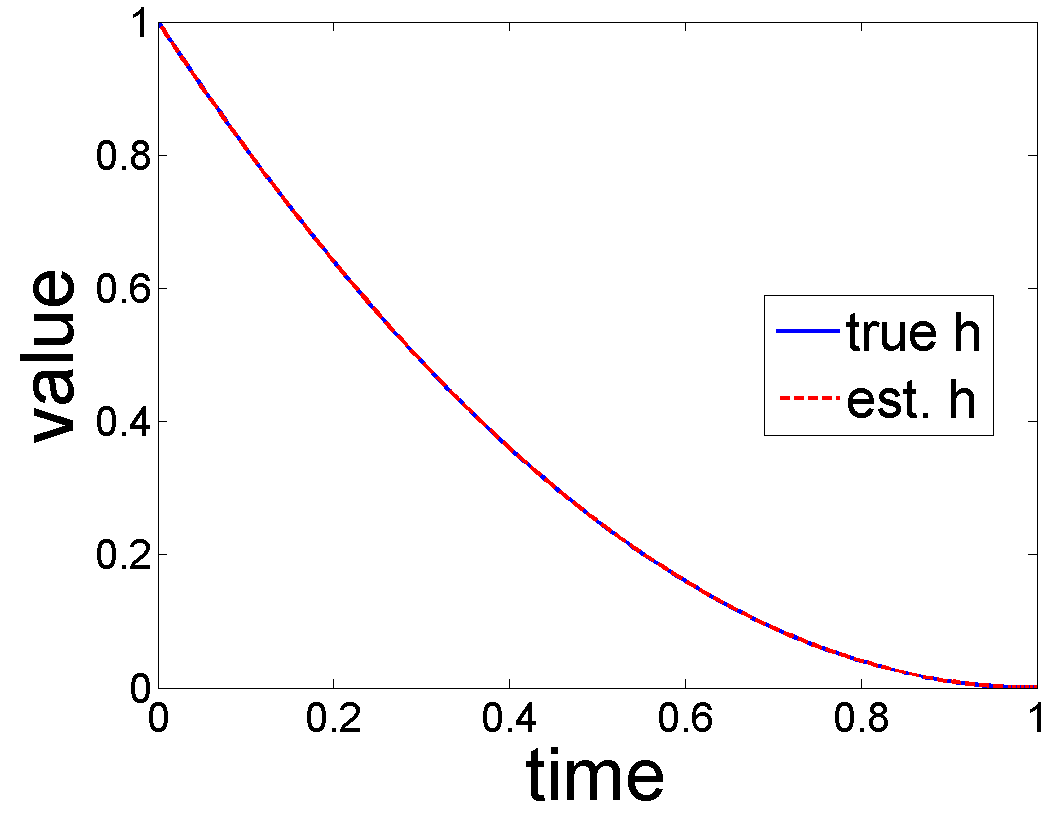}}
	 \subfloat[seasonality]
	    {\includegraphics[scale=0.23]{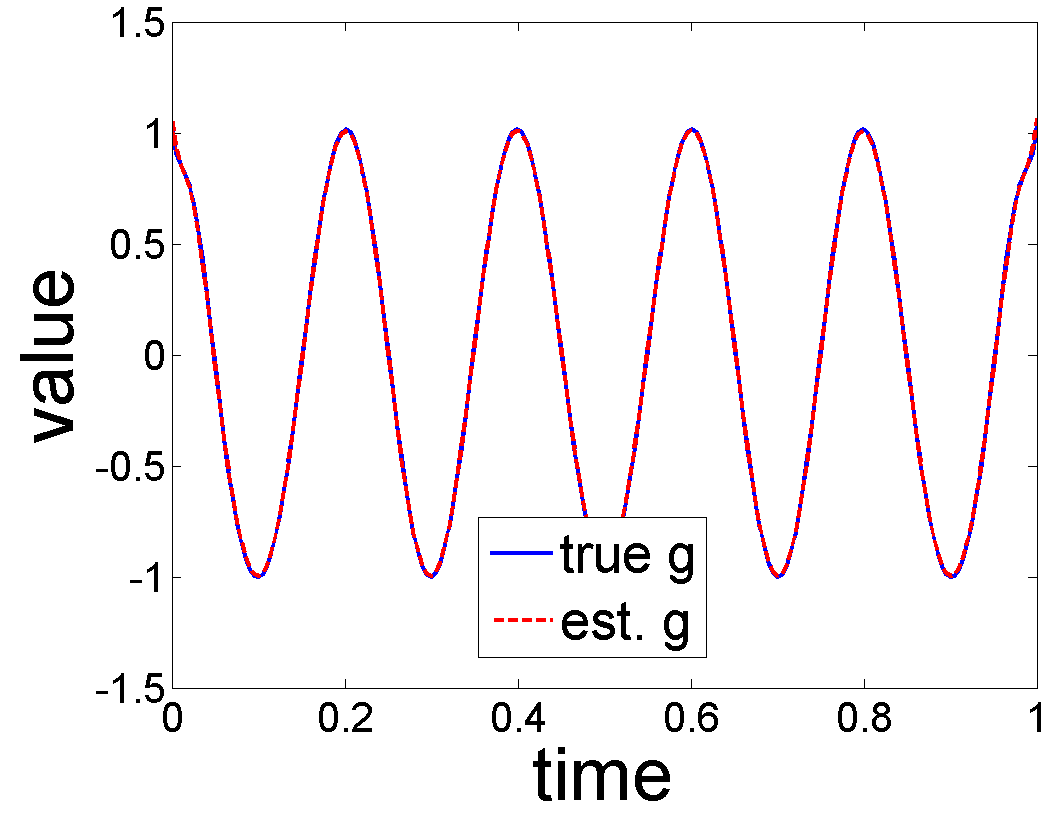}}
   \subfloat[warping functions]
	    {\includegraphics[scale=0.23]{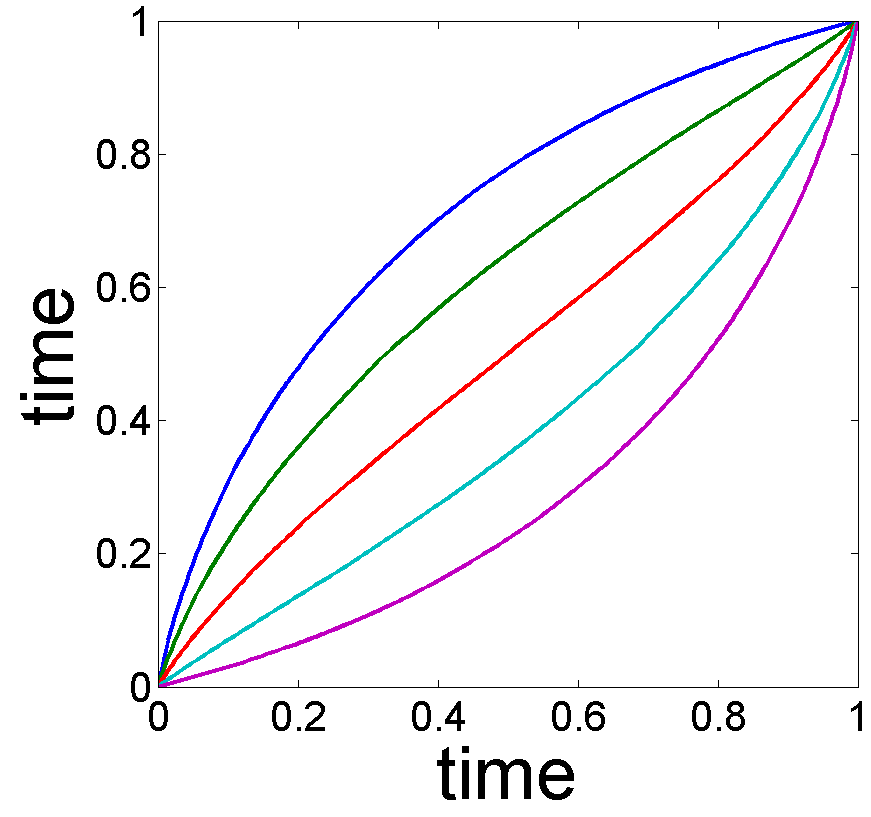}}
\caption{Illustration of trend and seasonality estimation of functional data under random time warpings.} 
\label{fig:Motivation1-f}
\end{center}
\end{figure}

With this model, the goal is to estimate $h$ and $g$ using a set of 
observations $\{f_i, i=1,2,\dots,n\}$. 
We illustrate this setup pictorially using 
Fig. \ref{fig:Motivation1-f}(a) which shows a set of observed $f_i$s. 
These functions show a periodic behavior, with roughly
the same number of peaks and valleys, and a general decreasing trend 
from left to right. However, the seasonal variations, denoted by high-frequency peaks and valleys, 
are quite misaligned, implying presence of phase variability in the seasonal components. 
Traditionally, the seasonal effects are removed by averaging (or smoothing or low-pass filtering) 
the observed functions across time or observations. A simple cross-sectional average 
($\bar{f}(t)=\frac{1}{n}{{\sum}_{i=1}^{n}}f_{i}(t)$) of the data results in Fig. \ref{fig:Motivation1-f}(b). 
Although this function displays a decreasing trend, 
it also contains some artifacts that result mainly from
the misalignment of seasonal components across individual observations. 
Therefore, one has to perform an alignment when estimating components in such data. 
Given the observed functions $\{f_i\}$, 
a more comprehensive solution is to recover the seasonality $g$, warping functions $\{\gamma_{i}\}$, 
and the underlying trend $h$ under a statistical model. 
We will call this problem {\it trend and variable phase seasonality estimation}, 
and a comprehensive solution will isolate the three components, as shown in Fig. \ref{fig:Motivation1-f}(c)-(e). 

%%%%%%%%%%%%%%%%%%%%%%%%%%%%%%%%%%%%%%%%%%%%%%%%%%%%%%%%%%%%%%%%%%%%%%%%%%%%%%%%%%%%
%%%%%%%%%%%%%%%%%%%%%%%%%%%%%%%%%%%%%%%%%%%%%%%%%%%%%%%%%%%%%%%%%%%%%%%%%%%%%%%%%%%%
\subsection{Past Approaches} \label{sec: Past Approaches}
Before presenting our model-based solution for trend and seasonality estimation, we summarize the main
ideas present in the literature,  and point out their limitations and shortcomings. The relevant 
literature can be divided into the following broad categories,  each representing a sub-model of the one  
presented Eqn. \ref{eqn:general-model}.

\begin{enumerate}
\item {\bf Estimation from Single Observation}: 
The problem of estimating trend and seasonality from a single time-series originated in Economics 
(\cite{nerlove1964spectral} and \cite{godfrey1964spectrum}), followed by more formal 
developments in statistics, 
see \cite{grether1970some}, \cite{cleveland1976decomposition}, \cite{box1978analysis}, 
\cite{hillmer1982arima}, \cite{harvey1983forecasting}. 
Please refer to the review paper by
\cite{alexandrov2012review} on this subject, and to U.S. Census Bureau's website 
\footnote{http://www.census.gov/srd/www/sapaper/} for a larger list of papers on this topic.
\cite{harvey1983forecasting}  formalized the  structural time-series model 
(also termed the classical decomposition model by \cite{brockwell2006introduction}) as: 
\begin{equation*}
	f(t)= h(t)  + g(t) + \epsilon(t)\ . 
	\label{eq: trend and seasonality estimation}
\end{equation*}
The goal is to  recover the trend $h$ and the seasonality $g$, from a single observation $f$. 
Further special cases of this model result from assuming either $h$ or $g$ to be zero. 
Common approaches for  estimating trend include
parametric least squares (\citeauthor{brockwell2006introduction}, 2006), 
moving averages (\citeauthor{brockwell2006introduction}, 2006), 
and local linear smoothing (\citeauthor{friedman2001elements}, 2009). 
On the other hand, the seasonality estimation is often handled by finding cycles (intervals of
cyclostationarity) and averaging over those cycles.   

For estimating both trend and seasonality, the popular approaches 
include autoregressive-integrated-moving average 
(\citeauthor{hillmer1982arima}, 1982, \citeauthor{harvey1983forecasting}, 1983), 
STL filtering (\citeauthor{cleveland1990stl}, 1990), 
small trend method (\citeauthor{brockwell2006introduction}, 2006), 
moving average estimation (\citeauthor{brockwell2006introduction}, 2006), 
and differencing at lag period (\citeauthor{brockwell2006introduction}, 2006). 
While these methods are based on the assumption of equally-spaced observations,  
\cite{eckner2012note} studied the case of unevenly-spaced observations.

Note that none of these models address the issue of temporal misalignment of seasonal effects across cycles. 

\item {\bf Trend Estimation Only}: 
In the case where multiple observations are available, one can 
use techniques from  functional data analysis. 
For instance, one can pose a model of the type: 
\begin{equation*}
	f_i(t) = h(t) + \epsilon_i(t),\,i=1,2, \dots,n .
\end{equation*}
Under the zero-mean assumption of $\epsilon_i(t)$, an unconstrained estimator of $h$ 
is the cross-sectional mean, $\hat{h}(t)=\frac{1}{n}{\sum_{i=1}^{n}}{f_{i}(t)}$. 
If $h$ is assumed to belong to a certain subspace, 
say ${\cal H}$,  then there are several techniques available to estimate $h$: 
B-Splines (\citeauthor{besse1997simultaneous}, 1997),
Smoothing Splines (\citeauthor{brumback1998smoothing}, 1998),
Basis Functions (\citeauthor{ramsay2006functional}, 2006), 
Least Squares (\citeauthor{ramsay2006functional}, 2006), 
Roughness Penalty (\citeauthor{ramsay2006functional}, 2006), 
and Local Polynomial Kernel (\citeauthor{zhang2007statistical}, 2007). This model will perform badly in 
situations where the data contains some seasonal effects. 
We mention in passing that although the model used in \cite{ghosh2001nonparametric}, 
($f_i(t) = h(t) + g_i(t) + \epsilon_i(t)$), 
seems different from the one stated above,
it is effectively the same given  the authors' assumption that $\sum g_{i}(t)=0$. 

\item {\bf Curve Alignment/Seasonal Effect Only}: 
The third related area is registration or alignment of functions. Here, the model 
does not have a trend component and only considers time warpings of  $g$ according to: 
\begin{equation*}
	f_i(t) = g(\gamma_i(t)) + \epsilon_i(t),\,i=1,...,n\ .
\end{equation*}
Given several observations $\{f_i\}$, the goal here is to remove the effect of 
$\{ \gamma_i\}$ and estimate $g$. The simplest case here is pairwise 
alignment, which was first studied by \cite{SaCh1978} in signal processing. 
Later on, the problem of aligning multiple functions gained substantial interest 
in the statistics community with a variety of solutions presented in
\cite{KnGa1992}, \cite{WaGa1997}, \cite{Ronn2001}, \cite{liu2004functional}, \cite{Gega2004},
\cite{ramsay2006functional}, \cite{Jame2007}, \cite{TaMu2008}, \cite{SSVV2010a}, 
\cite{srivastava2011registration,srivastava-klassen:2016}, 
\cite{kurtek2011signal}, \cite{raket2014nonlinear}, and \cite{cheng2015bayesian}. 

A large majority of these techniques formulate the alignment problem using the standard $\ltwo$ metric. 
As pointed out in \cite{Marron2015functional}, this leads to degeneracy in the form of the {\bf pinching effect}, 
and also {\bf asymmetry} in the solution. \cite{srivastava2011registration} 
(see also \cite{srivastava-klassen:2016}) presented a natural solution 
that extends the Fisher-Rao metric to general function spaces and 
uses a square-root velocity function (SRVF) representation of curves for alignment. 
This transformation is supported by a fundamental 
result that the Fisher-Rao Riemannian metric, with its nice invariance properties, 
transforms to the $\ltwo$ inner-product 
under the SRVF transformation. 
The cross-sectional mean of these aligned SRVFs results in estimation of $g$, 
see \cite{kurtek2011signal} and \cite{cleveland2016norm}.  
\end{enumerate}

In summary, very few of the past papers study the full model given in Eqn. 
\ref{eqn:general-model} and only estimate some subset of the three components of interest -- 
tread, seasonal effect, and time-warping. 
The current paper differs from this literature in its consideration of multiple curves, 
and in estimation of all three components.

%%%%%%%%%%%%%%%%%%%%%%%%%%%%%%%%%%%%%%%%%%%%%%%%%%%%%%%%%%%%%%%%%%%%%%%%%%%%%%%%%%%%%%%%%%%%%%%%%%%%%%
%%%%%%%%%%%%%%%%%%%%%%%%%%%%%%%%%%%%%%%%%%%%%%%%%%%%%%%%%%%%%%%%%%%%%%%%%%%%%%%%%%%%%%%%%%%%%%%%%%%%%%
\subsection{Our Approach} \label{Our Approach}
We take a comprehensive approach and explicitly estimate the three components -- trend, 
seasonal effect, and seasonal time warping -- using a statistical model. 
The key idea is to formulate the time-warping of the seasonal component 
in such a way that the well-known problems of {\it pinching} and {\it asymmetry} are avoided. 
This is accomplished by assuming the time warping to be $g \mapsto (g \circ \gamma) \sqrt{\dot{\gamma}}$, 
rather than the traditional $g \mapsto (g \circ \gamma)$, 
as suggested for SRVFs in \cite{srivastava2011registration} and \cite{srivastava-klassen:2016}. 
In other words, the model is posed in the SRVF space, rather than the original function space. 
This warping is norm-preserving, i.e. $\| g\| = \|(g \circ \gamma) \sqrt{\dot{\gamma}}\|$, with $\| \cdot \|$ 
denoting the $\ltwo$ norm, for any time warping function $\gamma$, 
and thus has fundamentally better mathematical and computational properties. 

Using this warping action, we formulate a statistical model where the observation is a superposition of the trend, 
the time-warped seasonal effects, and the observation noise. 
With this model, we formulate a maximum-likelihood estimation problem and solve it 
using a coordinate-descent algorithm. 
This requires a critical choice of orthogonal subspaces associated with the trend and 
(unwarped) seasonal components, and estimating coefficients of these components with respect to the respective bases. 
Furthermore, we use maximum likelihood to perform model selection in terms of subspace choices, 
and to compare different candidate models in this problem area. 
Finally, we use bootstrap to provide confidence bands around estimates of trend and seasonal effects. 
We illustrate the strength of this framework in formulating hypothesis tests associated with the estimated trends and 
seasonal effects. 

The rest of this paper is organized as follows. In Section \ref{sec: Model-Based Problem Formulation},
we formulate the trend and variable-phase seasonality estimation by a statistical model.
Section \ref{sec: Solution, Algorithm} presents a MLE solution to the model, 
followed by coordinate-descent optimization and bootstrap analysis. 
This algorithm requires a pre-determined subspace for the trend 
and Section \ref{sec: Subspace Selection} develops a rule for selecting this subspace automatically from the data.
Section \ref{sec: Experimental Results} illustrates synthetic and real data examples with a comparison between
our MLE algorithm and other models. The paper ends with a conclusion in Section \ref{sec: Conclusion}.

%%%%%%%%%%%%%%%%%%%%%%%%%%%%%%%%%%%%%%%%%%%%%%%%%%%%%%%%%%%%%%%%%%%%%%%%%%%%%%%%%
%%%%%%%%%%%%%%%%%%%%%%%%%%%%%%%%%%%%%%%%%%%%%%%%%%%%%%%%%%%%%%%%%%%%%%%%%%%%%%%%%
\section{Model-Based Problem Formulation} \label{sec: Model-Based Problem Formulation}
In order to formulate the problem of trend and seasonality estimation, under time warping of seasonal effects, 
we particularize Eqn. \ref{eqn:general-model} according to: for $t \in [0,1]$, 
\begin{equation}
f_{i}(t)  = h(t) + \left(g\circ\gamma_{i}\right)(t)\sqrt{\dot{\gamma}_{i}(t)}+\epsilon_{i}(t),\,i=1,...,n\ .
\label{eqn:our-model}
\end{equation}
For simplicity, we will assume that $\epsilon_{i}(t)$ is a white Gaussian noise process, 
with $\epsilon_i(t) \sim {\cal N}(0,\sigma^2)$ for each $t$ independently.
(The domain of all functions $\{f_{i}(t)\},$ $g(t),$ $h(t),$ and $\{\gamma_{i}(t)\}$ is $[0,1]$.)
The warping functions $\{\gamma_{i}\}$ are assumed to be elements of the set:
$$
	\Gamma=\left\{ \gamma:[0,1]\rightarrow[0,1]\,|\,\gamma(0)=0,\,\gamma(1)=1,\,\gamma\text{ is a diffeomorphism}\right\}.
$$ 
This set has been studied extensively for function and curve representation in shape and functional data analysis 
(\cite{SKJJ2011,srivastava2011registration,srivastava-klassen:2016}). 
$\Gamma$ is a group with composition as group operation
and the identity element $\gamma_{id}(t) = t$. For each $\gamma \in \Gamma$, there exists a unique 
element $\gamma^{-1}$ such that $\gamma \circ \gamma^{-1} = \gamma^{-1} \circ \gamma = \gamma_{id}$. 
We will denote the term $\left(g\circ\gamma_{i}\right)(t)\sqrt{\dot{\gamma}_{i}(t)}$ 
by $(g,\gamma_{i})(t)$ to reduce notation. 
Note that  $(g,\gamma_{i})(t)$ is the right group action of a group $\Gamma$ on $g$. 
We assume that each of the observations and components $h$ and $g$ are elements of $\ltwo([0,1],\real)$.

Having specified the model, Eqn. \ref{eqn:our-model}, we make some assumptions that ensure 
identifiability of trend and seasonality components.
\begin{itemize}

\item Consider a simpler case where  $\gamma_{i}=\gamma_{id}=t$, for all $i$ and
 the model reduces to $f_{i}(t) = h(t)+g(t) + \epsilon_{i}(t)$.   
In order to identify $h$ and $g$, we require two subspaces ${\cal H}, {\cal G}  \subset \ltwo$, 
such that ${\cal H} \perp {\cal G}$, and then we restrict to $h \in {\cal H}$ and $g \in {\cal G}$. 
Assuming, as earlier, that  $E[\epsilon_{i}(t)] = 0$, 
for all $i$ and $t$,  the estimates of $h$ and $g$ are given by 
$\hat{h} = \Pi_{\cal H}({\frac{1}{n}}\sum_{i=1}^n f_i)$ and $\hat{g} = \Pi_{\cal G}({\frac{1}{n}} \sum_{i=1}^n f_i)$, 
where $\Pi_{\cal H}$ and  $\Pi_{\cal G}$ are the projections onto ${\cal H}$ and ${\cal G}$, respectively. 
This, of course, requires the knowledge of ${\cal H}$ and ${\cal G}$ beforehand. 

\item Now consider the case where $\gamma_{i}$s are not identity. 
In this case, the functions $(g,\gamma_{i})$ are no longer guaranteed to be in the subspace ${\cal G}$, i.e.
$(g,\gamma_{i})$s may have nonzero components in ${\cal H}$. 
This underlines the {\bf main challenge in the trend and variable-phase seasonality estimation}. 
If $(g,\gamma_{i})$ had remained orthogonal to ${\cal H}$, then we could solve the problem using orthogonal 
projections (or some related smoothing methods). Note that the 
warping by itself can be handled using the alignment procedures developed 
in \cite{srivastava2011registration} and \cite{srivastava-klassen:2016}. 
However,  since $(g,\gamma_i)$ may not be guaranteed to be orthogonal to ${\cal H}$, 
a more sophisticated approach is required to perform the separation.

\item Another issue here is a lack of identification of $g$, due to its time warping. Since 
$(g, \gamma_i)  = (g, \gamma_0 \circ \gamma_0^{-1} \circ \gamma_i)$ 
$= ((g, \gamma_0), \gamma_0^{-1} \circ \gamma_i) = (\tilde{g}, \tilde{\gamma}_i)$, 
for any $\gamma_0  \in \Gamma$, there is a problem in representing $g$ uniquely. 
This problem can be avoided by assuming an additional constraint on $\{ \gamma_i\}$.  
\cite{srivastava2011registration} suggested forcing the Karcher mean of  the inverse warping functions 
to be the identity, or  $KM\left\{ \gamma_{i}^{-1}\right\} =\gamma_{id}$. 
(The concept of Karcher mean on $\Gamma$ is discussed later in Section \ref{sec: Optimization}.) 
The similar idea of constraining the mean of warping functions to be $\gamma_{id}$ was also used in \cite{TaMu2008}. 
\end{itemize}

%%%%%%%%%%%%%%%%%%%%%%%%%%%%%%%%%%%%%%%%%%%%%%%%%%%%%%%%%%%%%%%%%%%%%
%%%%%%%%%%%%%%%%%%%%%%%%%%%%%%%%%%%%%%%%%%%%%%%%%%%%%%%%%%%%%%%%%%%%%
\section{Maximum Likelihood Solution}\label{sec: Solution, Algorithm}

Given observations $\{f_i\}$, and the model stated in Eqn. \ref{eqn:general-model}, 
our goal is to recover the seasonality $g$, warping functions $\{\gamma_i\}$, and the trend $h$, 
under the assumptions and constraints stated earlier. We will use the maximum-likelihood approach 
for solving this problem. 

%%%%%%%%%%%%%%%%%%%%%%%%%%%%%%%%%%%%%%%%%%%%%%%%%%%%%%%%%%%%%%%%%%%%%%%%%%%%%%%%%%%
\subsection{Maximum Likelihood Formulation} \label{sec: Maximum Likelihood Estimation}

To develop a formal estimation setup, 
start by setting  $f_i^o(t) =h(t)+(g,\gamma_{i})(t)$. 
Let $\{t_i, t_2, \dots, t_m\}$ denote a finite partition of $[0,1]$ representing the observation times.
The discrete time samples follow the model  $f_i(t_j)  =f_i^o(t_j) +\epsilon_{i}(t_j),\,i=1,...,n,\,j=1,...,m$. 
Assuming $\epsilon_{i}(t_j) \sim {\cal N}(0,\sigma^2)$ for each $i$ and $j$, 
the conditional distribution of $f_i(t_j)$ given $f^o_i(t_j)$ is ${\cal N}(f_i^o(t_j),\sigma^2)$. 
Since $f^o$ is determined by $h$ and $\{(g, \gamma_i)\}$, 
the average log-likelihood of these components, given the observations $\{f_{i}(t_j)\}$,  is given by:
\begin{equation*}
	 - \log(2 \pi \sigma^2) - 
	{\frac{1}{2 mn \sigma^2}} \sum_{i=1}^n  \sum_{j=1}^m \left( f_{i}(t_j)-f_{i}^o(t_j) \right)^2.
\label{eq: huge MLE equation}
\end{equation*}
Maximizing the above term is the same as minimizing
\begin{equation}
	 {\frac{1}{mn}}  \sum_{i=1}^n  \sum_{j=1}^m \left( f_{i}(t_j)-f_{i}^o(t_j) \right)^2\ .
	 \label{eq: portion of MLE equation}
\end{equation}
In the limit,  $m \rightarrow \infty$, the above term becomes
\begin{equation*}
	{\frac{1}{n}} \lim_{m\rightarrow\infty} \sum_{i=1}^n  
		\left( \frac{1}{m} \sum_{j=1}^m \left( f_{i}(t_j)-f_{i}^o(t_j) \right)^2  \right)
	= {\frac{1}{n}}  \sum_{i=1}^n   \left( \int_0^1 \left( f_{i}(t)-f_{i}^o(t) \right)^2 dt \right)
	= {\frac{1}{n}} \sum_{i=1}^n \| f_{i}-h-(g,\gamma_{i})\|^2. 
\end{equation*}
Thus, finding MLE becomes a problem in constrained functional minimization: 
\begin{equation}
    (\hat{g},\{\hat{\gamma}_{i}\},\hat{h}) = \arg \inf_{g,\{\gamma_{i}\},h} C(g,\{\gamma_{i}\},h)
\label{eq: final cost function}
\end{equation}
where $C: {\cal G} \times \Gamma^{n} \times {\cal H} \rightarrow\mathbb{R}$ is given by $C(g,\{\gamma_{i}\},h)
   =\frac{1}{n}{\sum_{i=1}^{n}}\left\Vert f_{i} -h -(g,\gamma_{i}) \right\Vert^2$.

We reiterate the assumptions that  ${\cal H} \perp {\cal G}$ 
and Karcher Mean$(\{\gamma_{i}^{-1}\})=\gamma_{id}$ are associated with Eqn. \ref{eq: final cost function}. 
Of course, the optimization stated above requires the knowledge of ${\cal H}$ and ${\cal G}$. 
We will simplify a little bit by assuming that 
${\cal G} = {\cal H}^{\perp}$ and, thus, the two choices of $\cal H$ and $\cal G$ are unified into one.  
If the subspace ${\cal H}$ is spanned by an orthonormal basis, 
then the choice of ${\cal H}$ is same as the choice of its basis. 
Let $\Phi = \{\phi_k, k = 1, \dots, \infty\}$ denote a complete orthonormal basis of $\ltwo$. 
We can choose any subset of $\Phi$ and set ${\cal H}$ to be its span.  
Given $\Phi$, we will assume that ${\cal H} = \mbox{span}\{ \phi_k | k = 1, \dots, l\}$ 
for some positive integer $1 \leq l < \infty$. 
This choice is motivated by the fact that trend is often a slowly varying function over time 
and first few basis elements should suffice to estimate $h$. 
Thus, the choice of ${\cal H}$ boils down to the finding an appropriate $l$. 

%%%%%%%%%%%%%%%%%%%%%%%%%%%%%%%%%%%%%%%%%%%%%%%%%%%%%%%%%%%%%%%%%%%%%%%%%%%%%%%%%
\subsection{Optimization Using Coordinate-Descent} \label{sec: Optimization}
We will use a coordinate-descent method for solving Eqn. \ref{eq: final cost function}. 
This optimizes $C$ along one direction/variable at a time, and iterates 
until we reach a stationary point. 
To apply the coordinate-descent method to our problem, we need to derive each 
of the following items: 
%%%%%%%%%%%%%%%%%%%%%%%%%%%%%%%%%%%%%%%%%%%%%%%%%%%%%%%%%
\begin{enumerate}
\item \textbf{Update trend $h$:} Given the current estimates 
 $\hat g \in \cal G$ and $\{\hat \gamma_{i}\} \in \Gamma$, the estimate for $h$ is as follows: 
\begin{equation}
	\hat{h} = \Pi_{\cal H}\left[\frac{1}{n}{\sum_{i=1}^{n}}\left(f_{i}-(\hat g,\hat \gamma_{i})\right)\right] 
	= {\sum^{l}_{k=1}} \left\langle \frac{1}{n}{\sum_{i=1}^{n}}
	  \left(f_{i}-(\hat g,\hat \gamma_{i})\right),\phi_{k}\right\rangle \phi_k
\label{eq: minimizer-h}
\end{equation}
where $\{\phi_{1}(t),...,\phi_{l}(t)\}$ is an orthogonal basis of ${\cal H}$ and $\langle\cdot,\cdot\rangle$ 
denotes the standard $\ltwo$ inner product. We consider the following bases in this paper: 
Fourier basis $\left\{ 1, \sqrt{2}\sin\left(2n\pi t\right), \sqrt{2}\cos\left(2n\pi t\right),\,n=1,2,3,...\right\}$, 
sine basis $\left\{ \sqrt{2}\sin\left(n\pi t\right),\,n=1,2,3,...\right\}$, 
cosine basis $\left\{ 1, \sqrt{2}\cos\left(n\pi t\right),\,n=1,2,3,...\right\}$, 
and shifted Legendre basis (see \cite{kreyszig1989introductory}) 
\[ 
	\left\{ {\phi}_{k}(t)=\frac{1}{2k-1}(-1)^{k-1}{\sum^{k-1}_{j=0}}\begin{pmatrix}k-1\\j\end{pmatrix}
	     \begin{pmatrix}k+j-1\\j\end{pmatrix} (-t)^{j}, k=1,2,3...\right\}.
\]

%%%%%%%%%%%%%%%%%%%%%%%%%%%%%%%%%%%%%%%%%%%%%%%%%%%%%%%%%%
\item \textbf{Update seasonality $g$:} Given $\{\hat \gamma_{i}\}\in\Gamma$ and $\hat h\in {\cal H}$,
the update for $g$ is defined as follows: 
\begin{equation*}
   \hat g =\underset{g\in {\cal H}^{\perp}}{\arg\min} \left( \frac{1}{n}{\sum^{n}_{i=1}}
	    \left\Vert f_{i}-\hat h-(g,\hat \gamma_{i})\right\Vert^{2}  \right)
		  =\underset{g\in {\cal H}^{\perp}}{\arg\min} \left( \frac{1}{n}{\sum^{n}_{i=1}}
			  \left\Vert \left(\left(f_{i}-\hat h\right),\hat \gamma_{i}^{-1}\right)-g\right\Vert^{2} \right).
\label{eq: cost functional of g}
\end{equation*}
The last equality uses the fact that 
$\left\Vert f_{1}-f_{2}\right\Vert=\left\Vert \left(f_{1},\gamma\right)-\left(f_{2},\gamma\right)\right\Vert$, 
see \cite{srivastava2011registration}. 
Hence, the optimization problem is simplified into a vector space optimization under subspace 
${\cal H}^\perp \in \ltwo$ and the minimizer for $g$ is
\begin{equation}
	\hat g
	 =\frac{1}{n}{\sum^{n}_{i=1}}\left((f_{i}-\hat h),\hat \gamma_{i}^{-1}\right)
	    -\Pi_{\cal H}\frac{1}{n}{\sum^{n}_{i=1}}\left((f_{i}-\hat h),\hat \gamma_{i}^{-1}\right) 
	 =\frac{1}{n}{\sum^{n}_{i=1}}\left((f_{i}-\hat h),\hat \gamma_{i}^{-1}\right)
	   -{\sum^{l}_{k=1}}{\left\langle \frac{1}{n}{\sum^{n}_{i=1}}\left((f_{i}-\hat h),
		  \hat \gamma_{i}^{-1}\right),\phi_{k}\right\rangle} \phi_k. 
\label{eq: minimizer-g}
\end{equation}

%%%%%%%%%%%%%%%%%%%%%%%%%%%%%%%%%%%%%%%%%%%%%%%%%%%%%%%%%%%%%%%%%%%
\item \textbf{Update warping functions $\{\gamma_i\}$:} 
Given the estimates of $h$ and $g$, this problem can be rephrased as: 
\begin{equation}
   \left\{ \hat{\gamma}_{1},...,\hat{\gamma}_{n}\right\} 
	 =\underset{\left\{ \gamma_{1},...,\gamma_{n}\right\} \in\Gamma^{n}}{\arg\inf}\left(
	 \frac{1}{n}{\sum^{n}_{i=1}}\left\Vert f_{i}-\hat h-(\hat g,\gamma_{i}))\right\Vert^{2} \right)
\label{eq: Cost Function; update gamma}
\end{equation}
with Karcher mean constraint $KM\left\{ \gamma_{i}^{-1}\right\} =\gamma_{id}.$ Notice that $\{\gamma_{i}\}$ 
are independent of each other in their contributions 
to the cost function in Eqn. \ref{eq: Cost Function; update gamma}. 
Therefore, we solve the optimization problem as an unconstrained one and then impose the Karcher mean constraint. 
For each $i$, we solve for: 
\[
	\left\{ \check\gamma_{i}\right\}=\underset{\gamma_{i}\in\Gamma}{\arg\inf} \left(
	 \left\Vert \left[f_{i}-\hat h\right]-(\hat g,\gamma_{i})\right\Vert^{2} \right).
\] 
This can be solved using Dynamic Programming technique (\citeauthor{bellman1954theory}, 1954). 
The computation of the Karcher mean of warping functions $\{\gamma_i\}$, 
under the Fisher-Rao metric, has been provided in Algorithm 1 of the paper \cite{srivastava2011registration} and
 Section 7.5 of the textbook \cite{srivastava-klassen:2016}. 
By using Lemma 4 in \cite{srivastava2011registration}, the mean constraint 
$KM\left\{ \gamma_{i}^{-1}\right\} =\gamma_{id}$ is imposed by setting 
$\gamma_{i}=\left[\check\gamma_{i}^{-1}\circ\left(\check\gamma_{i}^{-1}\right)_{KM}^{-1}\right]^{-1}$
$=\left(\check\gamma_{i}^{-1}\right)_{KM}\circ\check\gamma_{i}.$ 
The procedure for solving the constrained functional optimization (Eqn. \ref{eq: Cost Function; update gamma}) 
with the condition $KM\left\{ \gamma_{i}^{-1}\right\} =\gamma_{id}$ is summarized in 
Algorithm \ref{alg: optimization over warping functions}. 

\begin{algorithm}[!htb]
\SetAlgoLined
	\KwData{observations $\{f_i\}$, seasonality $g^{(j-1)}$ and trend $h^{(j-1)}$. 
	        Requires Algorithm 1 in \cite{srivastava2011registration}} 
	\KwResult{warping functions $\gamma_{1}^{(j)},...,\gamma_{n}^{(j)}$ with $KM\{\gamma_{i}^{-1}\}=\gamma_{id}.$} 

	\For{$i=1,...,n$}{
		Use Dynamic Programming to solve 
		$\check{\gamma}_{i}^{(j)}=\underset{\gamma_{i}\in\Gamma}{\arg\min}
		   \left\Vert \left[f_{i}-h^{(j-1)}\right]-(g^{(j-1)},\gamma_{i})\right\Vert^{2}$ \;
		
		Compute $\left(\check{\gamma}_{i}^{(j)}\right)^{-1}$ \;
	}
	Use Algorithm 1 in \cite{srivastava2011registration} to compute
	Karcher mean of $\left(\check{\gamma}_{1}^{(j)}\right)^{-1},...,\left(\check{\gamma}_{n}^{(j)}\right)^{-1},$
	denoted by $\left(\check{\gamma}_{i}^{(j)}\right)_{KM}^{-1}$ \;
	For {$i=1,...,n$}, update 
		$\gamma_{i}^{(j)}=\left(\check{\gamma}_{i}^{(j)}\right)_{KM}^{-1}\circ\check{\gamma}_{i}^{(j)}$ \;	
\caption{Optimization over $\gamma_{1},...,\gamma_{n}$}
\label{alg: optimization over warping functions}
\end{algorithm}

Given updates for variables $h$, $g$, and $\{\gamma_i\}$ in 
 Eqn. \ref{eq: minimizer-h}, Eqn. \ref{eq: minimizer-g}, 
and Algorithm \ref{alg: optimization over warping functions}, 
 the coordinate-descent method yields Algorithm \ref{alg: CD for Signal Separation}.

\begin{algorithm}[!htb]
\SetAlgoLined
	\KwData{observations $\{f_{i}\}$. Requires Algorithm \ref{alg: optimization over warping functions}
            (which in turn requires Algorithm 1 in \cite{srivastava2011registration}) } 
	\KwResult{$\{\hat{\gamma}_{i}\},$ $\hat{g},$ $\hat{h}$}
	
	Initialization $g^{(0)}=f_{\tilde i}$ where $\tilde i=\underset{i}{\arg\min}
	          \left\Vert f_{i}-\bar{f}\right\Vert$ , $h^{(0)}=0$, $\{\gamma_{i}^{(0)}\}=\gamma_{id}$\;

	\For{$j=1,...,max$}{
		Update $\gamma_{1}^{(j)},...,\gamma_{n}^{(j)}$ with 
		    Karcher mean condition using Algorithm \ref{alg: optimization over warping functions} \;
		
		Compute $\bar{f}^{(j)}=\frac{1}{n}{\sum^{n}_{i=1}}
		      \left(\left(f_{i}-h^{(j-1)}\right),\left(\gamma_{i}^{(j)}\right)^{-1}\right)$ 
					and $g^{(j)}=\bar{f}^{(j)}-\Pi_{\cal H}\bar{f}^{(j)}$ \;
		
		Compute $\check{f}^{(j)}=\frac{1}{n}{\sum^{n}_{i=1}}
		    \left[f_{i}-(g^{(j)},\gamma_{i}^{(j)})\right]$ and $h^{(j)}=\Pi_{\cal H}\check{f}^{(j)}$ \;
	}
	Output $\hat{\gamma}_{i}=\gamma_{i}^{(Max)},$ $\hat{g}=g^{(Max)},$ and $\hat{h}=h^{(Max)}$
	
\caption{Coordinate-descent optimization for the trend and seasonality estimation problem} 
\label{alg: CD for Signal Separation}
\end{algorithm}
\end{enumerate}

\begin{figure}[!htb]
\begin{center}
   \subfloat[observations $\{f_i\}$]
	      {\includegraphics[scale=0.18]{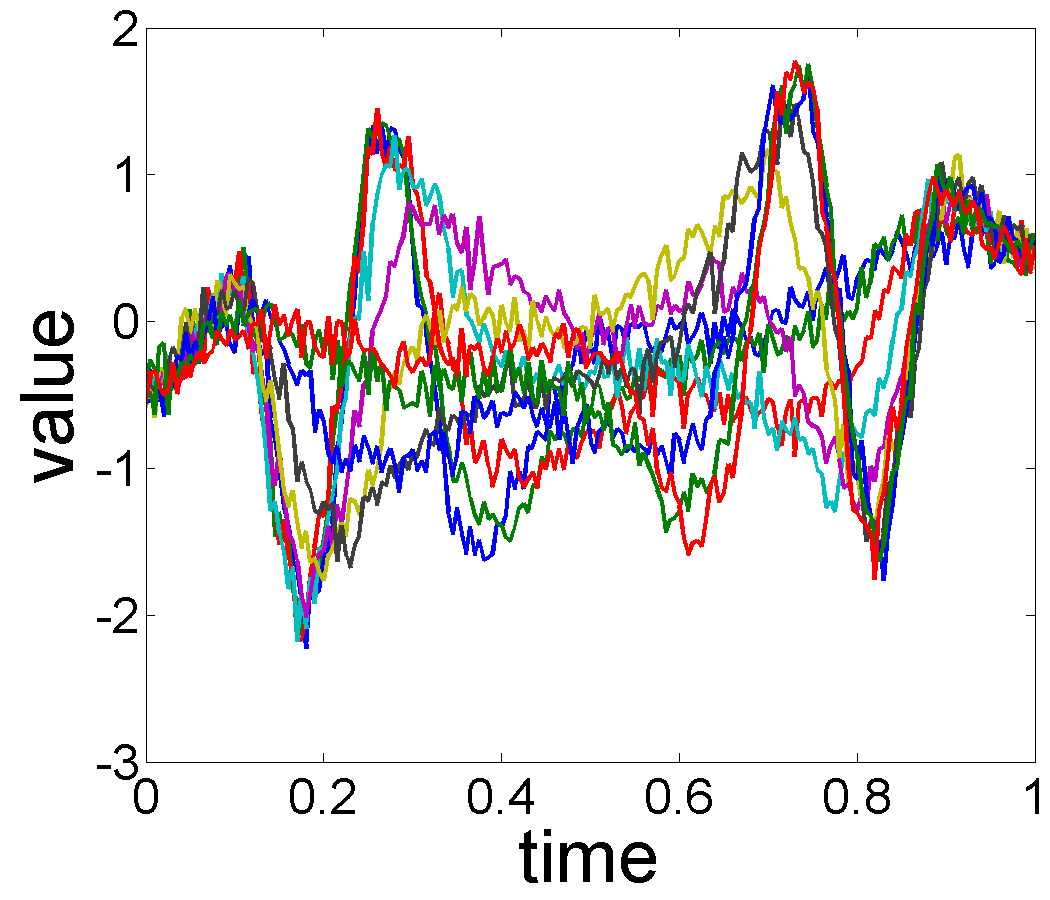}}
   \subfloat[recovered trend $\hat h$]
	      {\includegraphics[scale=0.18]{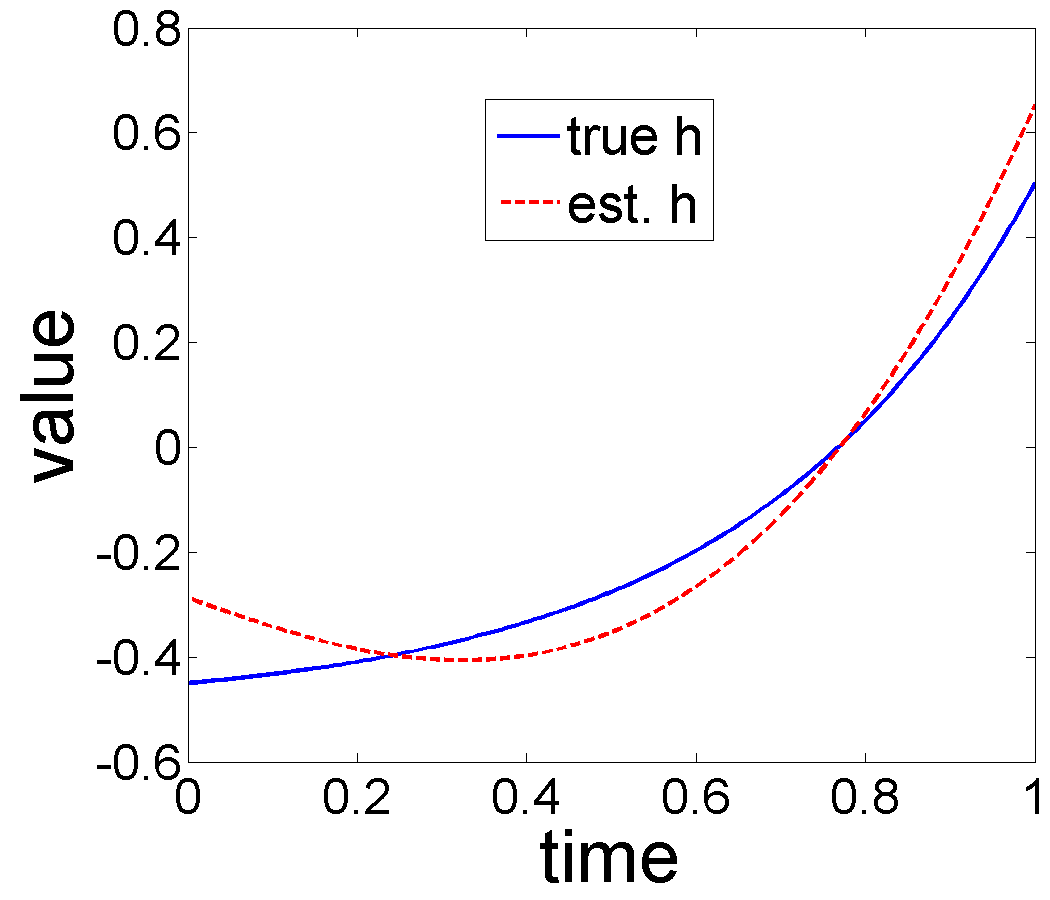}} 
   \subfloat[recovered seasonality $\hat g$]
	      {\includegraphics[scale=0.18]{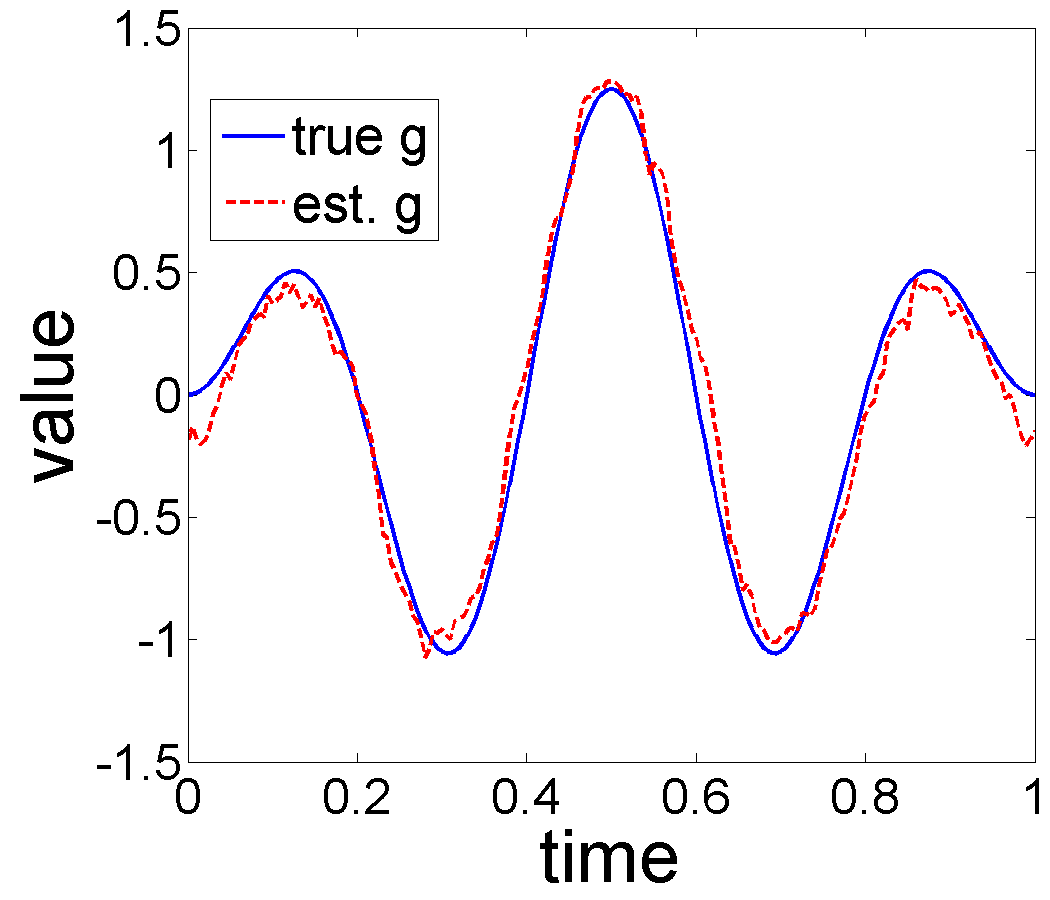}} 
   \subfloat[recovered $\{\hat \gamma_i\}$]
	      {\includegraphics[scale=0.18]{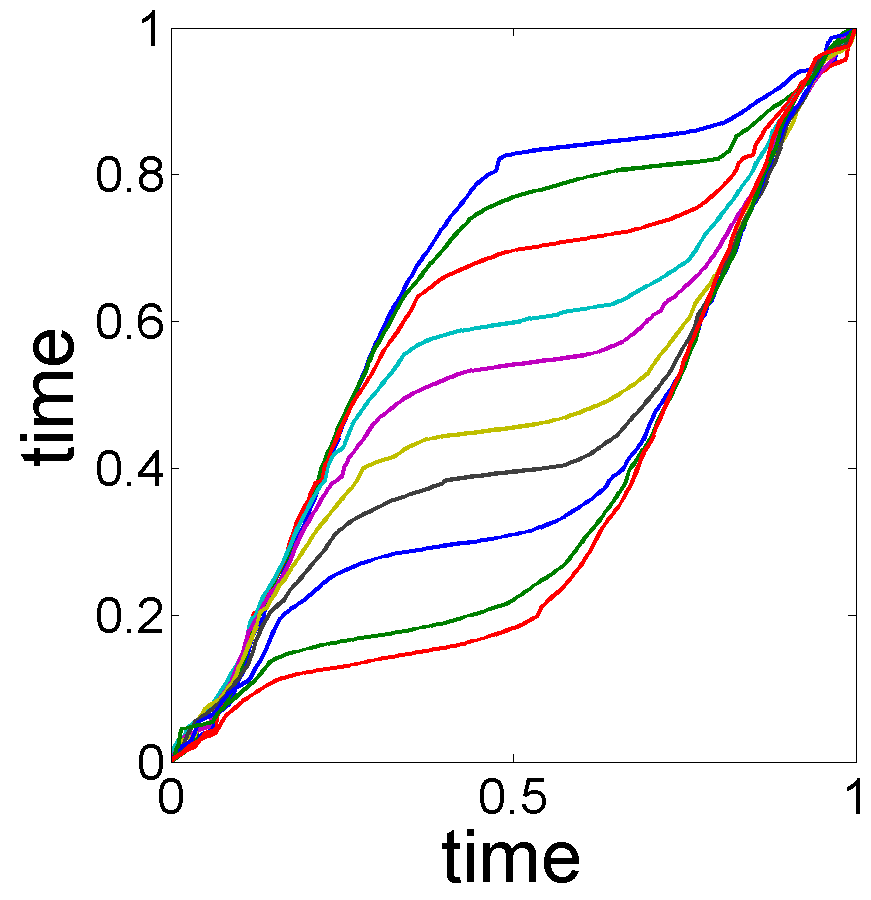}} 
   \subfloat[iterations of Eqn. \ref{eq: final cost function}]
	      {\includegraphics[scale=0.18]{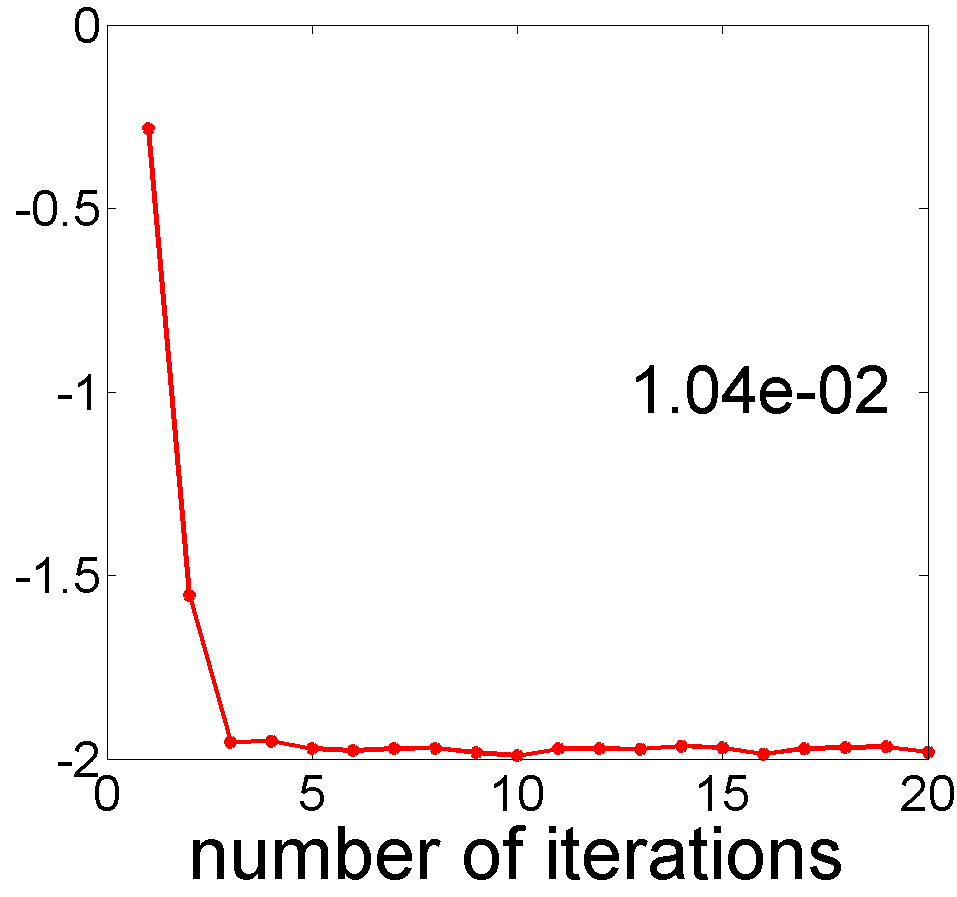}} \\ 
   \subfloat[observations $\{f_i\}$]
	      {\includegraphics[scale=0.18]{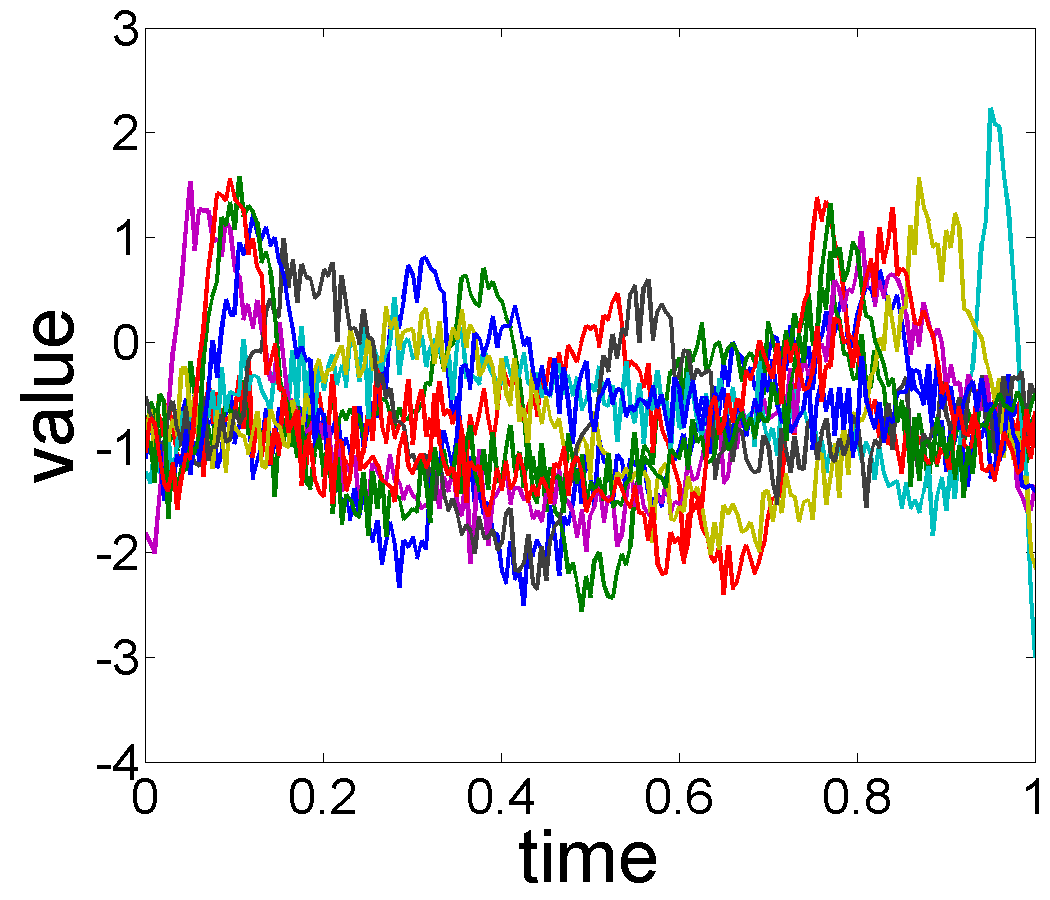}}
   \subfloat[recovered trend $\hat h$]
	      {\includegraphics[scale=0.18]{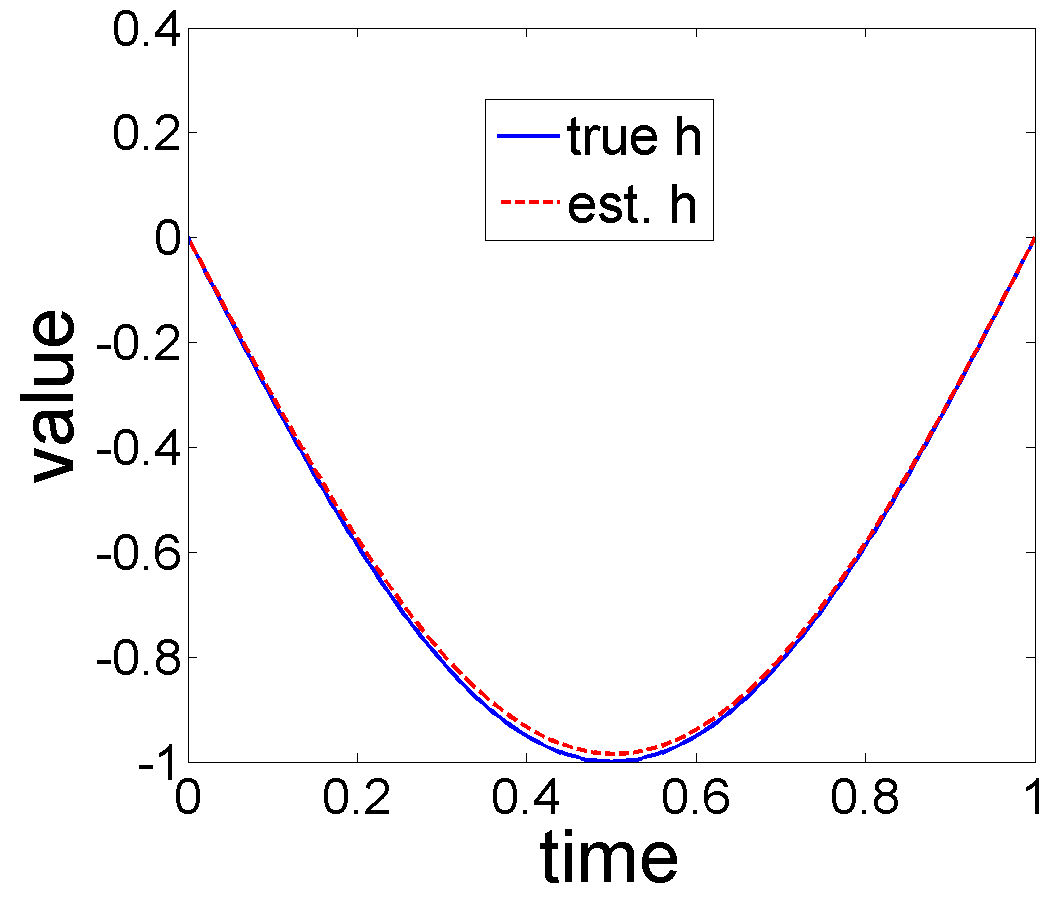}} 
   \subfloat[recovered seasonality $\hat g$]
	      {\includegraphics[scale=0.18]{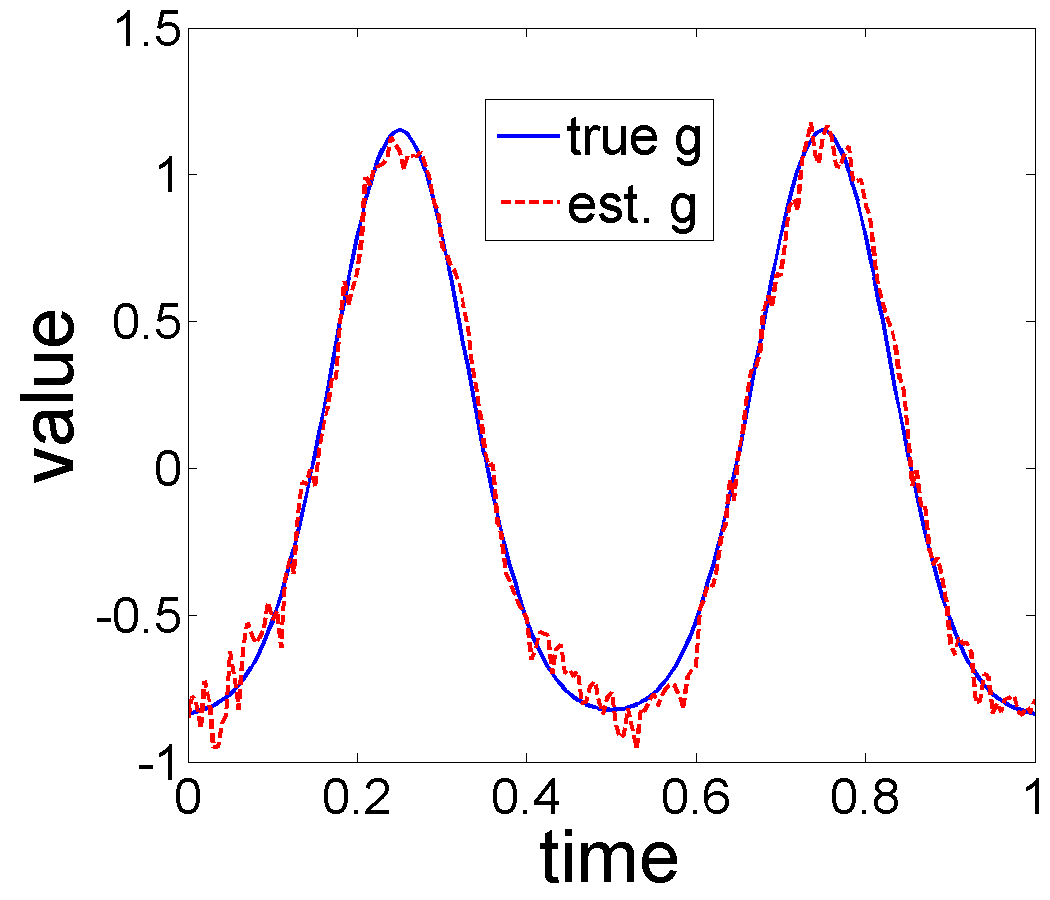}} 
   \subfloat[recovered $\{\hat \gamma_i\}$]
	      {\includegraphics[scale=0.18]{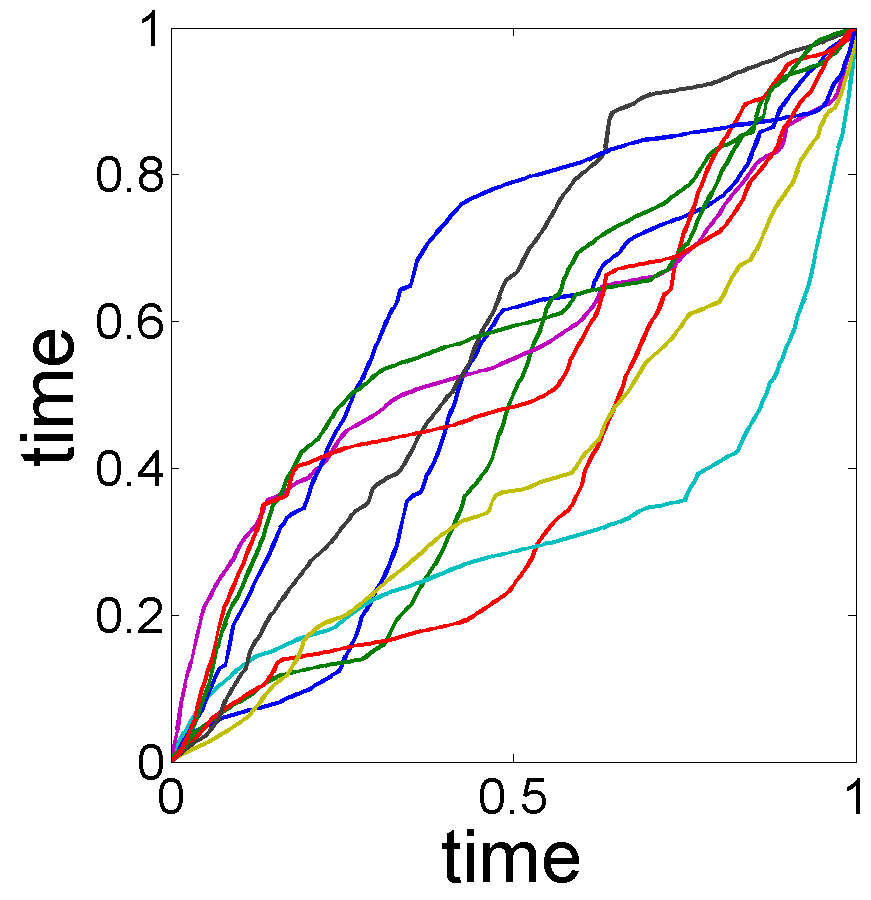}} 
   \subfloat[iterations of Eqn. \ref{eq: final cost function}]
	      {\includegraphics[scale=0.18]{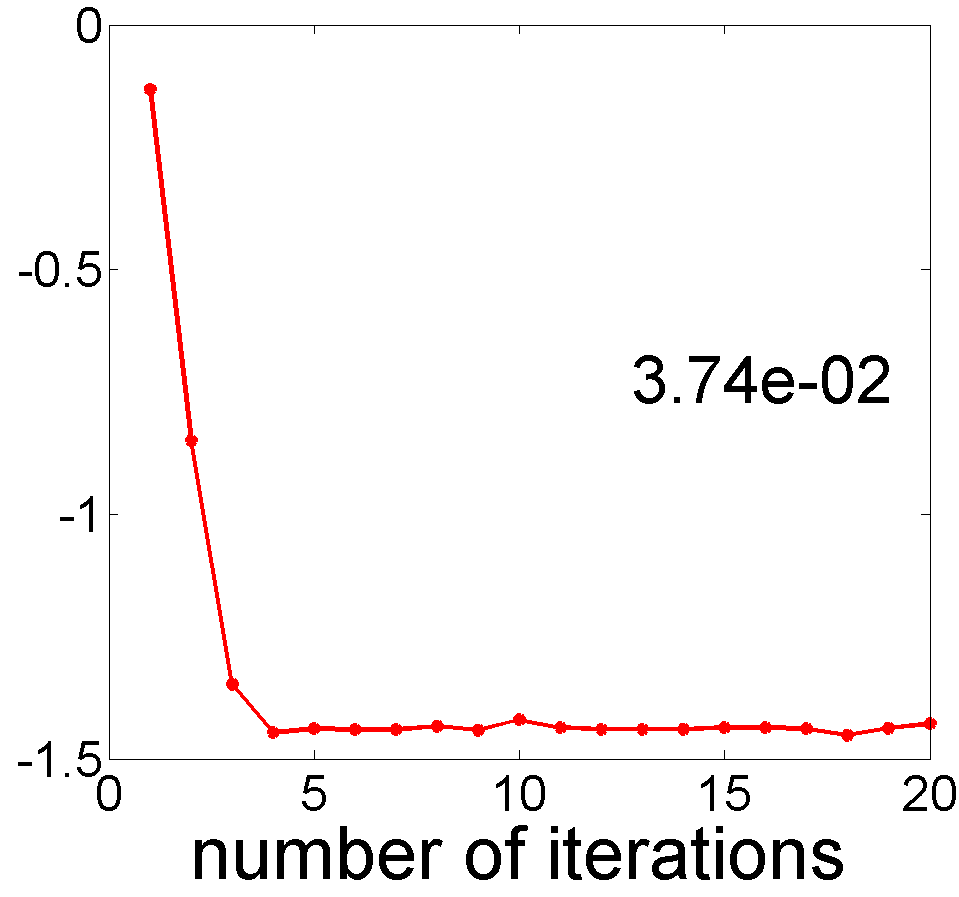}}  

\caption{Two illustrations of Algorithm \ref{alg: CD for Signal Separation}. 
				Note that the function $C$ in Eqn. \ref{eq: final cost function},
         negative log-likelihood, is plotted in a log scale in (e) and (j). 
					Values inside the rightmost panels are minimized negative log-likelihood.} 
\label{fig: synthetic illustration}
\end{center}
\end{figure}

Fig. \ref{fig: synthetic illustration} shows two examples of using Algorithm \ref{alg: CD for Signal Separation}
on simulated data. In the top case the estimated trend is monotonic increasing while in the bottom case
it first decreases and then increases.

%%%%%%%%%%%%%%%%%%%%%%%%%%%%%%%%%%%%%%%%%%%%%%%%%%%%%%%%%%%%%%%%%%%%%%%%%%%%%%%%%
\subsection{Bootstrap Analysis: Confidence Regions and Testing} \label{sec: Bootstrap}

Given the complexity of the data model, and the subsequent estimators, it is difficult to derive analytical 
expressions for asymptotic distributions of the estimated quantities. Therefore, we take a bootstrap approach
and compute estimator statistics using random replication, and use these statistics
for testing hypotheses about trend and seasonality. In the following, we present 
details for computing bootstrap estimates of standard deviations of certain test statistics. 
These statistics, in turn, can be used to test important hypotheses, such as the presence or absence 
of a trend or a seasonality component in the observed data. \\

\noindent  \textbf{Hypothesis Testing for Trend and Seasonality}: 
An important and challenging problem in functional data analysis 
is to test the presence of a trend in the given data. 
That is, given a set of functions $\{f_i,i=1,...,n\}$, we can pose the question: Is $h = 0$, or not?
This leads to a formal binary hypothesis test, with null hypothesis
$H_0$:  $h = 0$  and the alternative hypothesis
$H_1$:  $h \neq 0$. In view of the implicit assumptions about continuity of $h$ 
(due to the condition that $h \in {\cal H}$, a subspace of smooth functions), we have that 
$h = 0$ is equivalent to $\| h\| = 0$.  Therefore, we define a test statistic
$\rho_{h_0} = \| \hat{h} \|$ and rewrite the hypothesis test as:  
null hypothesis
$H_0$:  $\rho_{h_0} = 0$  and the alternative hypothesis
$H_1$:  $\rho_{h_0} > 0$. 
Let $\hat{h}_b$ denote the bootstrap replicate of the estimator $\hat{h}$, and let ${\rho}_{h_0, b}$ 
denote its $\ltwo$ norm. Furthermore,  let $\hat{se}_B$ be the standard error of ${\rho}_{h_0, b}$ using 
$B$ replicates. 
Then, we can compute the $p$ value of the test statistic assuming a normal distribution 
${\cal{N}}(0,\hat{se}_B)$ under the null hypothesis.

In fact, one can use the bootstrap procedure to test any specific shape pattern 
of the trend and seasonality function. For instance, one can test the trend function for being
constant, linear, or monomial of certain order. 
As an example, we can test if the trend $h$ is a constant function by modifying the test statistic to be 
${\rho}_{h_c}=\left\Vert \hat{h}-\int_{0}^{1}\hat{h}\,dt\right\Vert$.
Note that there are other choices possible for the test statistic in this case (for example 
${\rho}_{h_c} =  \| \dot{h}\|$) but we have chosen one arbitrarily here. 
A test statistic for testing the linearity of the trend $h$ is 
$\rho_{h_l}= \left\Vert \dot{\hat h} - \int_{0}^{1} \dot{\hat h}\,dt \right\Vert$. \\

\noindent  \textbf{Bootstrap Cross-Sectional Confidence Band:} 
In addition to providing point estimates of $h$ and $g$ 
in their respective subspaces, one can use bootstrap to provide a confidence region associated with these
estimates. The basic idea is to take bootstrap replicates of the estimator 
and use the $\ltwo$ norm to build confidence 
regions around the estimate, for either $h$ and $g$. Since under the $\ltwo$ metric, the mean of functions
corresponds to a cross-sectional mean, this task simplifies to building a confidence interval at each time $t$. 
We use the bootstrap replicates to arrive at these confidence intervals. 
Let  $\bar{h}$ and $\bar{g}$ be the  bootstrap averages, and $\hat{se}_h$ and $\hat{se}_g$ be the
bootstrap estimates of the standard errors, as functions of $t$, of $\hat{h}$ and $\hat{g}$, respectively.   
For a significance level $\alpha$, 
the confidence interval for $\hat{h}(t)$ is simply $[\bar{h}(t) \pm z^{1-\alpha/2}\cdot \hat{se}_h(t)]$,
 where $z^{1-\alpha/2}$ is the $100\cdot(1-\alpha/2)$th percentile point of a standard normal distribution. 
Similarly, the confidence interval for the estimated 
seasonal effect $\hat{g}(t)$ is simply $[\bar{g}(t) \pm z^{1-\alpha/2}\cdot \hat{se}_g(t)]$.
Examples of bootstrap-based analysis are shown later in this paper.

%%%%%%%%%%%%%%%%%%%%%%%%%%%%%%%%%%%%%%%%%%%%%%%%%%%%%%%%%%%%%%%%%%%%%%
%%%%%%%%%%%%%%%%%%%%%%%%%%%%%%%%%%%%%%%%%%%%%%%%%%%%%%%%%%%%%%%%%%%%%%
\section{Trend Subspace Selection} \label{sec: Subspace Selection}

So far in this framework we have assumed that ${\cal H}$, the subspace of $\ltwo$ associated with the trend, is 
known. Since ${\cal G} = {\cal H}^{\perp}$ need not choose $\cal G$ separately, we only need to choose $\cal H$.
The next question is:  
How to infer the subspace ${\cal H}$  automatically from the data? It turns out that the 
current framework also provides a criterion for choosing between potential candidates, by simply 
maximizing the likelihood under each candidate subspace 
and selecting the one that results in the highest maximized-likelihood. 
Earlier we assumed that ${\cal H} = \mbox{span}\{ \phi_k | k = 1, \dots, l\}$ 
for some positive integer $1 \leq l < \infty$. 
Thus, the choice of ${\cal H}$ boils down to the finding an appropriate $l$. 
With this setting,  we can try each potential value of $l$, up to a certain large value, 
maximize the likelihood under each choice of $l$, and selecting the one with the highest value of the likelihood 
(or, correspondingly the smallest value of negative log-likelihood in Eqn. \ref{eq: final cost function}). 

\begin{remark} \label{remark of separation model}
We point out that this approach of selecting ${\cal H}$ will not work if we ignore the phase variability in the 
seasonal component, or set $\gamma_i = \gamma_{id}$ for all $i$. 
For instance, we assume the model $f_i(t) = h(t) + g(t) + \epsilon_i(t)$, and
use the natural estimators $\Pi_{\cal H}(\bar{f})$ and $\Pi_{{\cal H}^{\perp}}(\bar{f})$ for $h$ and $g$, 
respectively, then the negative log-likelihood will remain unchanged with the changes in ${\cal H}$.  
Subspace selection using log-likelihood will work only when we have non-trivial warpings.   
\end{remark}

\begin{figure}[!htb]
\begin{center}
   \subfloat[true trend]
	     {\includegraphics[scale=0.22]{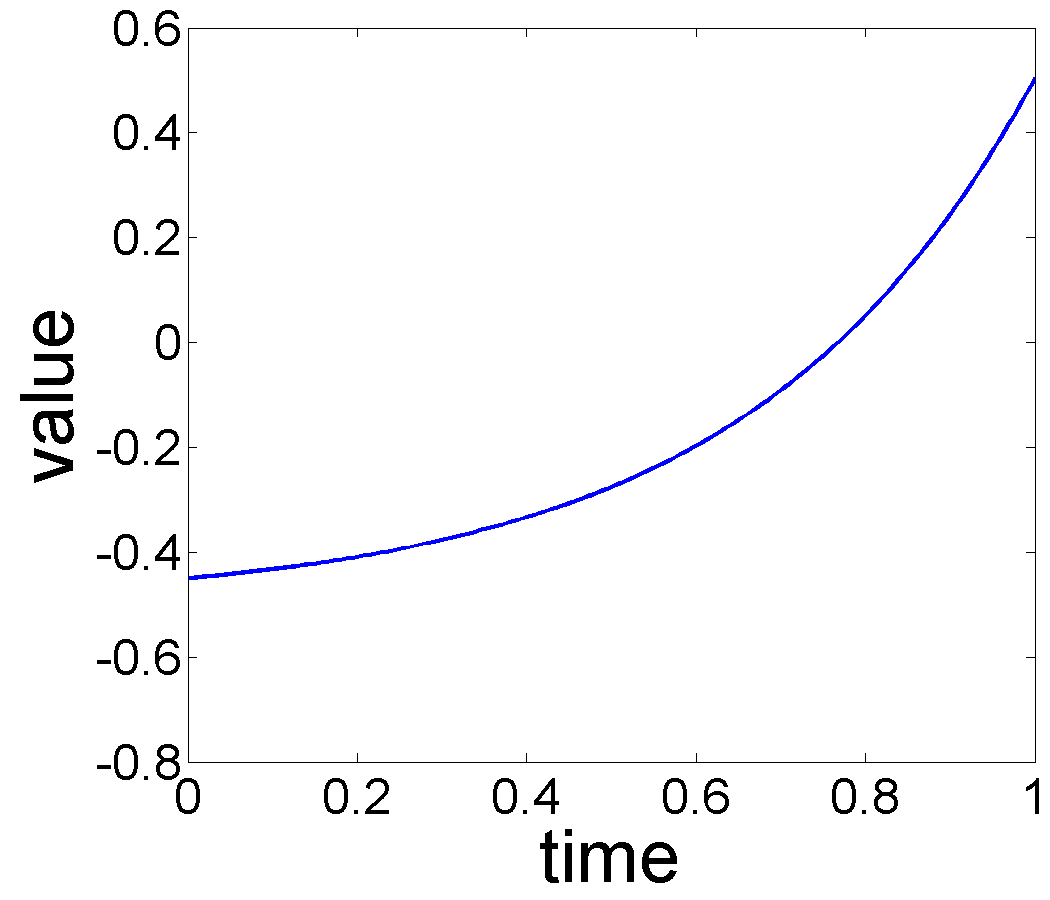}}
   \subfloat[true seasonality]
	     {\includegraphics[scale=0.22]{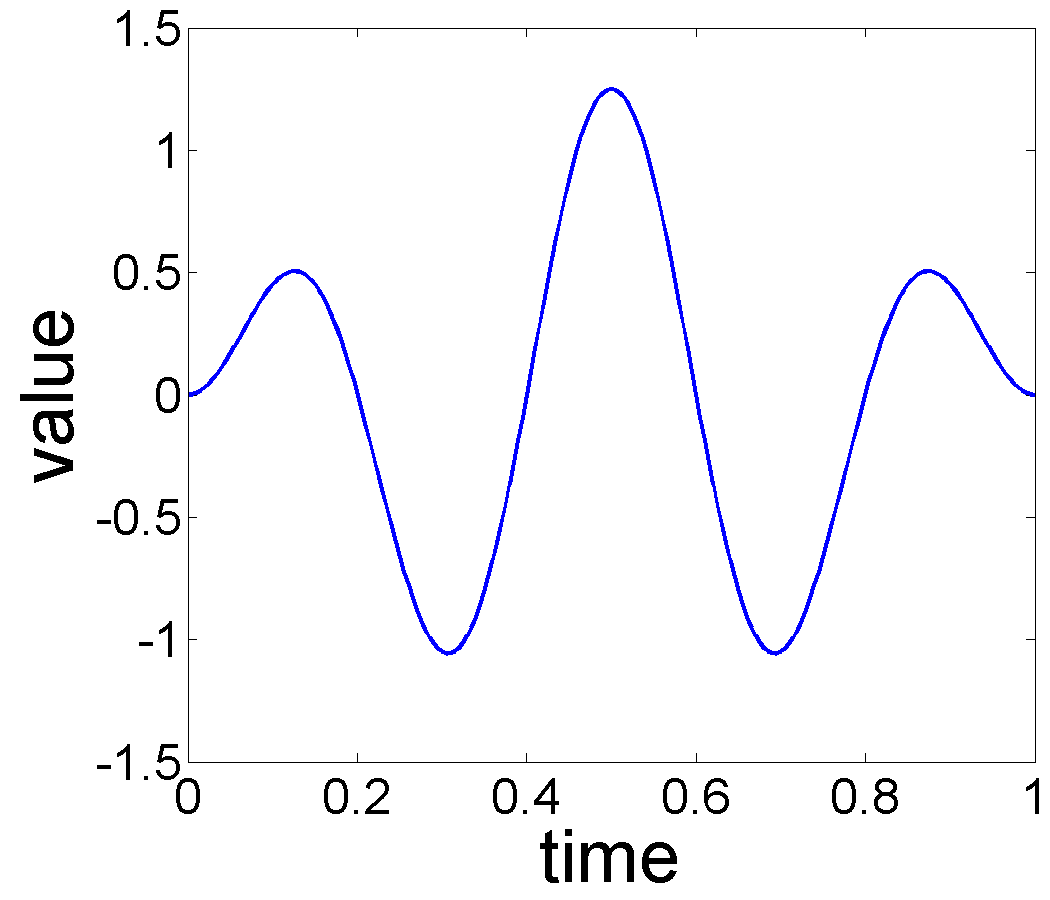}}
   \subfloat[warping functions]
	     {\includegraphics[scale=0.22]{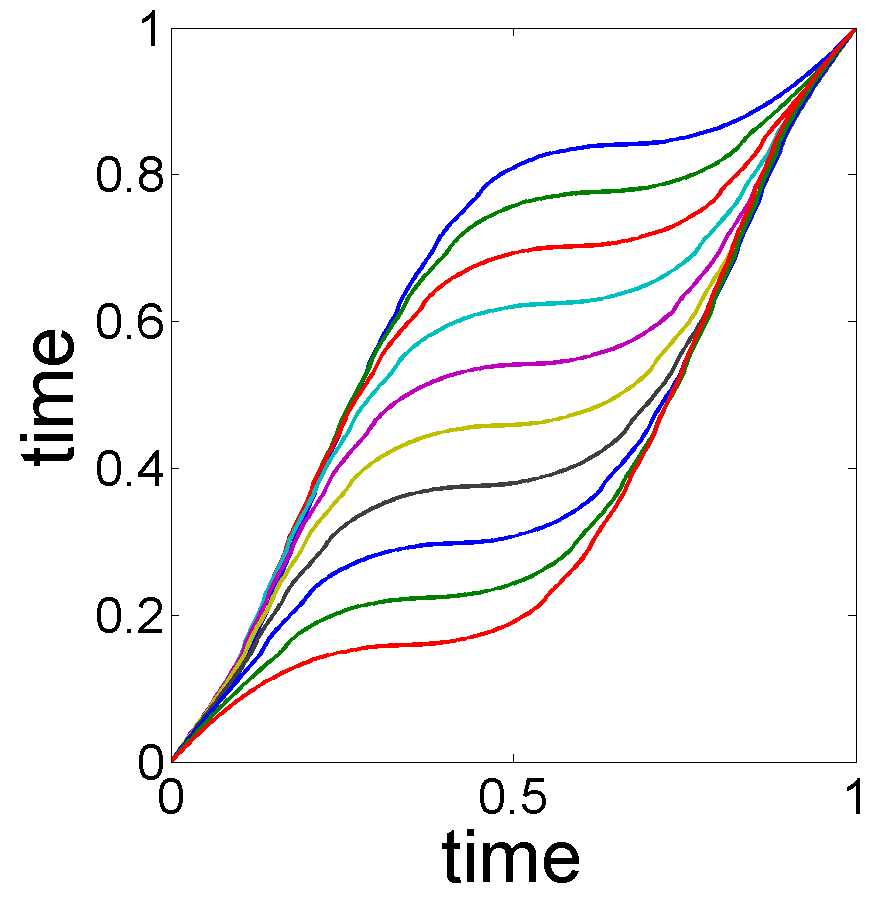}}
   \subfloat[observed functions]
	     {\includegraphics[scale=0.22]{graph/Trend_Extraction_Experiment/Artificial_data_experienment/Basis_Number_Selection/new/1_4/f}}
\caption{Synthetic ground truth data for trend subspace selection experiment.}
\label{fig:Synthetic-Data.-Basis selection-data} 
\end{center}
\end{figure}

We demonstrate this idea using a simulated example. 
In this experiment, we generate data using \\$g=5\left(0.25-(t-0.5)^{2}\right)\sin(5\pi t)$, $h=0.05e^{3t}-0.5$,  
and $\gamma_{i}=\int_{0}^{t}\check{\gamma}_{i}dt/\int_{0}^{1}\check{\gamma}_{i}dt$ where
$\check{\gamma}_{i}=\left(3\cos(\pi t-0.5+\frac{i}{n})\right)^{2}+0.1$ for $i=1,...,n$, 
with the additional constraint that $KM\{\gamma_{i}^{-1}\}=\gamma_{id}$. 
We add noise according to $\epsilon_{i}(t)\sim {\cal N}(0,\sigma^2)$, $\sigma=0.1$ in this experiment. 
The resulting data are shown in Fig. \ref{fig:Synthetic-Data.-Basis selection-data}(d).

\begin{figure}[!htb]
\begin{center}
\begin{tabular}[t]{>{\centering}m{2.6cm}>{\centering}m{3.1cm}>{\centering}m{3.1cm}>{\centering}m{2.5cm}>{\centering}m{3.1cm}}
Basis Range Selection 
& Trend
& Seasonality
& Warping Functions 
& Negative Log-likelihood
\tabularnewline
\hline 
\hline 

\footnotesize{$h={\sum^{1}_{k=1}}\tilde{d}_{k}\phi_{k}$ nonparametric $g$} 
& \includegraphics[scale=0.17]{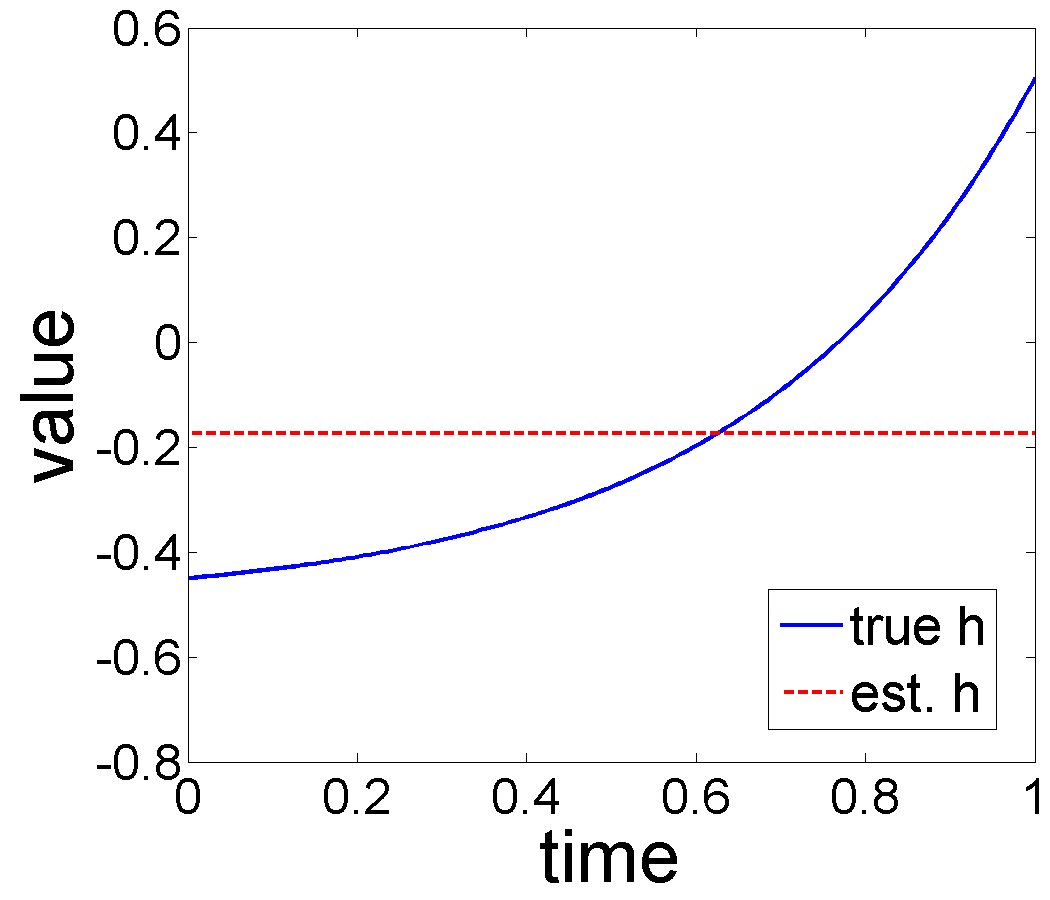} 
& \includegraphics[scale=0.17]{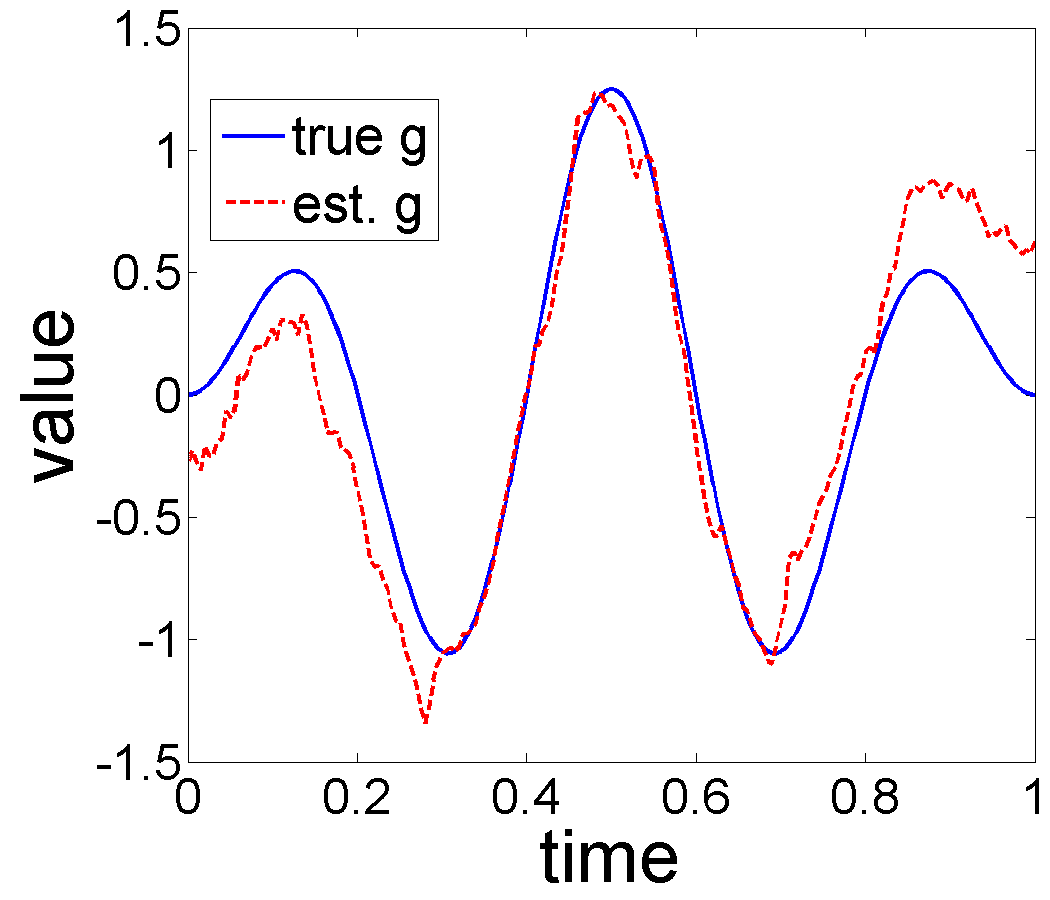} 
& \includegraphics[scale=0.17]{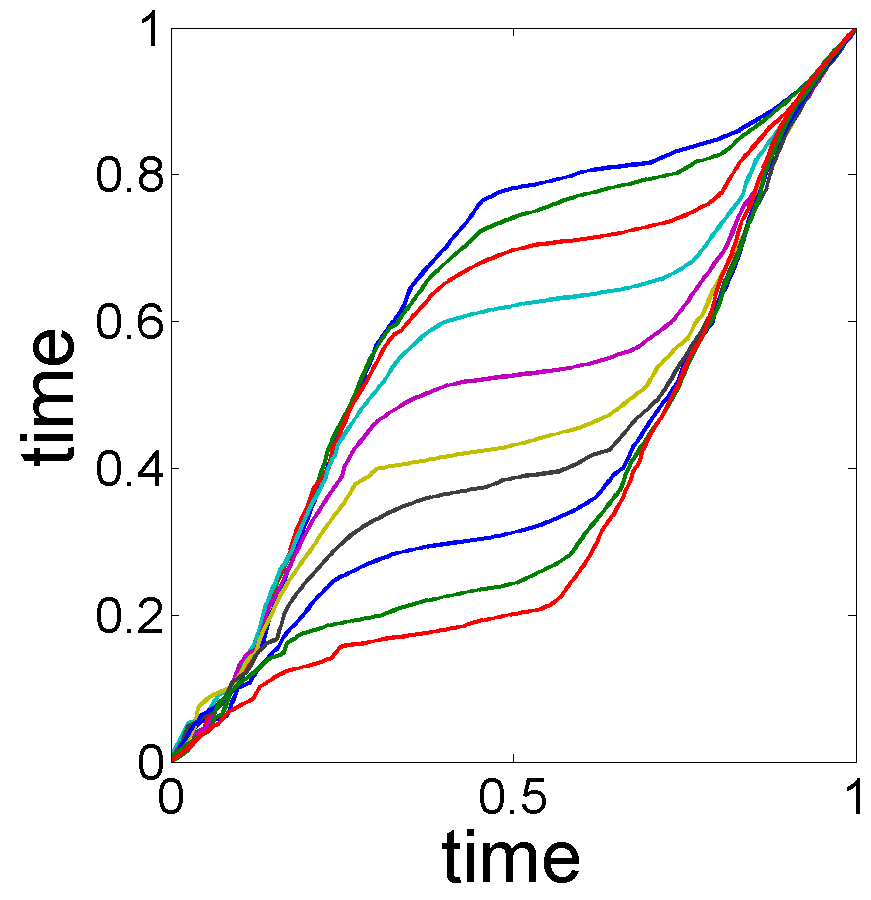} 
& \includegraphics[scale=0.17]{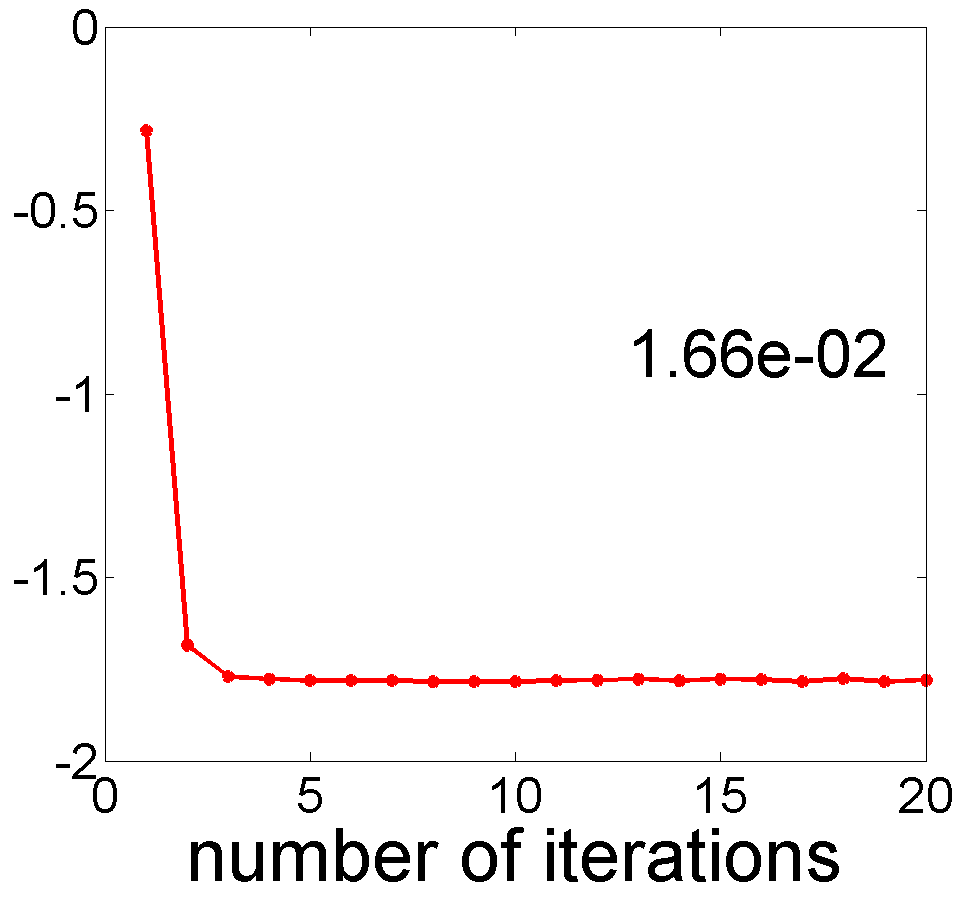}
\tabularnewline

\footnotesize{$h={\sum^{2}_{k=1}}\tilde{d}_{k}\phi_{k}$ nonparametric $g$}
& \includegraphics[scale=0.17]{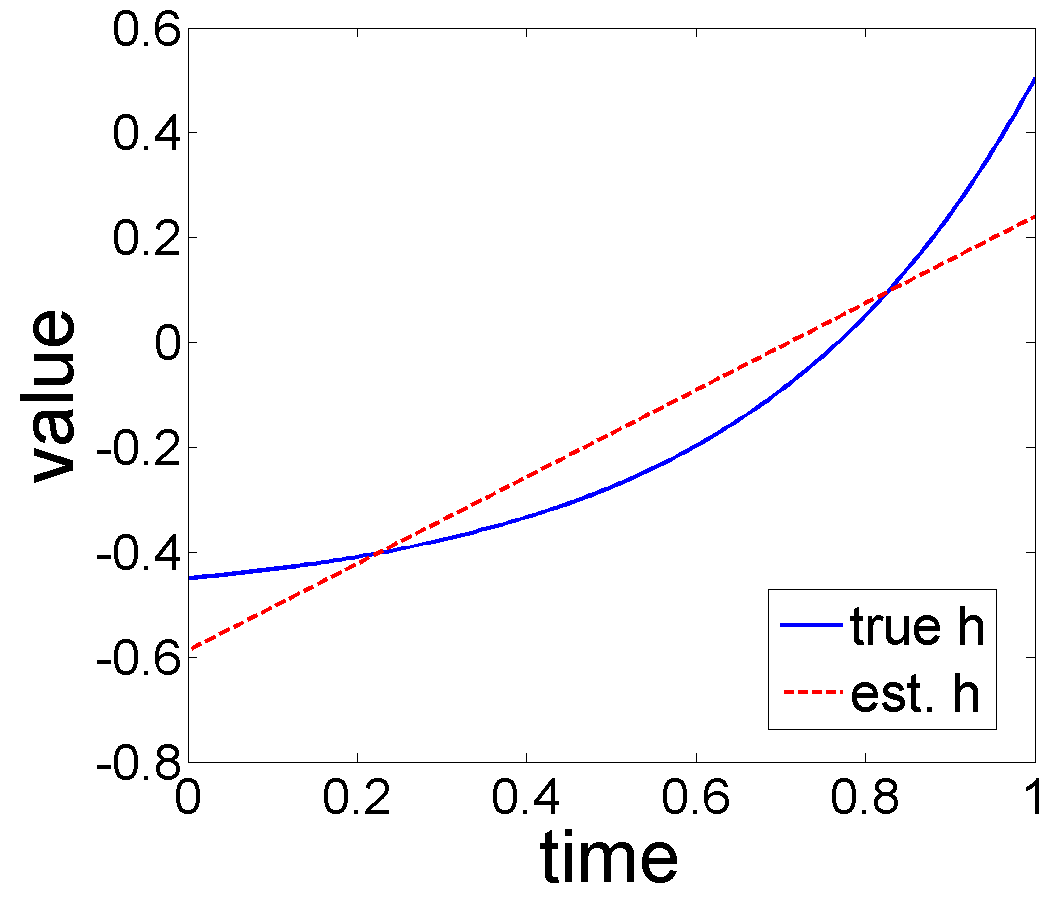} 
& \includegraphics[scale=0.17]{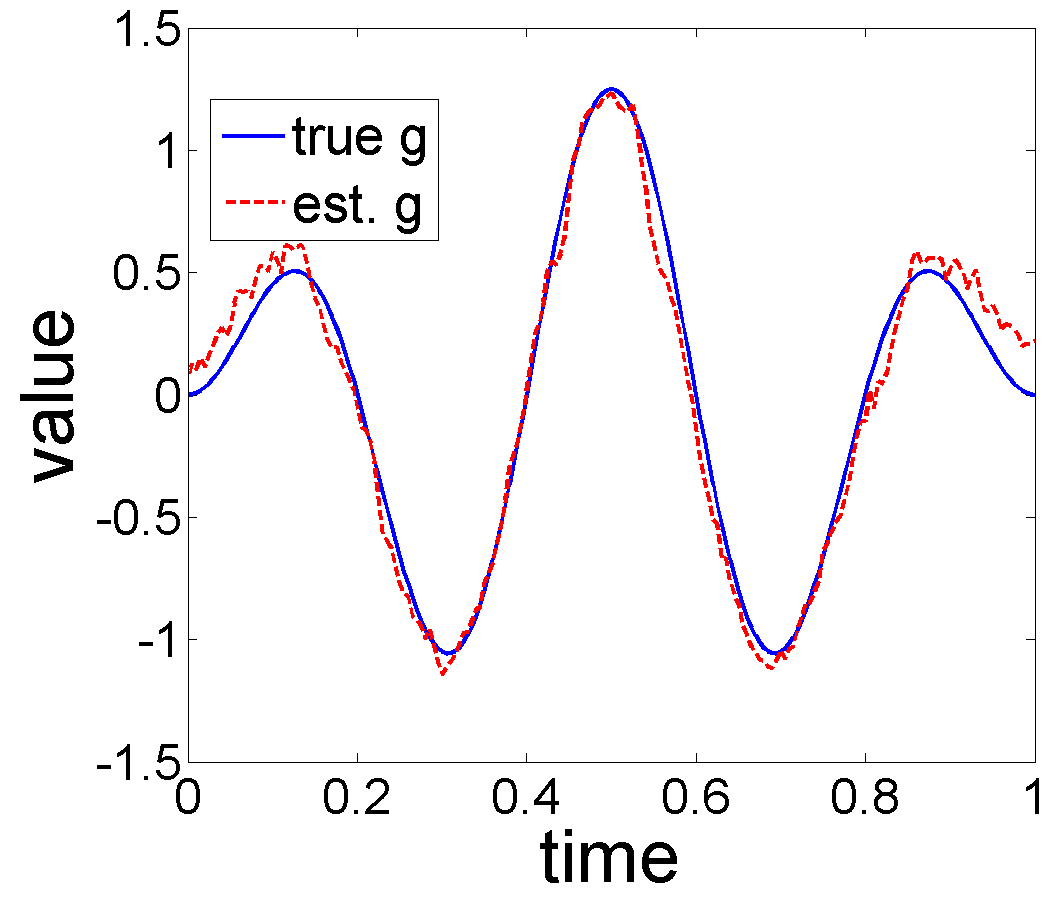} 
& \includegraphics[scale=0.17]{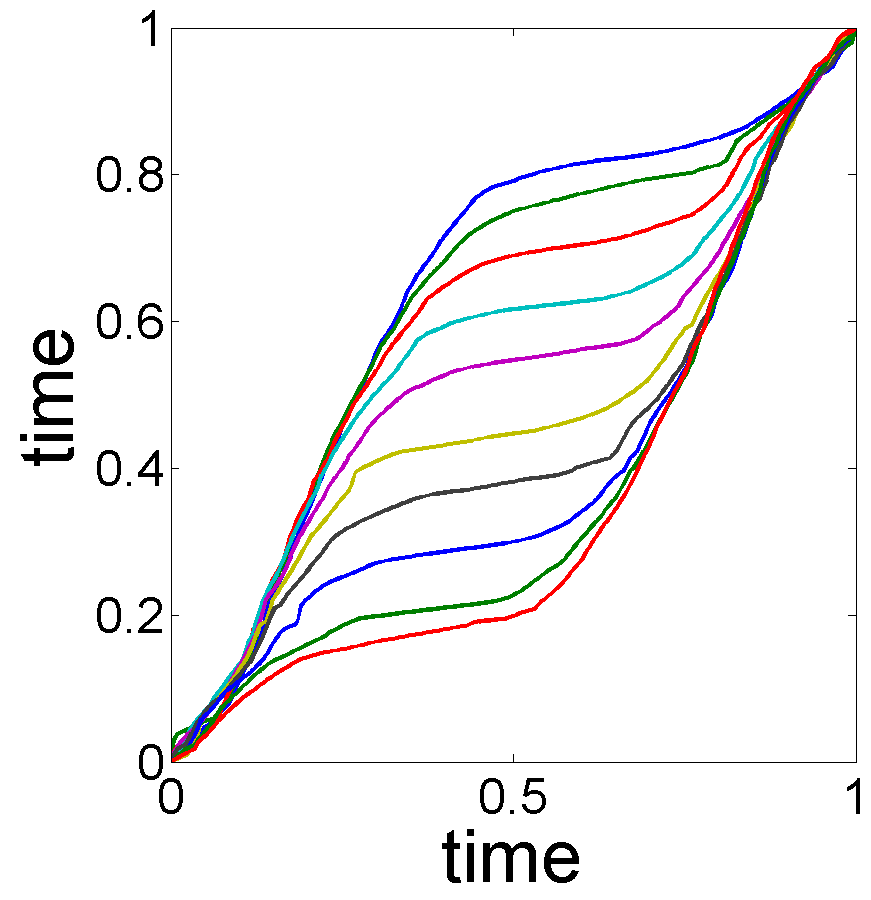} 
& \includegraphics[scale=0.17]{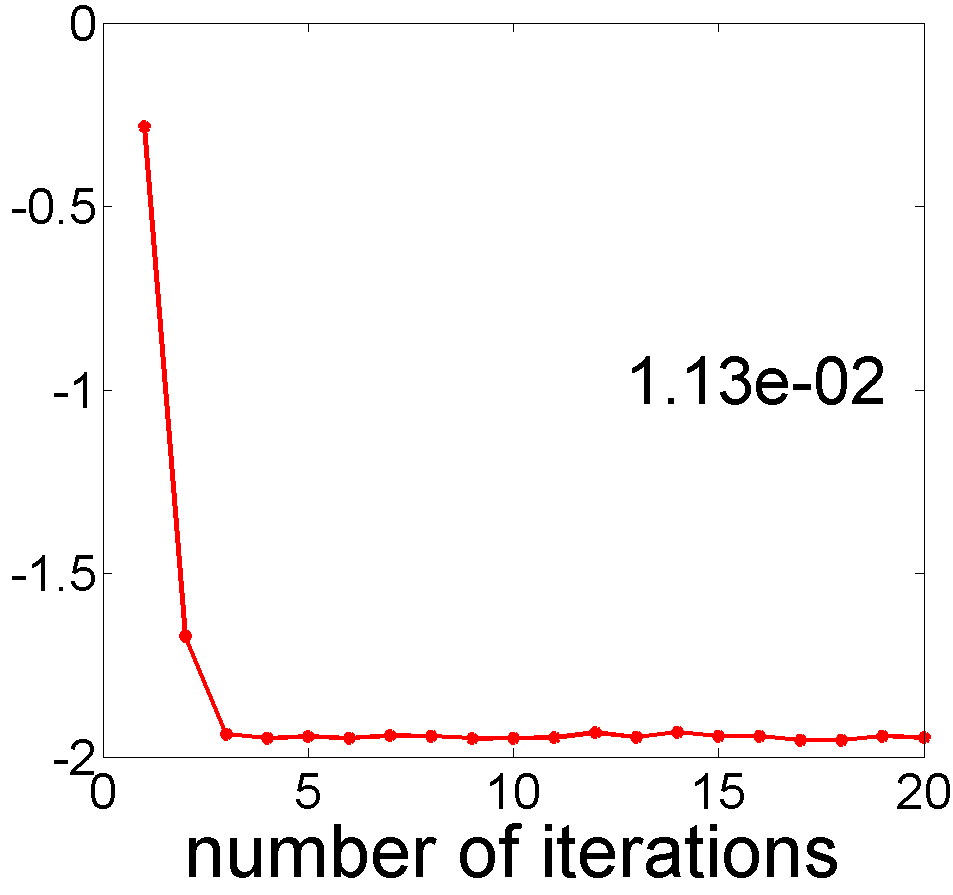}
\tabularnewline

\footnotesize{$h={\sum^{3}_{k=1}}\tilde{d}_{k}\phi_{k}$ nonparametric $g$}
& \includegraphics[scale=0.17]{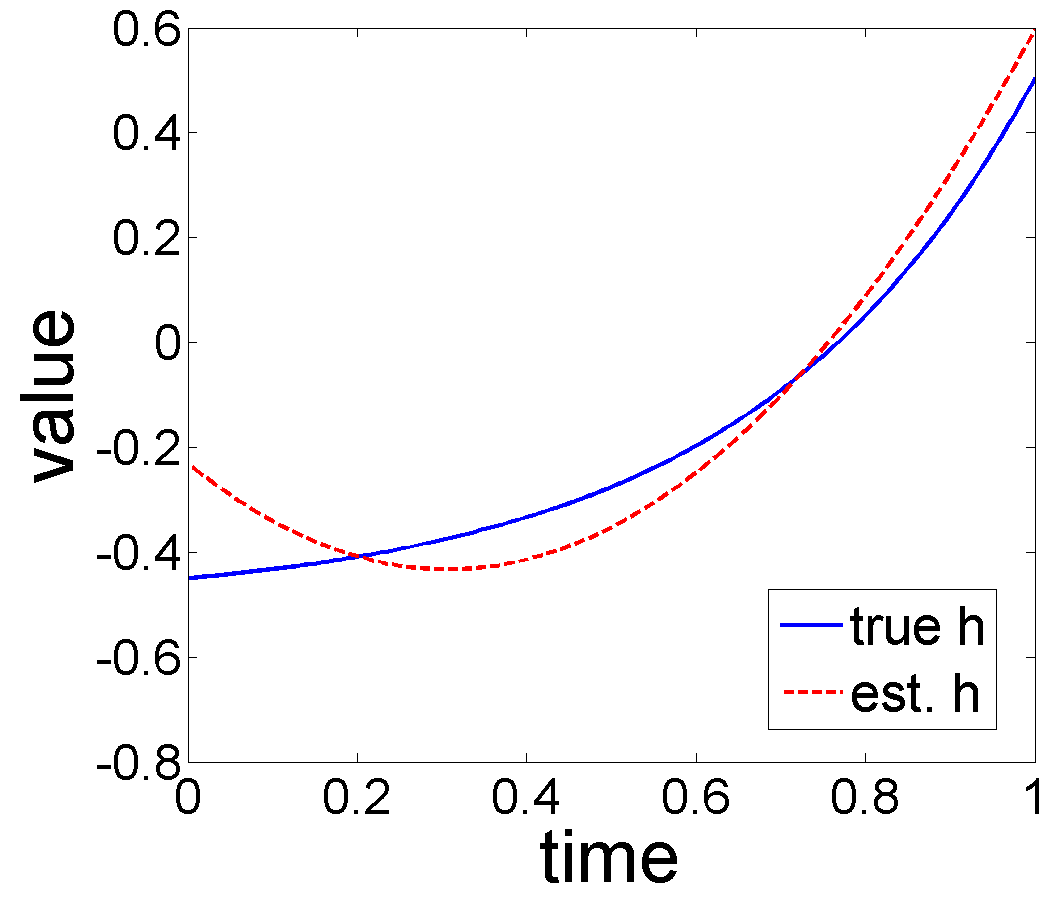} 
& \includegraphics[scale=0.17]{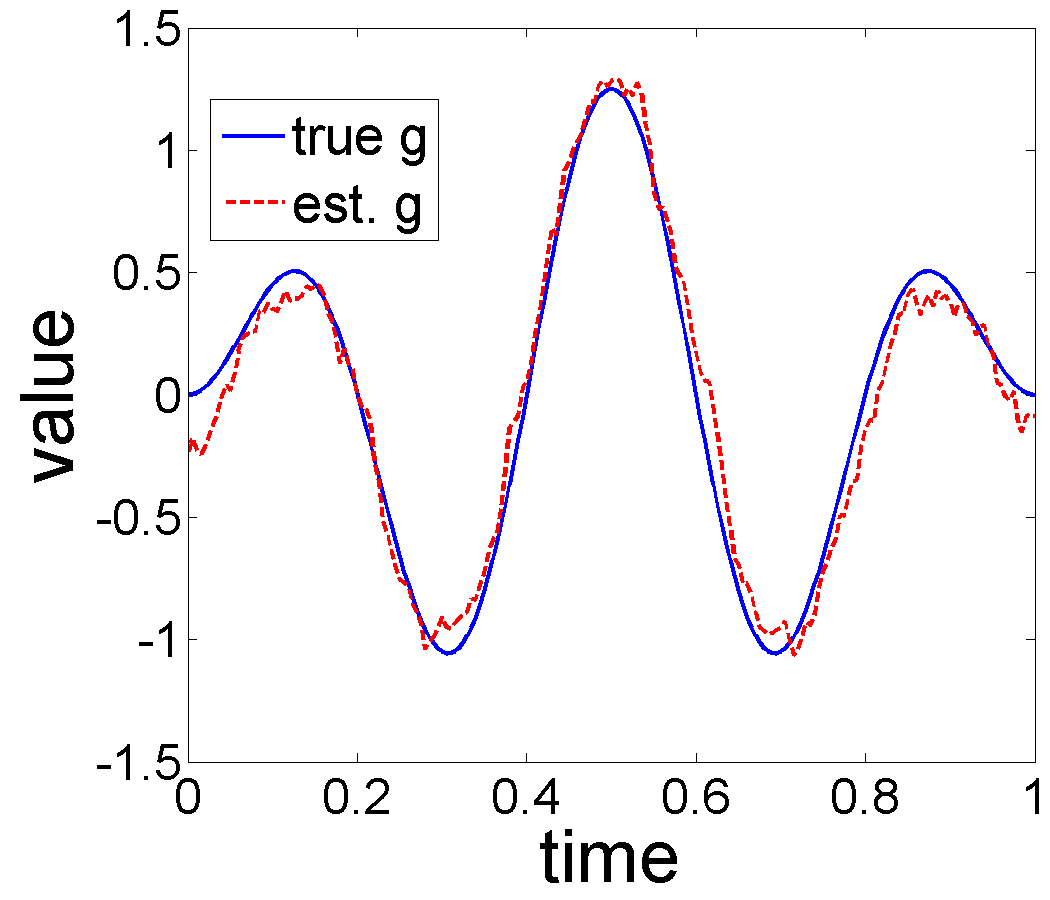} 
& \includegraphics[scale=0.17]{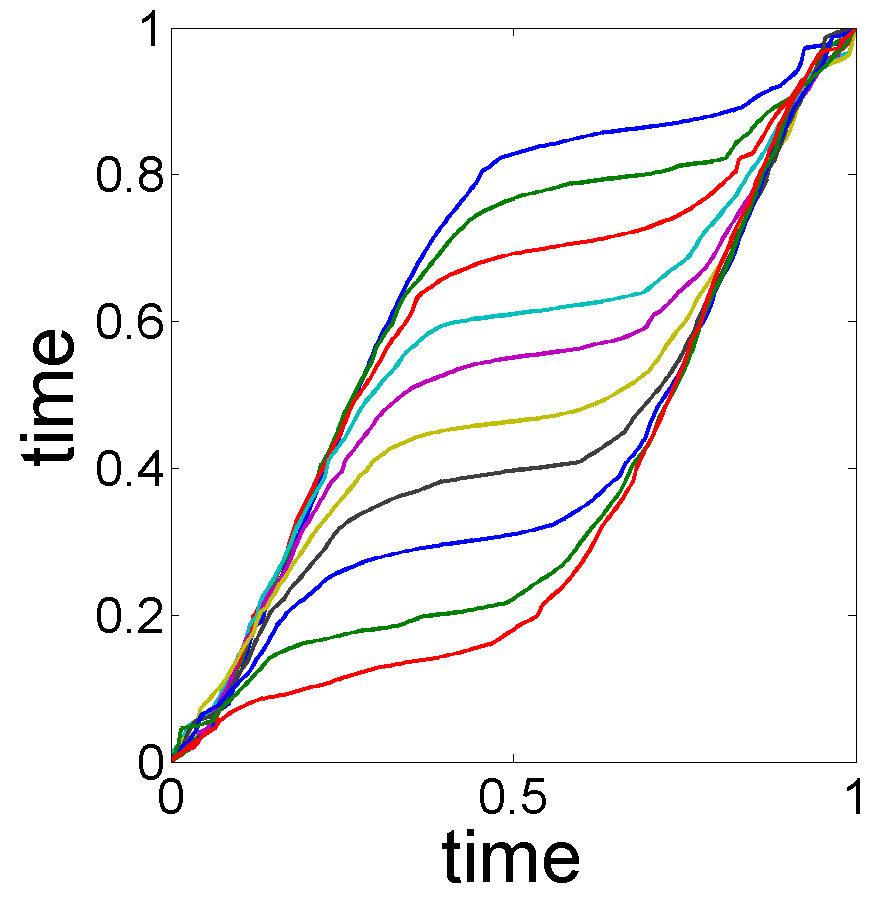} 
& \includegraphics[scale=0.17]{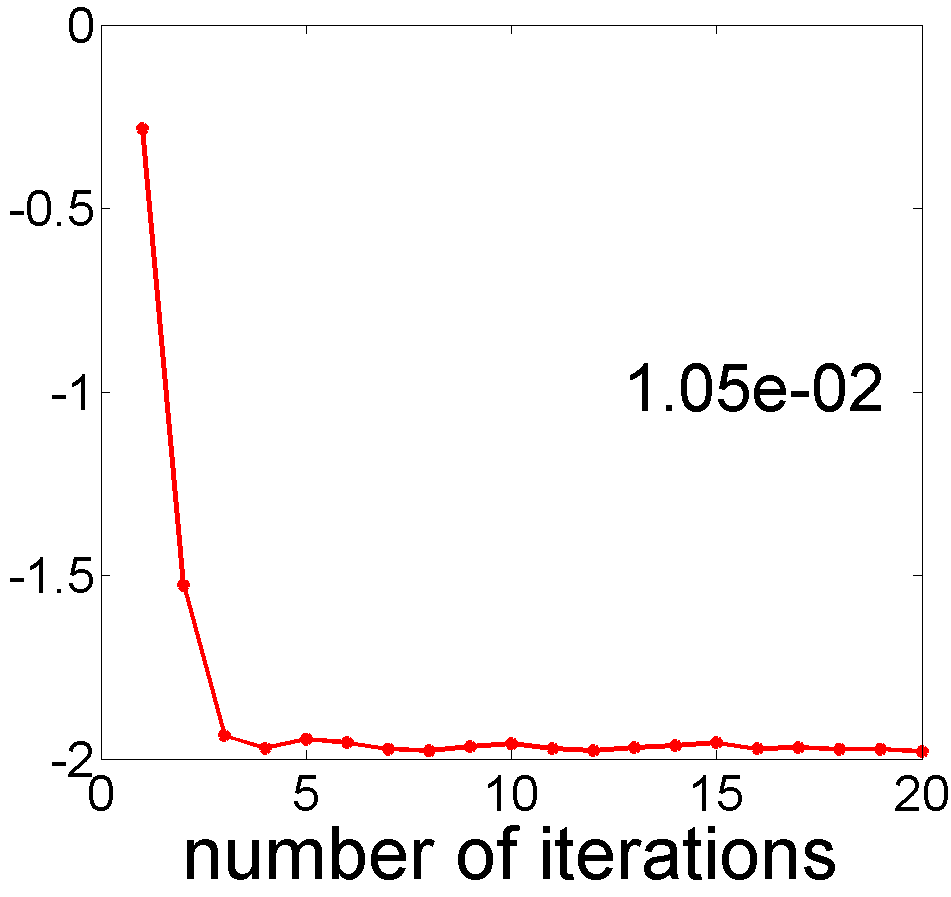}
\tabularnewline

\footnotesize{$h={\sum^{4}_{k=1}}\tilde{d}_{k}\phi_{k}$ nonparametric $g$}& 
\includegraphics[scale=0.17]{graph/Trend_Extraction_Experiment/Artificial_data_experienment/Basis_Number_Selection/new/1_4/est_h} & 
\includegraphics[scale=0.17]{graph/Trend_Extraction_Experiment/Artificial_data_experienment/Basis_Number_Selection/new/1_4/est_g} & 
\includegraphics[scale=0.17]{graph/Trend_Extraction_Experiment/Artificial_data_experienment/Basis_Number_Selection/new/1_4/gam} & 
\includegraphics[scale=0.17]{graph/Trend_Extraction_Experiment/Artificial_data_experienment/Basis_Number_Selection/new/1_4/L2cost}
\tabularnewline

\footnotesize{$h={\sum^{5}_{k=1}}\tilde{d}_{k}\phi_{k}$ nonparametric $g$}& 
\includegraphics[scale=0.17]{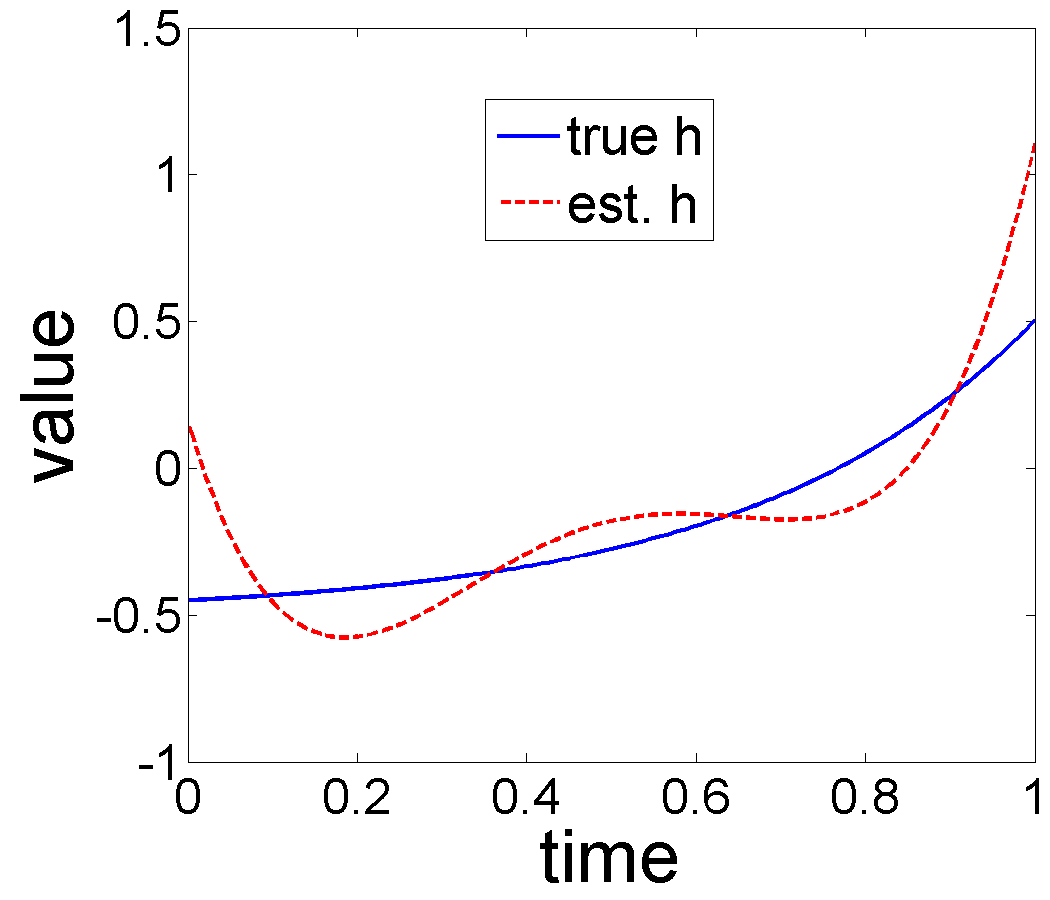} & 
\includegraphics[scale=0.17]{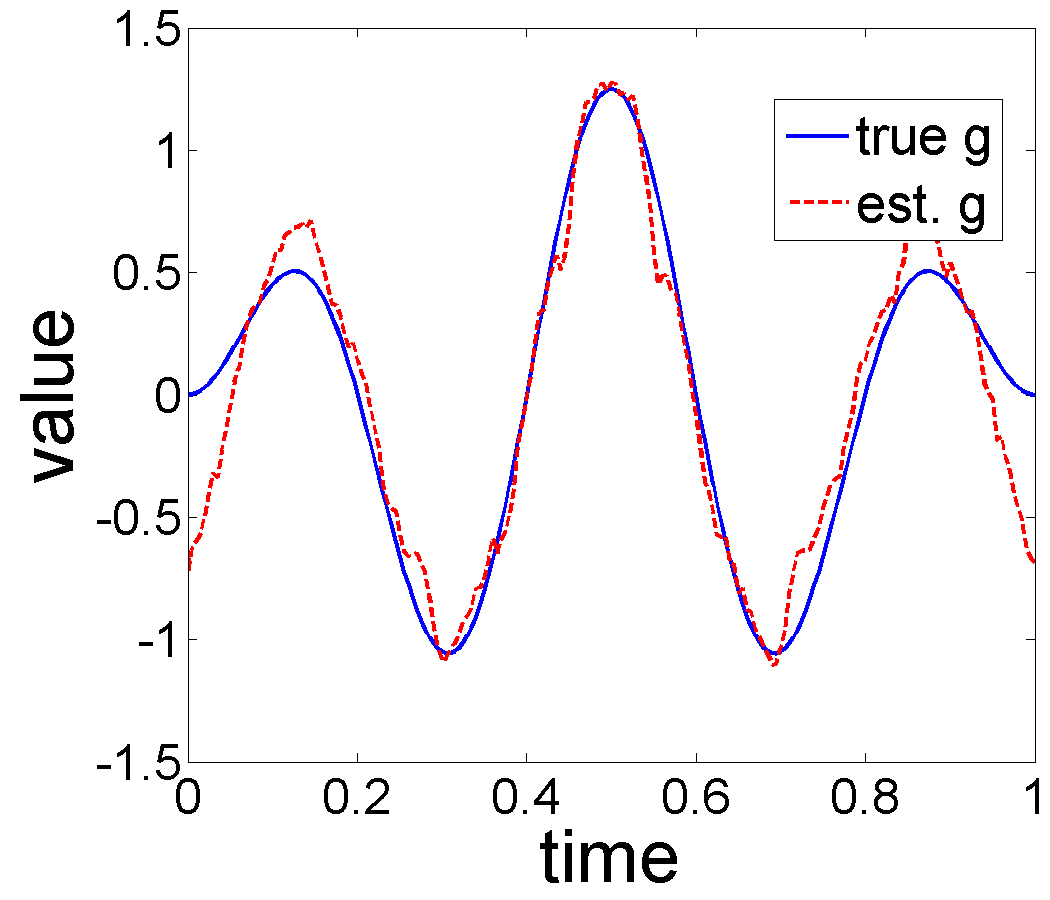} & 
\includegraphics[scale=0.17]{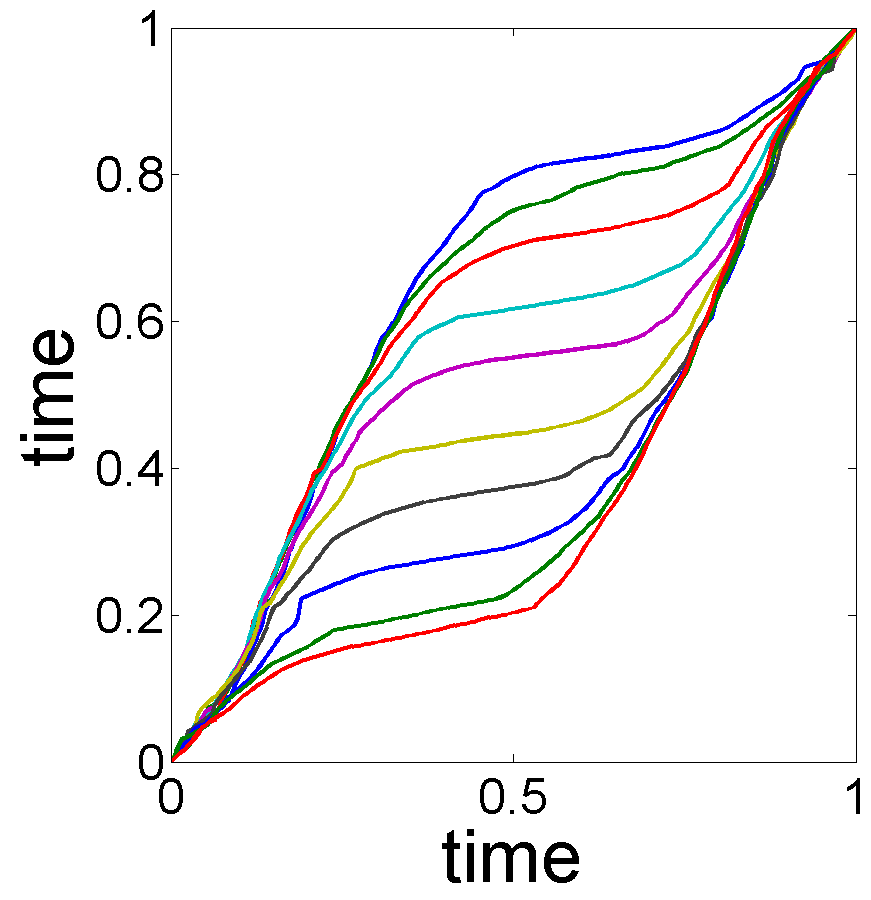} & 
\includegraphics[scale=0.17]{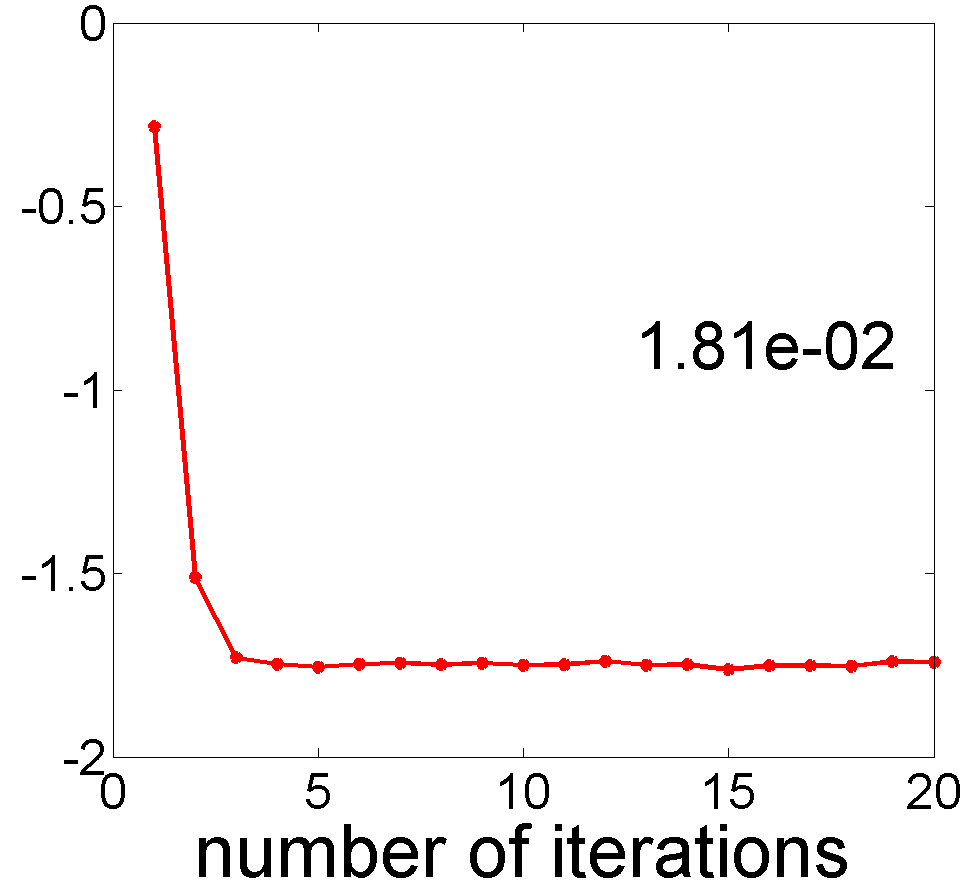}
\tabularnewline
\end{tabular}
\caption{Numerical results for basis range selection experiment. 
				Figures in the column of negative negative log-likelihood 
       are plotted in a log scale and the number inside the figures 
				are the minimized negative log-likelihood at the 20th iteration.}
\label{fig:Synthetic-Data.-Basis selection-graph} 
\end{center}
\end{figure}

\begin{figure}[!htb]
\begin{center}
\begin{tabular}{c}
   \includegraphics[scale=0.30]
		{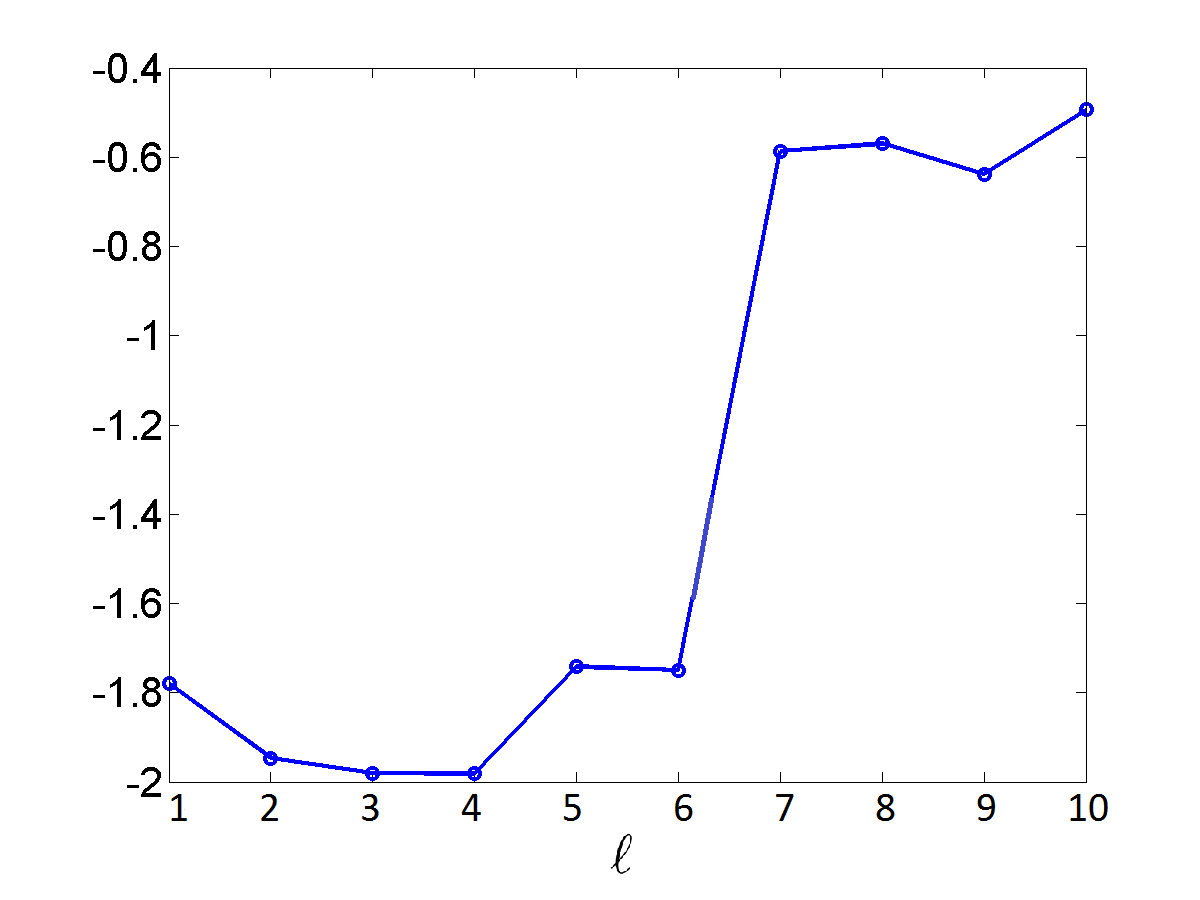} 
\tabularnewline
\end{tabular}
\caption{Minimized negative log-likelihood for $l=1,2,...,10$, plotted on a log scale.}
\label{fig:Synthetic-Data.-Basis selection-best-result} 
\end{center}
\end{figure}

We apply Algorithm \ref{alg: CD for Signal Separation} to this data for $l =  1, 2, 3, 4$, and $5$, 
and the results are presented in the Fig. \ref{fig:Synthetic-Data.-Basis selection-graph}. 
The $\phi$s used in this experiment come from the shifted Legendre polynomial basis
 and the trend $\hat h$ is thus a polynomial of degree $l$. 
From a visual perspective, setting $l=4$ yields the best estimates of $g$ and $h$.
Fig. \ref{fig:Synthetic-Data.-Basis selection-best-result} displays the minimized negative log-likelihood 
for $l=1$ to $l=10$ and the optimal is obtained when $l = 4$, supporting our approach for selecting $l$. 
This selection rule can be generalized to the situations 
when several potential basis types (polynomial, sine, cosine, Fourier) are given. 

\section{Experimental Results} \label{sec: Experimental Results}
In this section we present some results for estimating the trend and seasonal components using the 
MLE algorithm specified earlier. We will use both the synthetic and real datasets 
to illustrate the ideas. 

%%%%%%%%%%%%%%%%%%%%%%%%%%%%%%%%%%%%%%%%%%%%%%%%%%%%%%%%
\subsection{Synthetic Data} \label{sec: Synthetic Data}
%%%%%%%%%%%%%%%%%%%%%%%%%%%%%%%%%%%%%%%%%%%%%%%%%%%%%%%
\begin{enumerate}
\item \textbf{Performance Under Different Noise Levels}:
In this experiment, we select a specific form of the trend and seasonal components, 
and increase the variance, $\sigma^2$, of the additive noise $\epsilon_i$ to 
study the effect of noise on estimation performance.  
In this experiment we used $g=2\exp\left(-0.8(10t-7.5)^{2}\right)+2\exp\left(-0.8(10t-2.5)^{2}\right)$,
 $h=\cos(\pi t+\frac{\pi}{2})$, and warping functions 
$\gamma_{i}=\int_{0}^{t}\check{\gamma}_{i}dt/\int_{0}^{1}\check{\gamma}_{i}dt$, where
$\check{\gamma}_{i}=3\sin(2\pi a_{i}t)+2\cos(a_{i}t), a_{i}=-2+i\frac{4}{n}$.
For each $t$, we consider Gaussian noise $\epsilon_{i}(t)\sim {\cal N}(0,\sigma^2)$ 
where $\sigma$ is $0,0.2,0.4,0.8,$ and $1.6$ 
for the noise level experiments.
Each observation $f_i$ is generated from Eqn. \ref{eqn:our-model} and the number of time samples is taken to be $T=200$. 
Fig. \ref{fig:Synthetic-Data.-Noise perturbation-1} shows the true trend and the seasonal components, when no noise is added. 
Fig. \ref{fig:Synthetic-Data-noise perturbation-2} displays estimation results for the additive noise levels. 

\begin{figure}[!htb]
\begin{center}
   \subfloat[true trend]
	     {\includegraphics[scale=0.22]{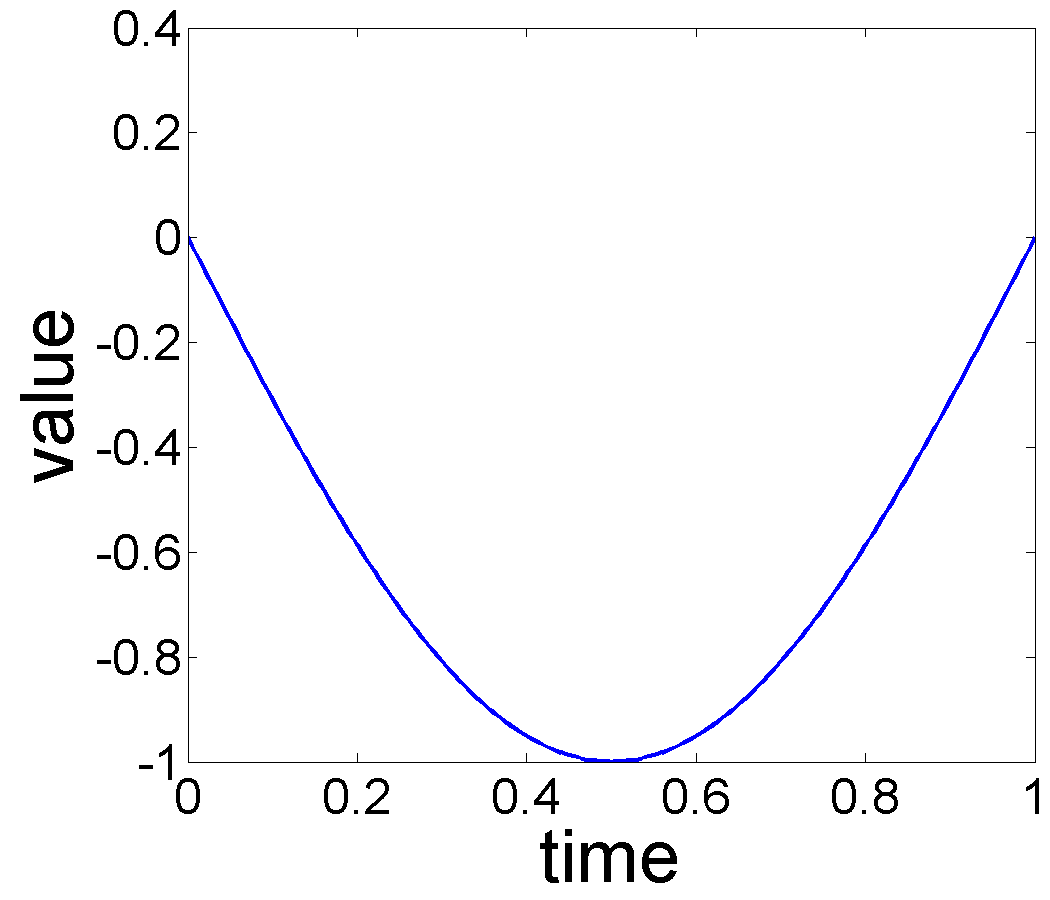}}
   \subfloat[true seasonality]
	     {\includegraphics[scale=0.22]{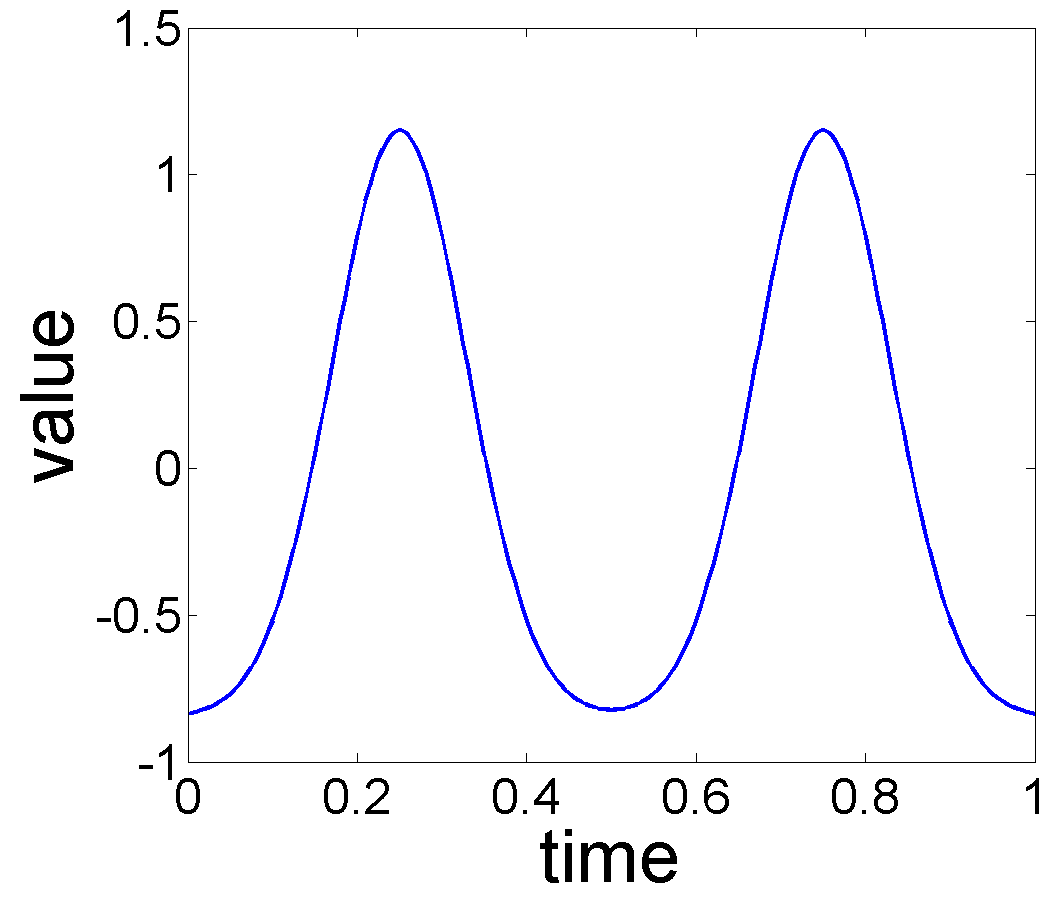}}
   \subfloat[warping functions]
	     {\includegraphics[scale=0.22]{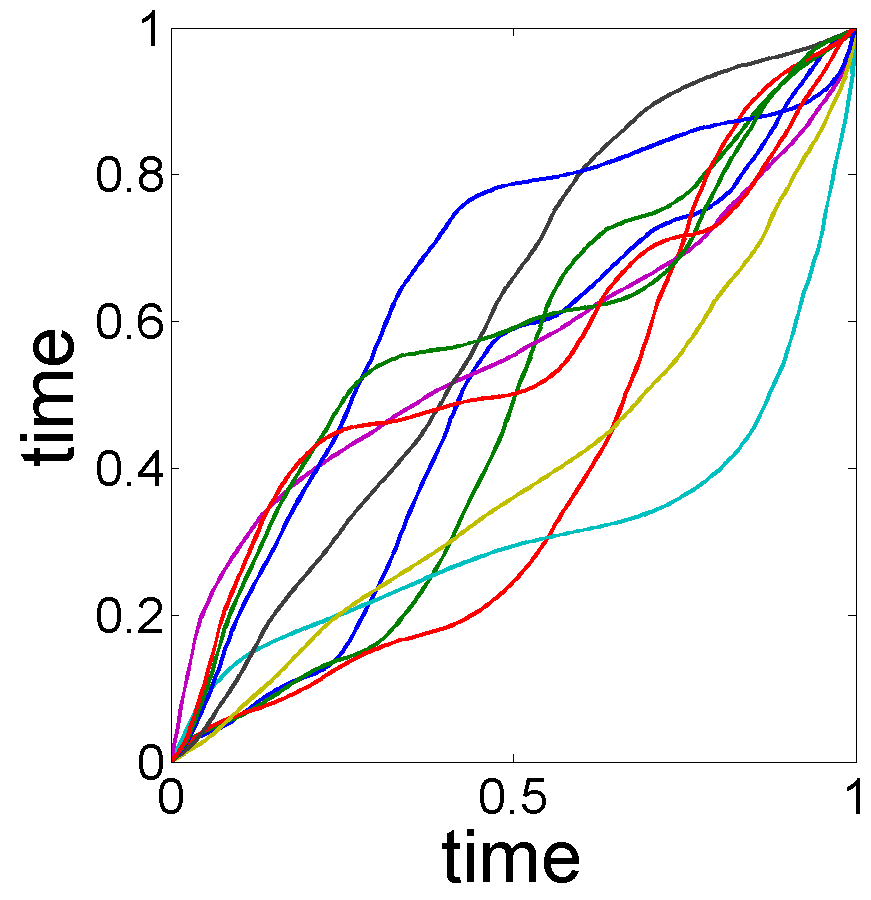}}
   \subfloat[observed functions]
	     {\includegraphics[scale=0.22]{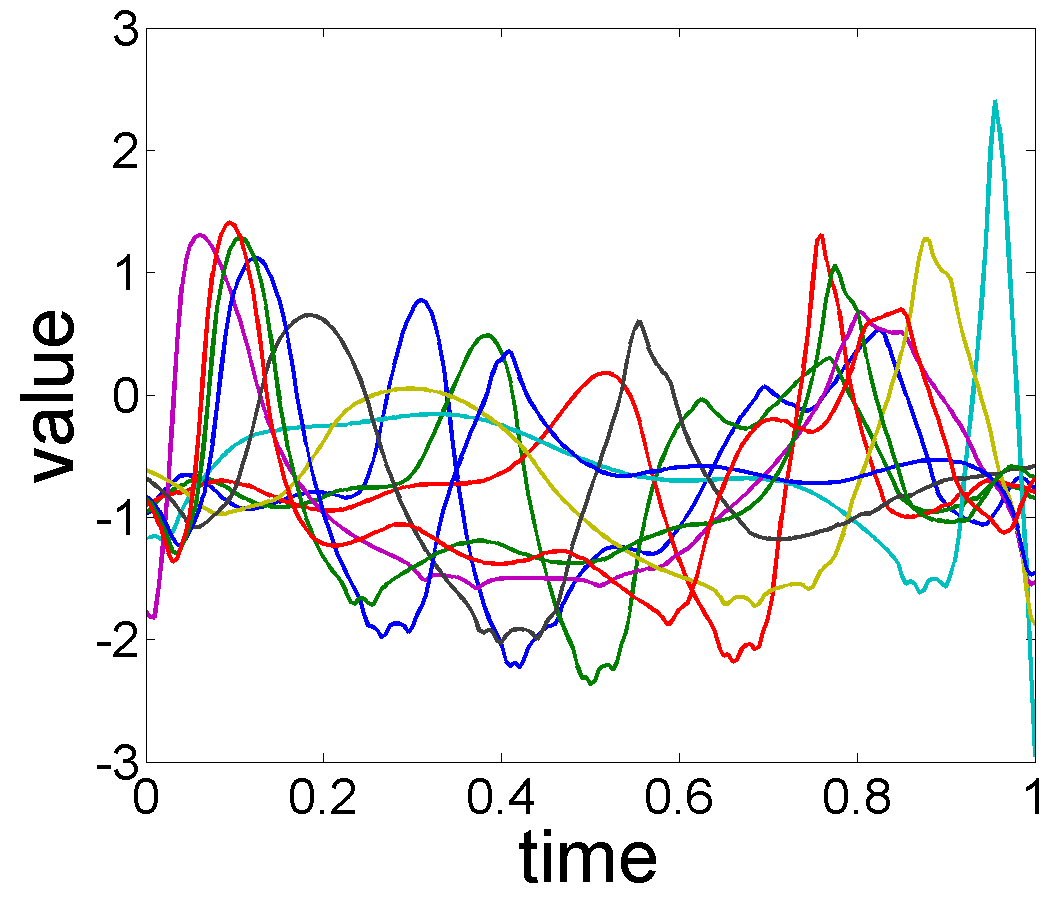}}
\caption{Synthetic truth data of noise perturbation experiment.}
\label{fig:Synthetic-Data.-Noise perturbation-1} 
\end{center}
\end{figure}

\begin{figure}[!htb]
\begin{center}
\begin{tabular}{>{\centering}m{0.6cm}>{\centering}m{3.0cm}>{\centering}m{3.0cm}>{\centering}m{3.0cm}>{\centering}m{2.5cm}>{\centering}m{3.0cm}}
\small{$\sigma$} & 
\small{Observed Functions} &
\small{Trend} &
\small{Seasonality} &  
\small{Warping Functions} & 
\small{Negative Log-likelihood} 
\tabularnewline

{\footnotesize{$0.2$}} &
\includegraphics[scale=0.165]{graph/Trend_Extraction_Experiment/Artificial_data_experienment/Noicy_and_Smoothing/new/02/f} &
\includegraphics[scale=0.165]{graph/Trend_Extraction_Experiment/Artificial_data_experienment/Noicy_and_Smoothing/new/02/est_h} &
\includegraphics[scale=0.165]{graph/Trend_Extraction_Experiment/Artificial_data_experienment/Noicy_and_Smoothing/new/02/est_g} &
\includegraphics[scale=0.165]{graph/Trend_Extraction_Experiment/Artificial_data_experienment/Noicy_and_Smoothing/new/02/gam}  &
\includegraphics[scale=0.16]{graph/Trend_Extraction_Experiment/Artificial_data_experienment/Noicy_and_Smoothing/new/02/L2cost}
\tabularnewline

{\footnotesize{}$0.4$} &
\includegraphics[scale=0.165]{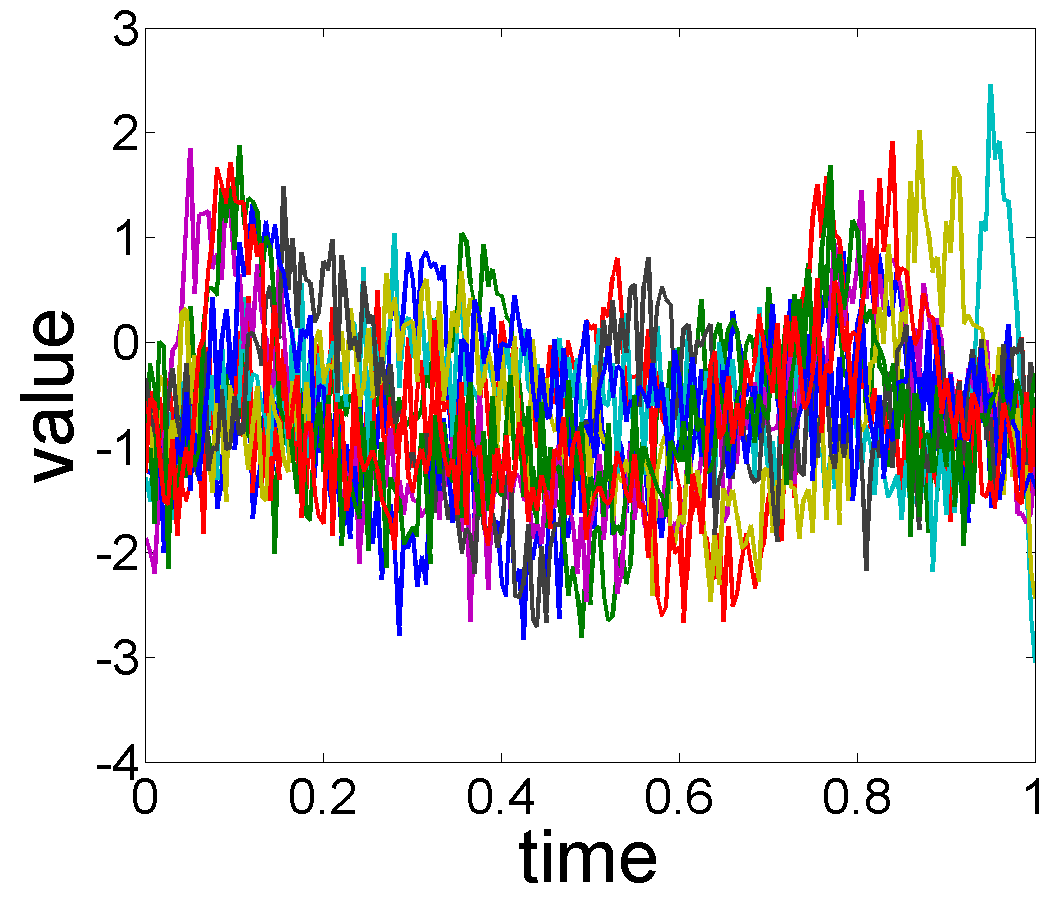} &
\includegraphics[scale=0.165]{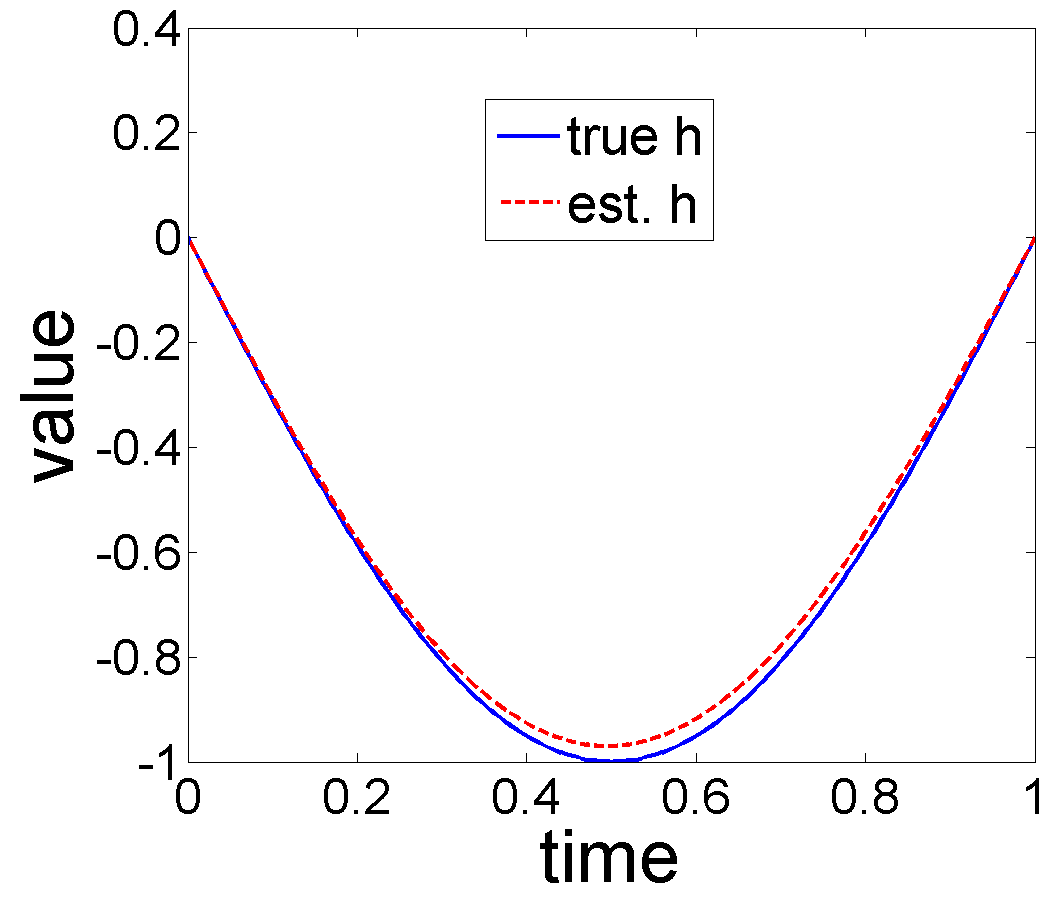} &
\includegraphics[scale=0.165]{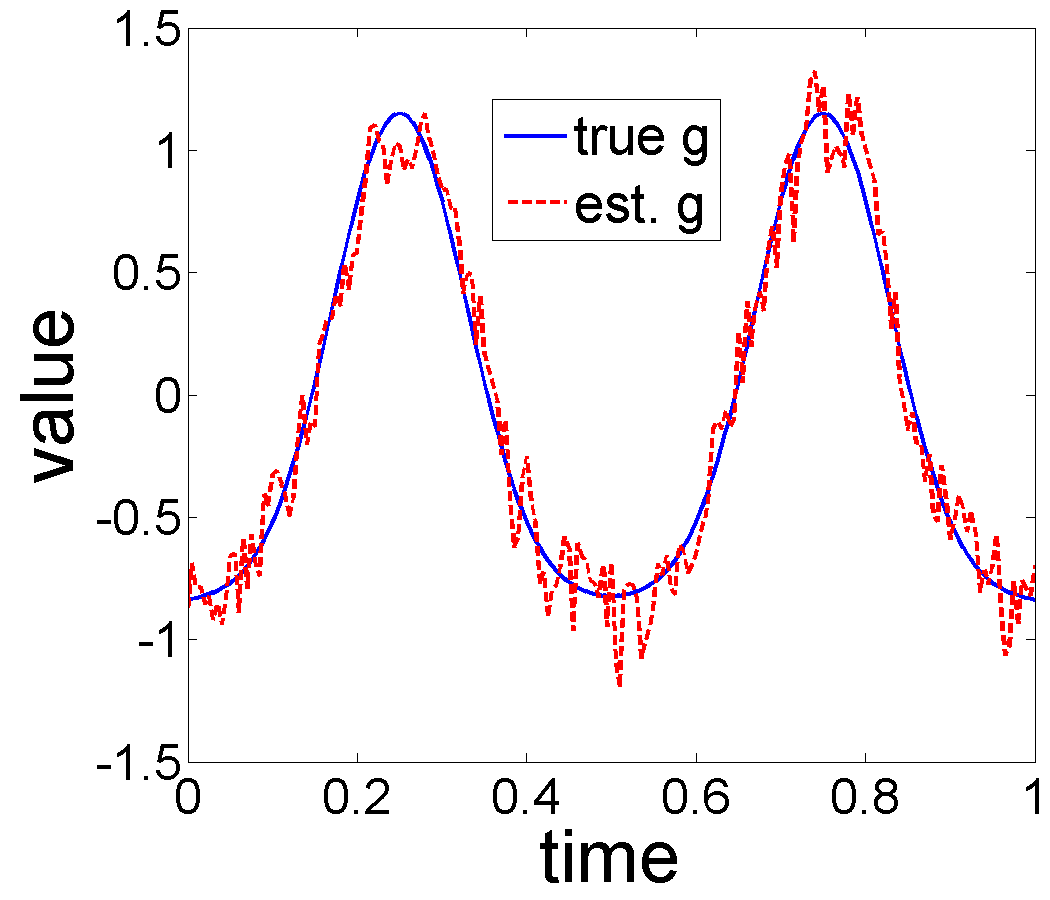} & 
\includegraphics[scale=0.165]{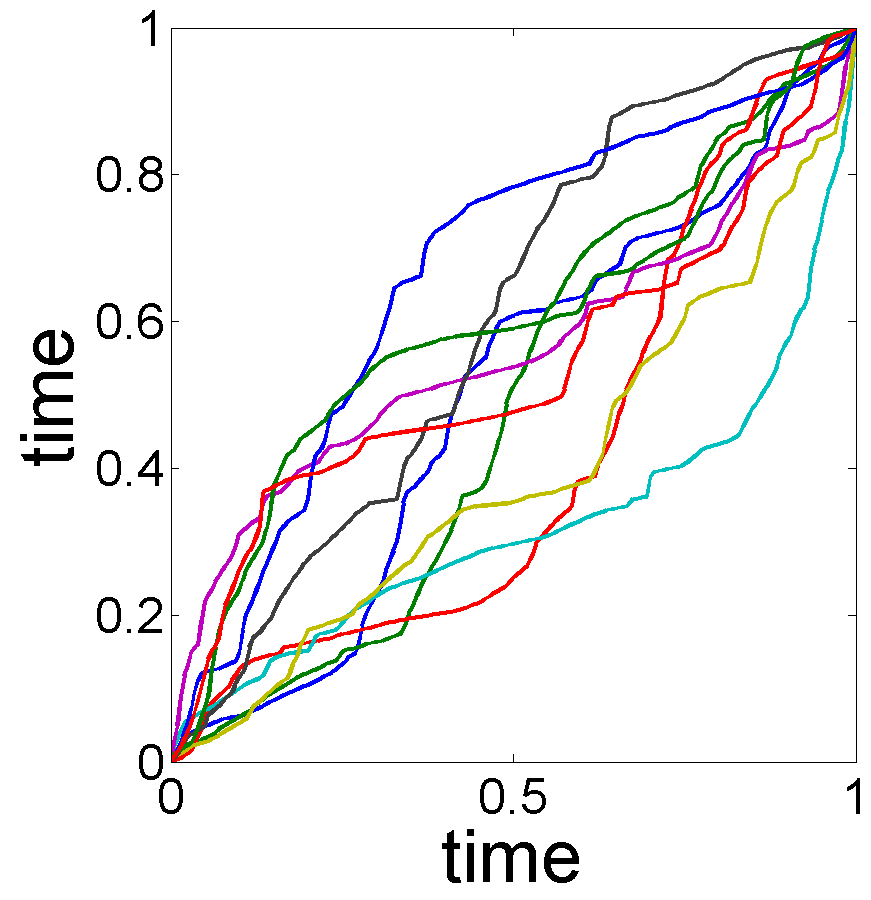}  &
\includegraphics[scale=0.165]{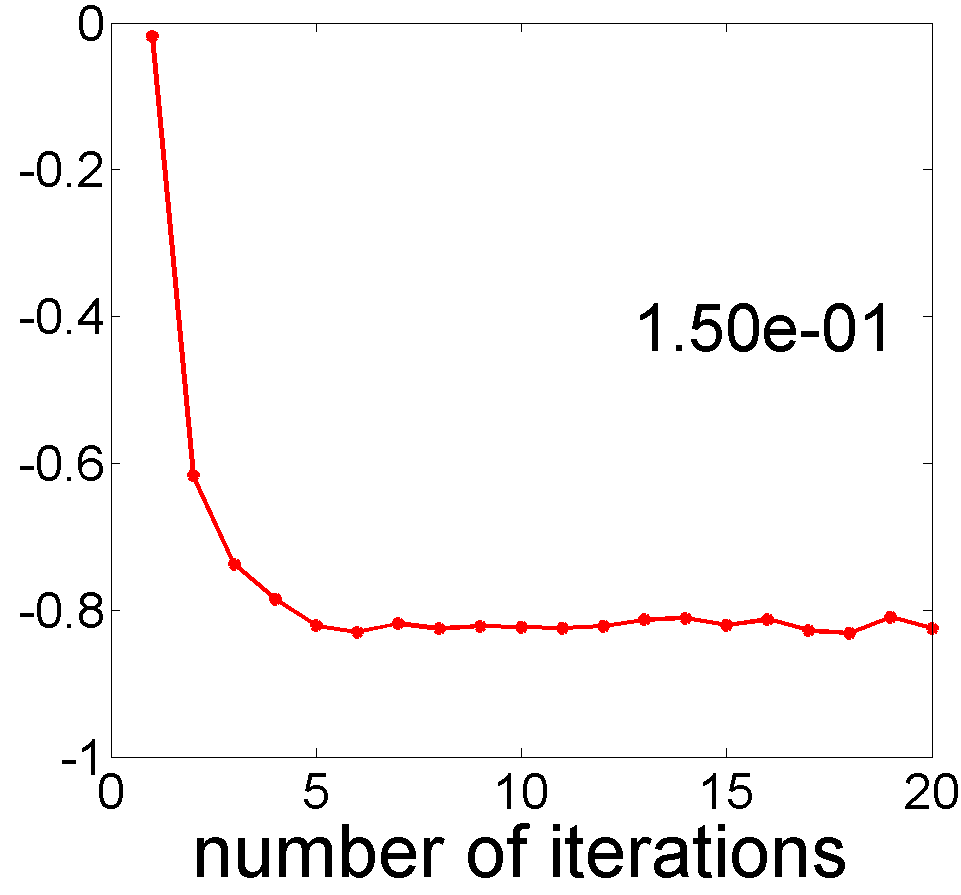}
\tabularnewline

{\footnotesize{}$0.8$} & 
\includegraphics[scale=0.165]{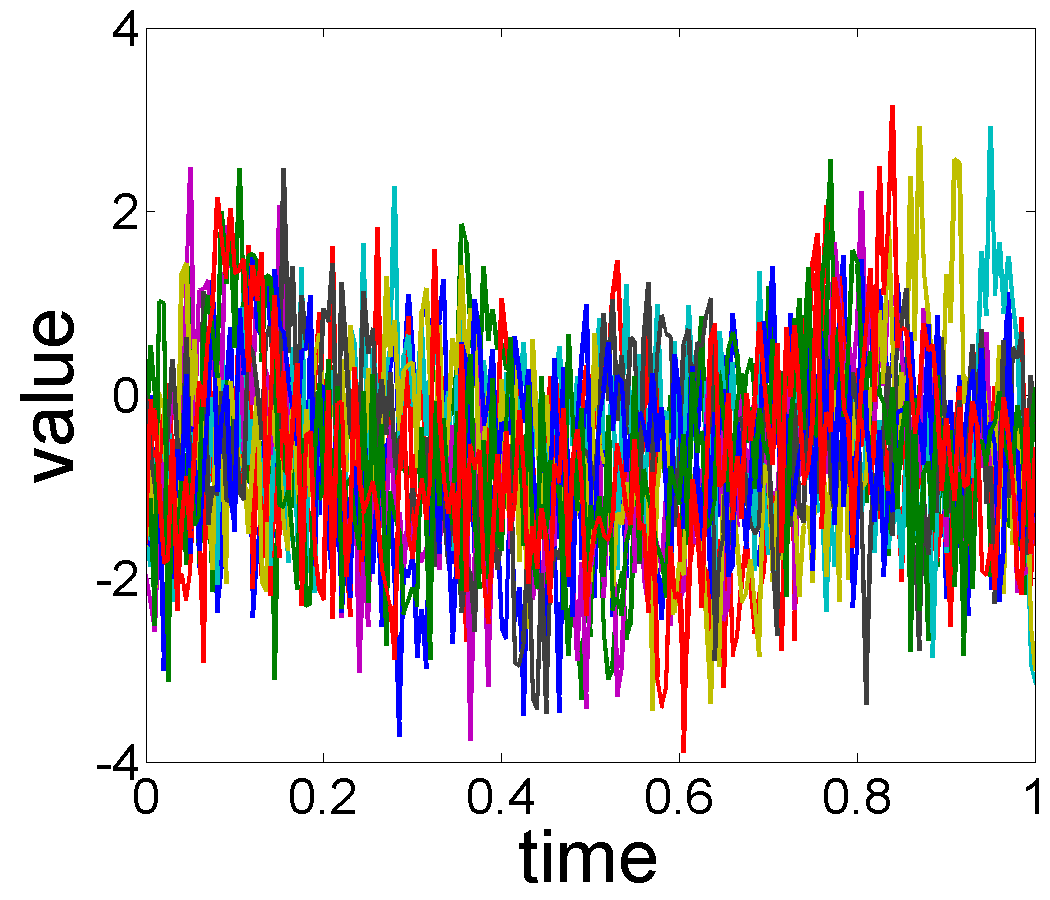}  &
\includegraphics[scale=0.165]{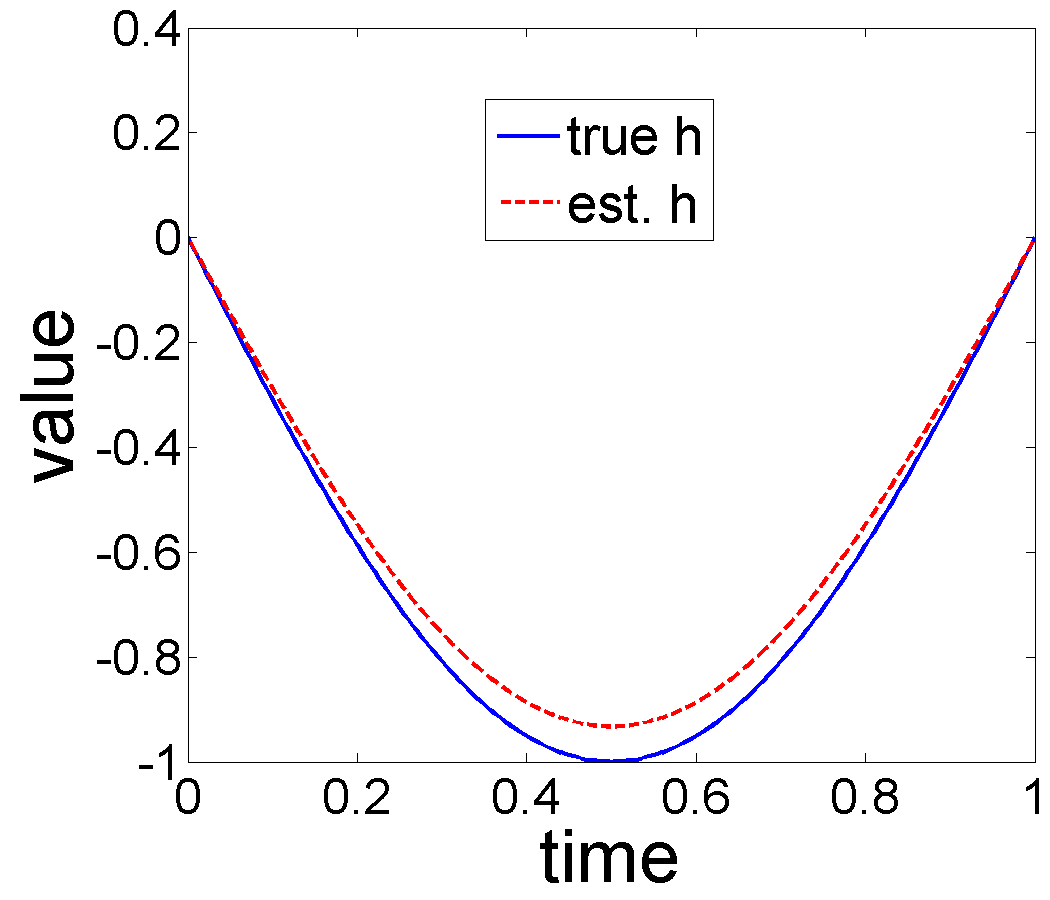} & 
\includegraphics[scale=0.165]{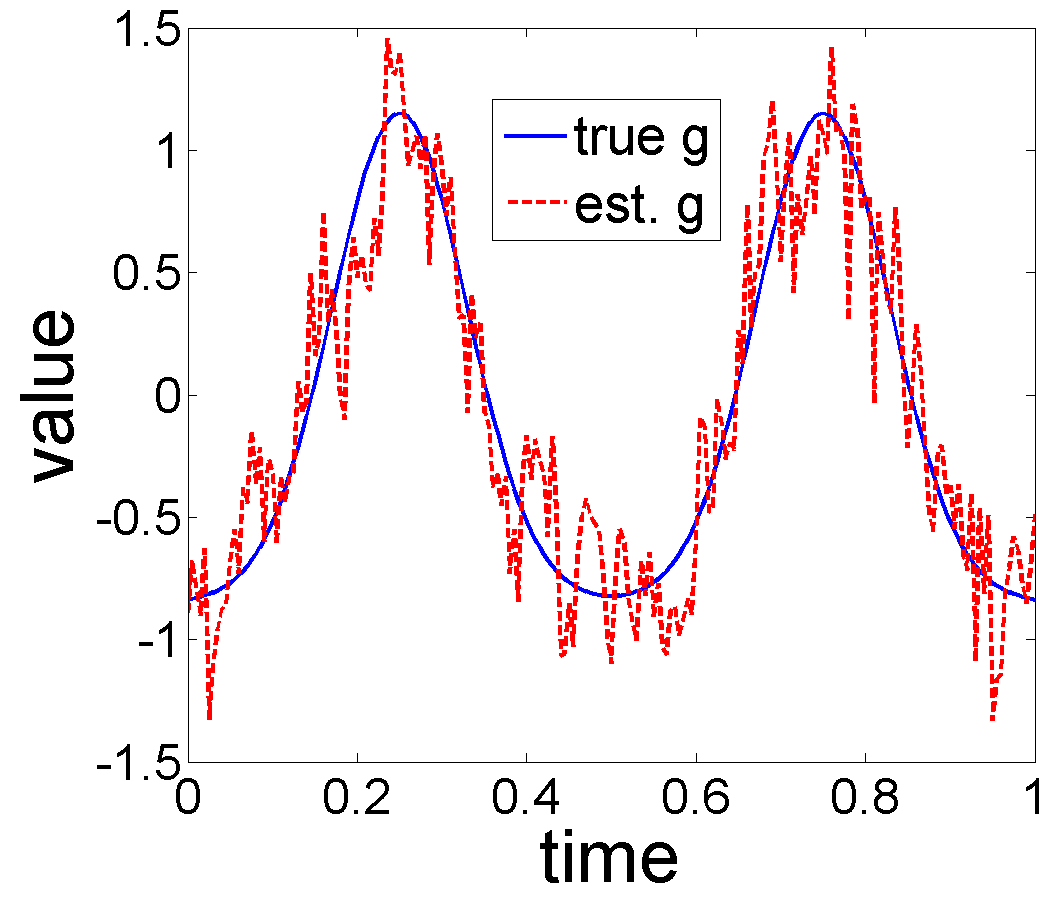} & 
\includegraphics[scale=0.165]{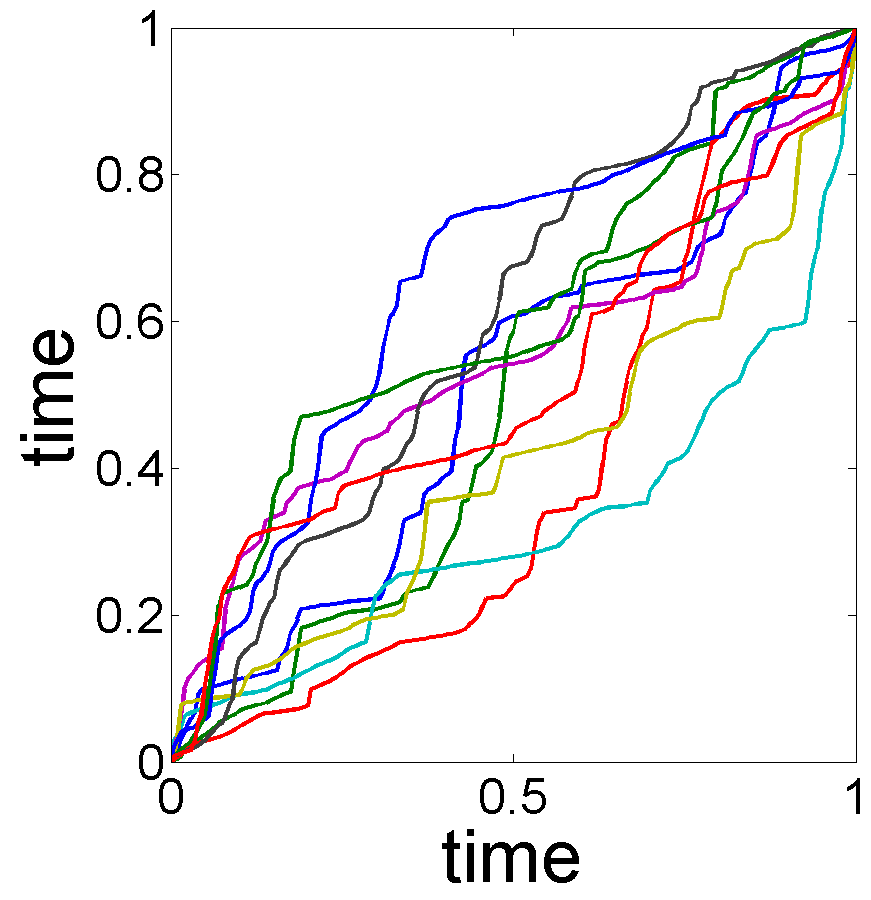}  &
\includegraphics[scale=0.165]{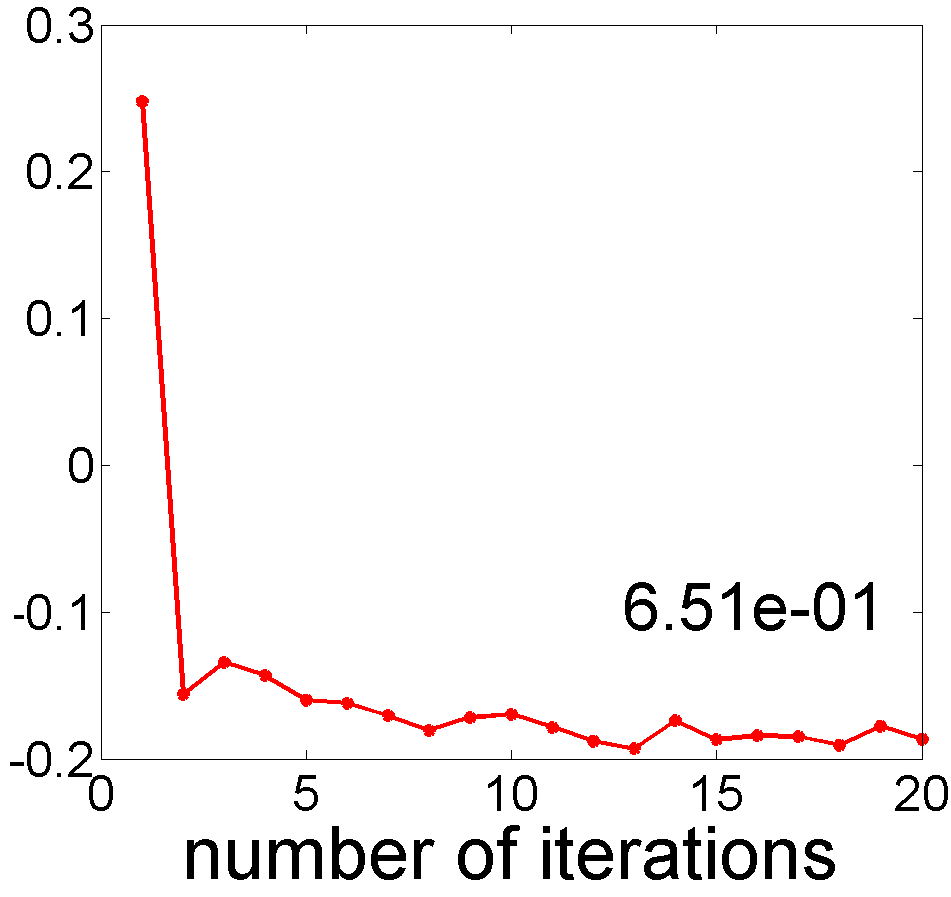}
\tabularnewline

{\footnotesize{}$1.6$}  &
\includegraphics[scale=0.165]{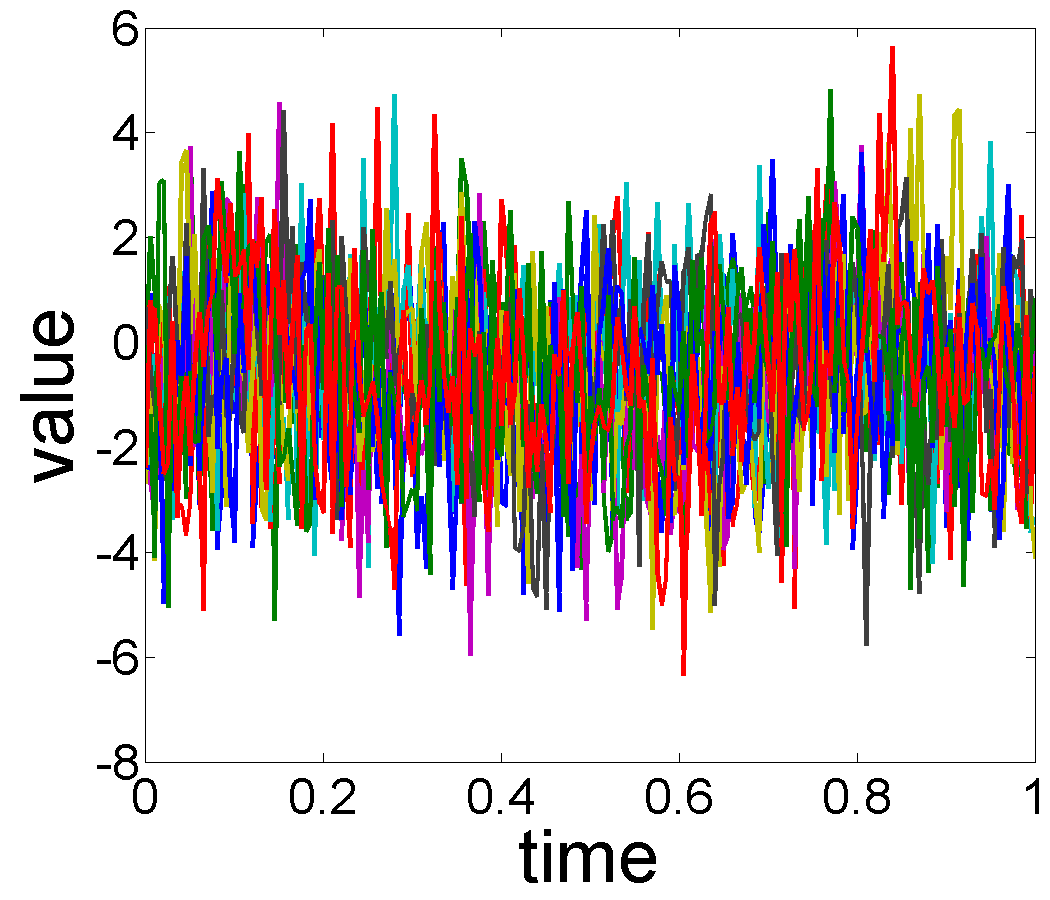}  &
\includegraphics[scale=0.165]{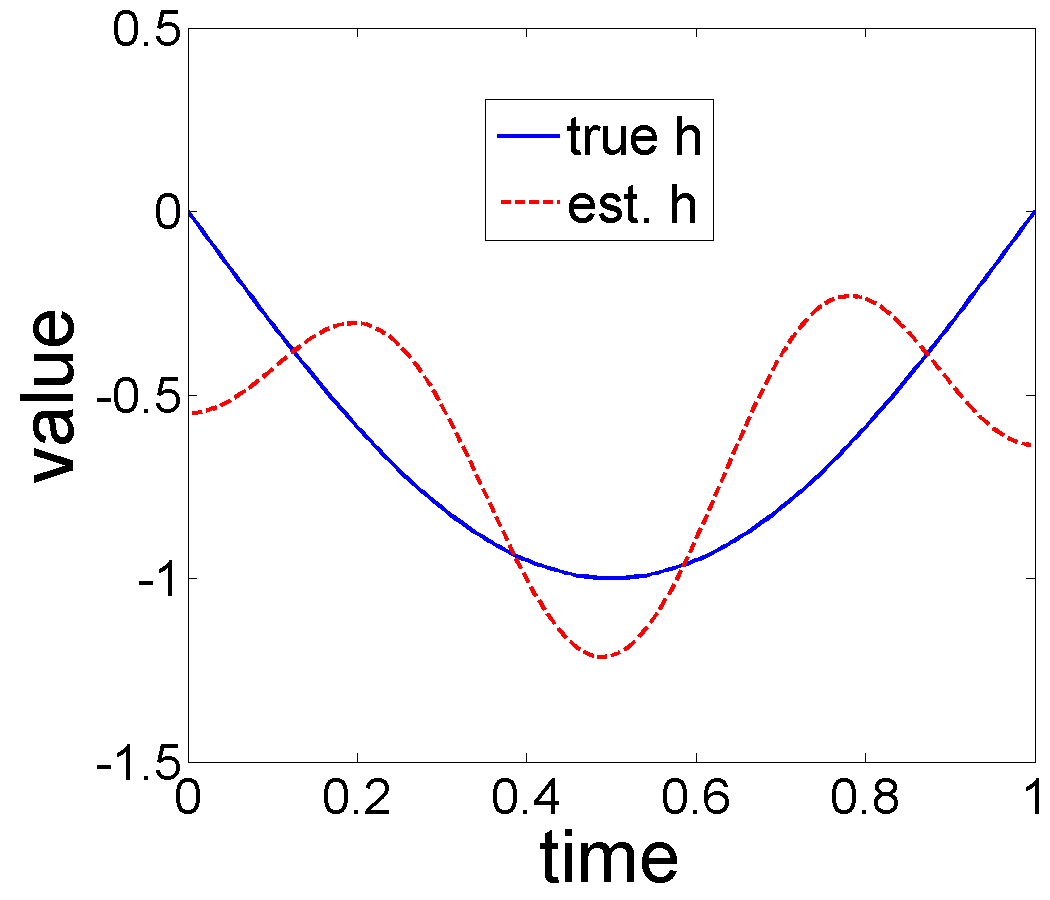} & 
\includegraphics[scale=0.165]{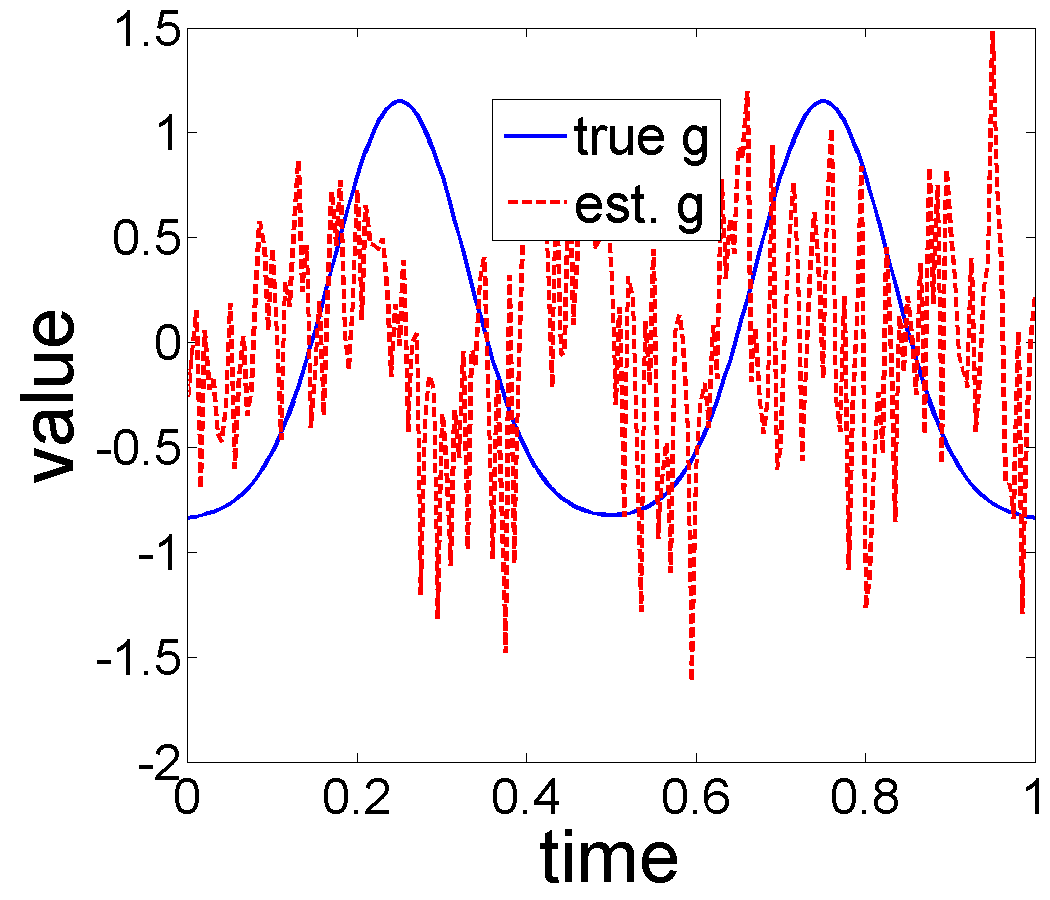} &
\includegraphics[scale=0.165]{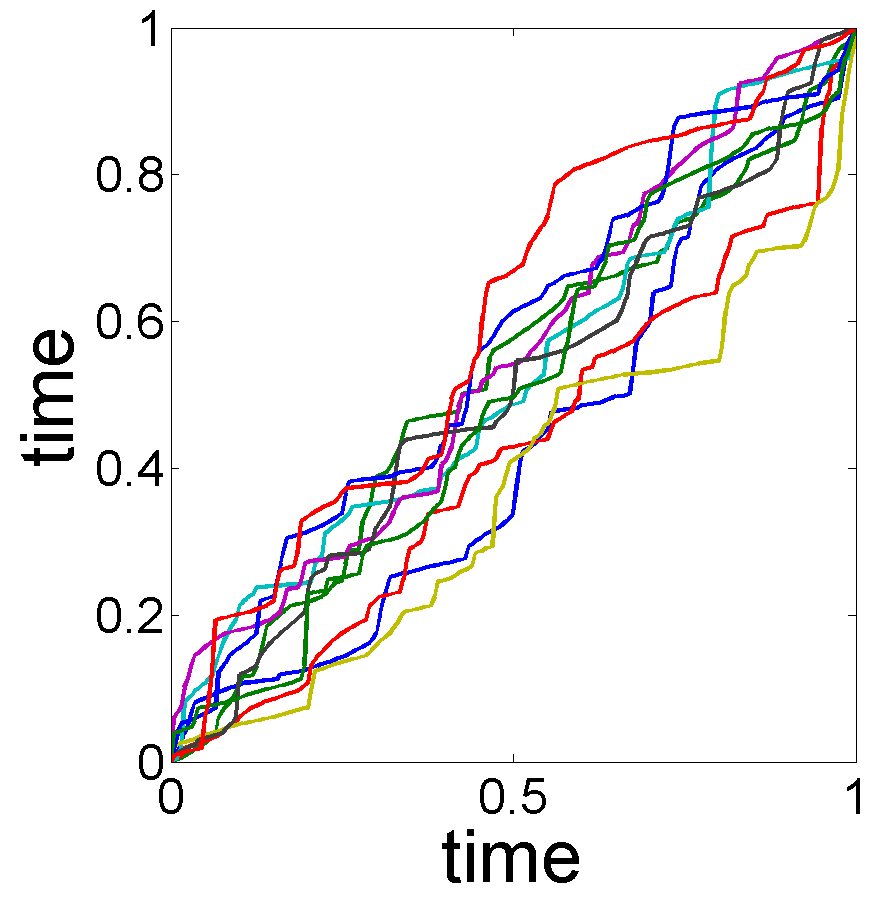}  &
\includegraphics[scale=0.175]{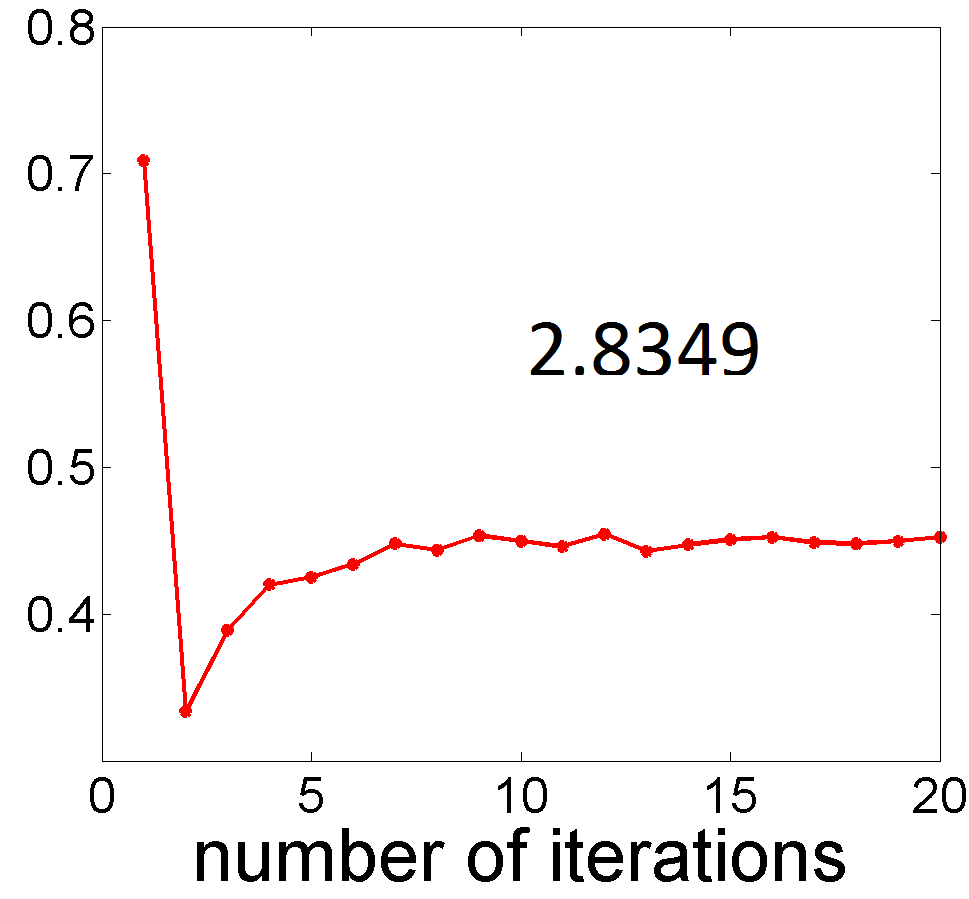}
\tabularnewline
\end{tabular}
\caption{Numerical results of noise perturbation experiment. 
				 Note that figures in the column of negative log-likelihood are plotted in a log scale.
         The number inside their panels are minimized negative log-likelihood at the 20th iteration.}
\label{fig:Synthetic-Data-noise perturbation-2}
\end{center}
\end{figure}

As shown in  Fig. \ref{fig:Synthetic-Data-noise perturbation-2}, the estimation results are very good when the
 noise level is relatively low ($\sigma \leq 0.4)$), with the relative $\ltwo$ error $1.04\times 10^{-1}$,
$1.58\times 10^{-2}$ and $1.19\times 10^{-2}$ for $h$, $g$, and $\{\gamma_i\}$ respectively. 
When the noise increases to $\sigma=0.6$ and $\sigma=0.8$, 
the reconstructed trends still have the desired pattern over $[0,1]$ 
but the recovered seasonality results are not good as noise level $(\sigma\leq 0.4)$. 
With large noise, $\sigma=1.6$, all the estimated results are far from their truth values. 
These experiments provide evidence that the MLE algorithm 
can recover good estimates of the trend, seasonality and warping functions in the presence of some levels of noise. 

%%%%%%%%%%%%%%%%%%%%%%%%%%%%%%%%%%%%%%%%%%%%%%%%%%%%%%%%%%%%%%%%%%%%%%%%%%%%%%%%%%%%%
\item \textbf{Illustration of Testing Using Bootstrap}: 
In this experiment, we illustrate the use of a bootstrap technique for testing different 
hypotheses associated with the shapes of estimated trend and seasonal effect. 
The general idea was described in Section \ref{sec: Bootstrap} 
and it is applied here to the data shown in Fig. \ref{fig:Motivation1-f}. 
In this data, we have used $g(t)=\cos(10\pi t)$,  $h(t) =1.5e^{-3t}$, 
and $\gamma_{i}(t)= \frac{e^{a_{i}t}-1}{e^{a_{i}}-1}, a_i= -3+i\frac{6}{n}$ for $i=1,...,n$.
Additionally, we set $\sigma=0$.  
The estimated trend and seasonality components are in Fig. \ref{fig:Motivation1-f}.
Fig. \ref{fig:Synthetic-Bootstrap-replicate-g-h}, (a) and (b), shows the bootstrap replicates 
$\{\hat{h}_{b}\}$ and $\{\hat{g}_{b}\}$ of the trend and seasonality estimates for $B=500$. 
As these replicates indicate, there is a significant phase variation in the replicates $\{\hat{g}_b\}$
and relatively small variability in the replicates $\{\hat{h}_b\}$. 
The bootstrap technique yields the following results:
 
\begin{figure}[!htb]
\begin{center}
   \subfloat[bootstrap replicates of $\hat{h}$]
	    {\includegraphics[scale=0.22]{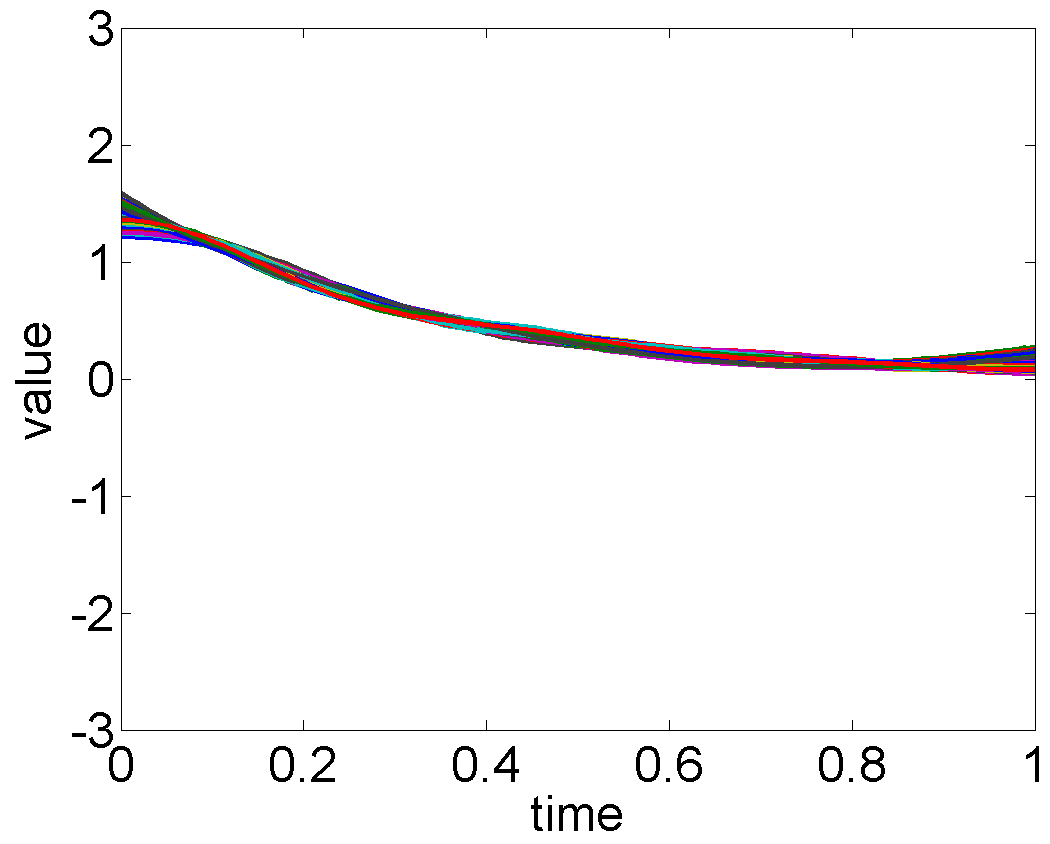}}
   \subfloat[bootstrap replicates of $\hat{g}$]
	    {\includegraphics[scale=0.22]{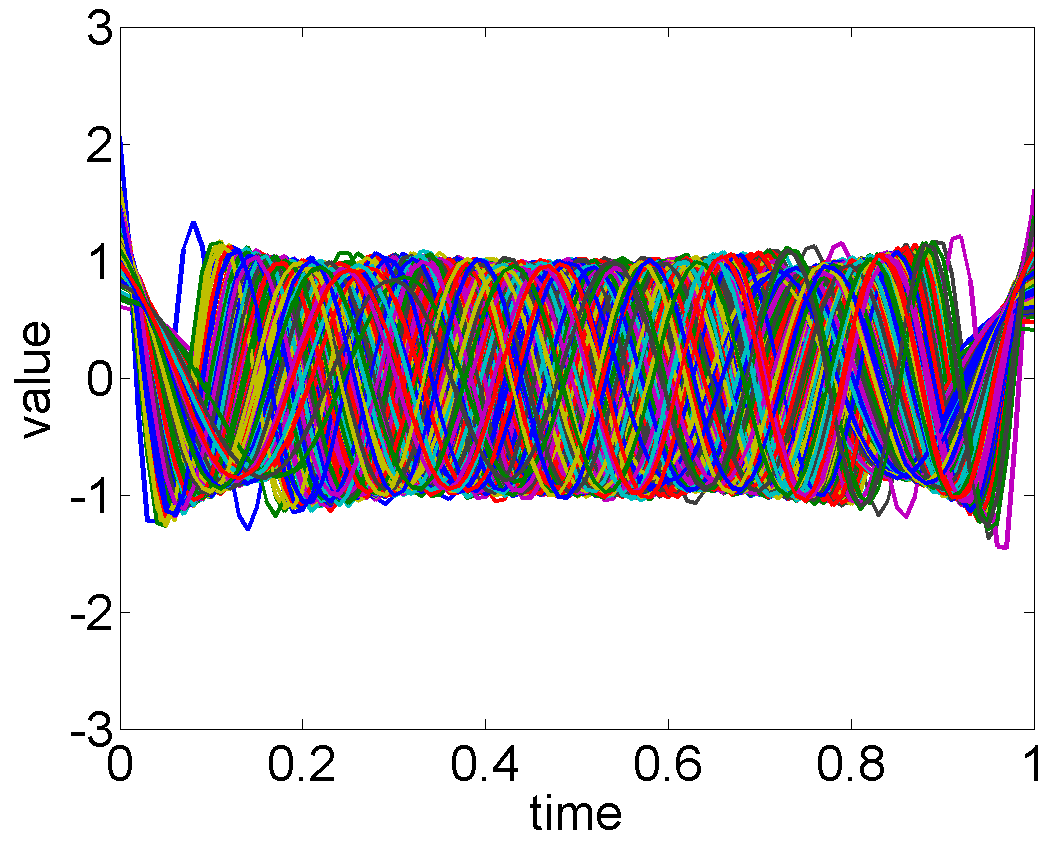}}
	 \subfloat[confidence band of $\hat{h}$]
	    {\includegraphics[scale=0.25]{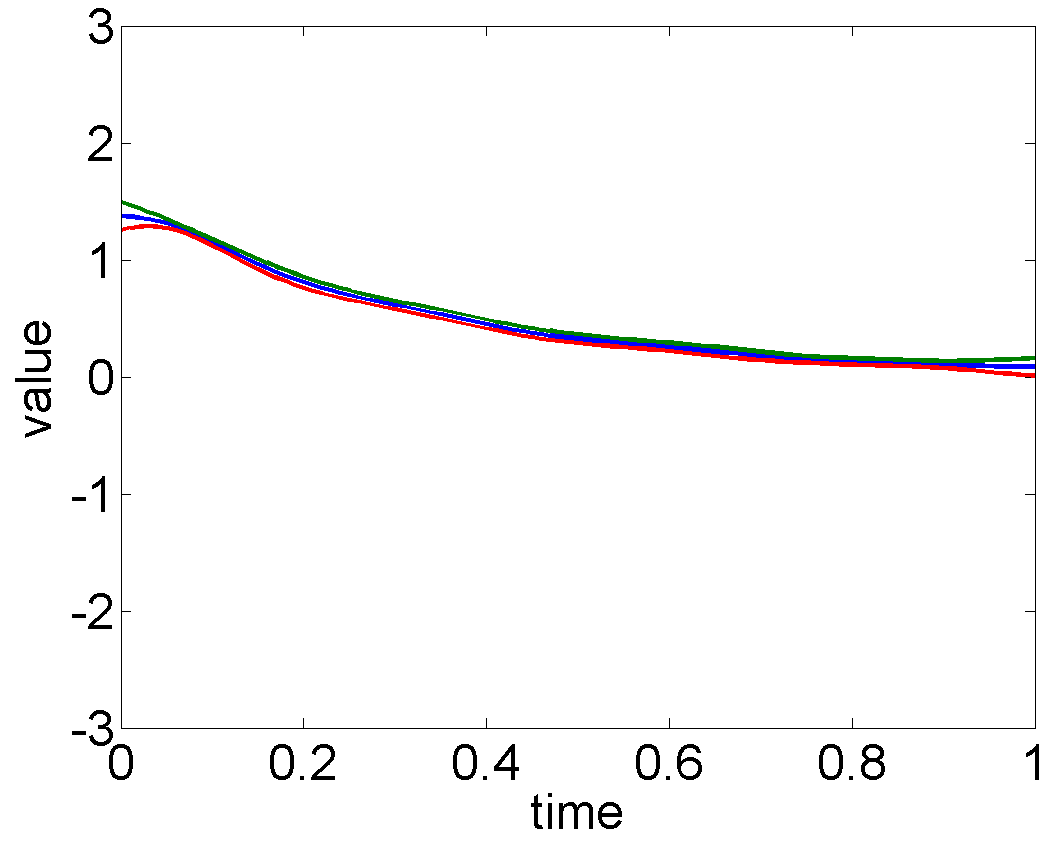}}
   \subfloat[confidence band of $\hat{g}$ ]
	    {\includegraphics[scale=0.25]{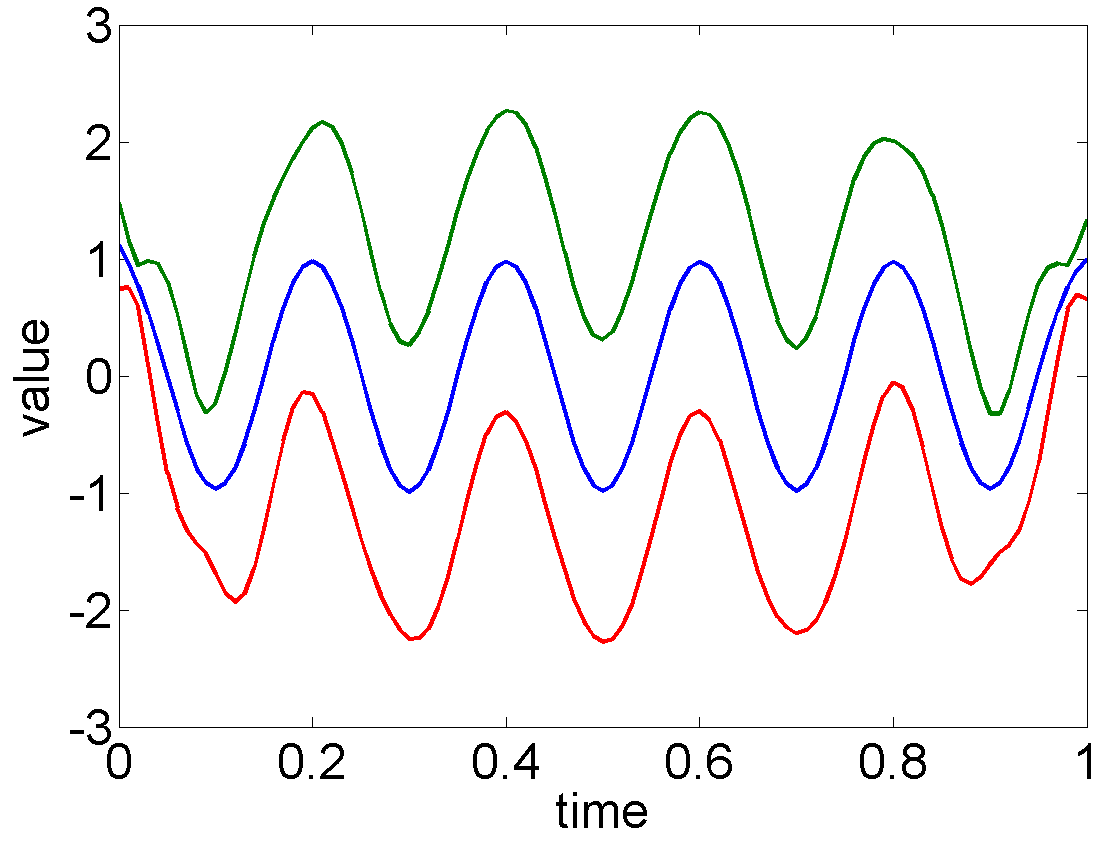}}
\caption{Five hundred bootstrap replicate with its cross-sectional confidence band to Fig. \ref{fig:Motivation1-f} data.}
\label{fig:Synthetic-Bootstrap-replicate-g-h}
\end{center}
\end{figure}

\begin{itemize}
	\item \textbf{Testing presence of a trend:} For testing $h=0$,
	      the test statistic using bootstrap is $\rho_{h_0}=0.61$ with $\hat{se}_B=3.5\times 10^{-3}$  
	      and a $p$ value of 0. The null hypothesis $h=0$ is therefore rejected. 
	\item \textbf{Testing constant shape for a trend:} 
	      For testing $h=c$, the test statistic using bootstrap is $\rho_{h_c}=0.38$ 
				with $\hat{se}_B=5.2\times 10^{-3}$ and a $p$ value of 0.
	      Therefore, we reject the null hypothesis: $h=c$. 
	\item \textbf{Testing linear shape for a trend:} For the linearity of a trend, 
				we obtain $\rho_{h_l}=1.05$, $\hat{se}_B=5.2\times 10^{-3}$, 
	      and a $p$ value of 0. Thus, we reject the null hypothesis: $h$ is a linear function. 
\end{itemize}

The cross-sectional confidence bands with $95\%$ confidence level in 
Fig. \ref{fig:Synthetic-Bootstrap-replicate-g-h}, (c) and (d), 
indicate that the estimator $\hat{h}$ is much better than the $\hat{g}$.
\end{enumerate}

%%%%%%%%%%%%%%%%%%%%%%%%%%%%%%%%%%%%%%%%%%%%%%%%%%%%%%%%%%%%%%%%%%%%%%%%%%%%%%%%%%%%%
%%%%%%%%%%%%%%%%%%%%%%%%%%%%%%%%%%%%%%%%%%%%%%%%%%%%%%%%%%%%%%%%%%%%%%%%%%%%%%%%%%%%%
\subsection{Real Data} \label{sec: Real Data}

%%%%%%%%%%%%%%%%%%%%%%%%%%%%%%%%%%%%%%%%%%%%%%%%%%%%%%%%%%%%%%%%%%%%%%%%%%%%%%%%%%%%%
\subsubsection{Berkeley Male Growth Velocity} \label{sec: Berkeley Male Growth Velocity}
\begin{figure}[!htb]
\begin{center}
   \subfloat[height]
	     {\includegraphics[scale=0.34]{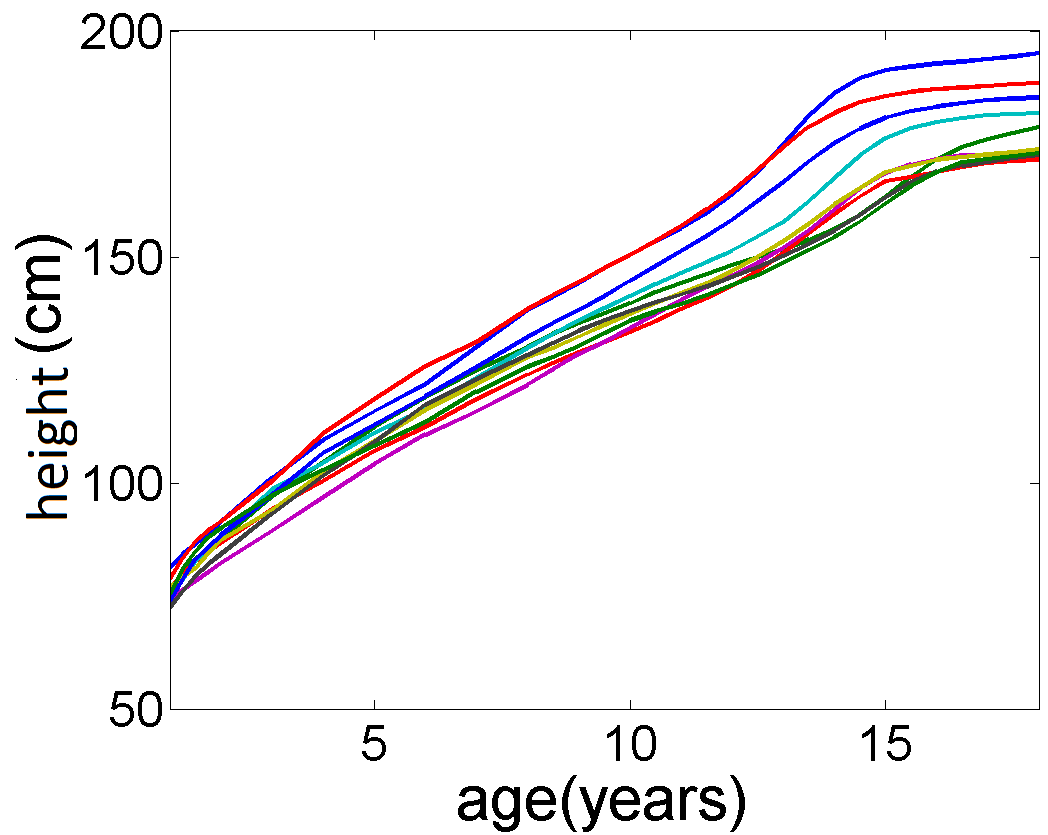}}
   \subfloat[growth velocity]
	     {\includegraphics[scale=0.3]{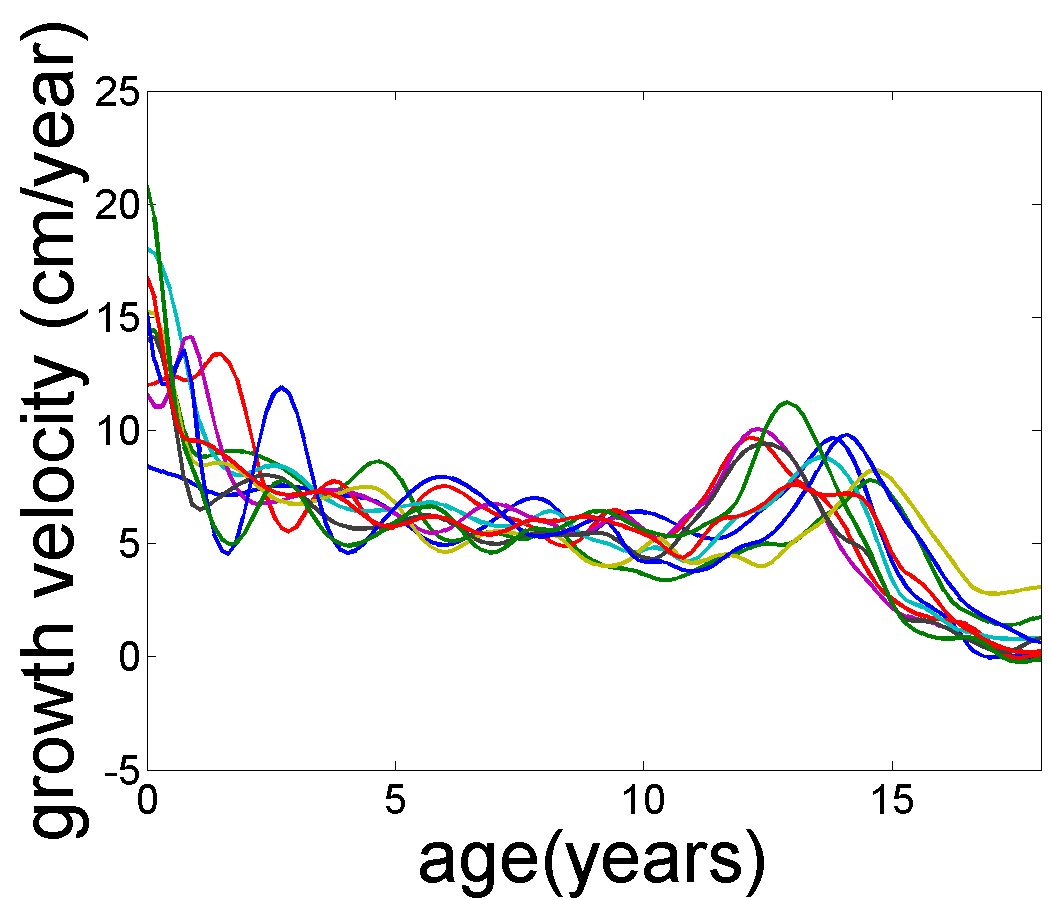}}
\caption{Ten curves data of Berkeley Male Growth Data.}
\label{fig:Male-Growth-Velocity-data}
\end{center}
\end{figure}

The Berkeley Growth Study of 54 females and 39 males was performed by \cite{tuddenham1954physical} 
and further discussed by \cite{ramsay2006functional}. 
As an example, 10 out of 39 male growth curves are shown in Fig. \ref{fig:Male-Growth-Velocity-data}(a). 
Due to the monotonic nature of the curve, researchers often analyze the time derivative, termed the
{\it growth velocity}, shown in Fig. \ref{fig:Male-Growth-Velocity-data}(b). 
Looking at these growth curves, one can discern some 
patterns of high growth, termed growth spurts, for all subjects. 
However, naturally these spurts are not synchronized across subjects, i.e. 
they occur at different times for different subjects. Additionally, there
is a downward trend in the growth velocities across all subjects. Our goal is to estimate the overall trend and 
the seasonal effect in this velocity data, and to test their significance. 

\begin{figure}[!htb]
\begin{center}
   \subfloat[recovered trend]
	     {\includegraphics[scale=0.22]{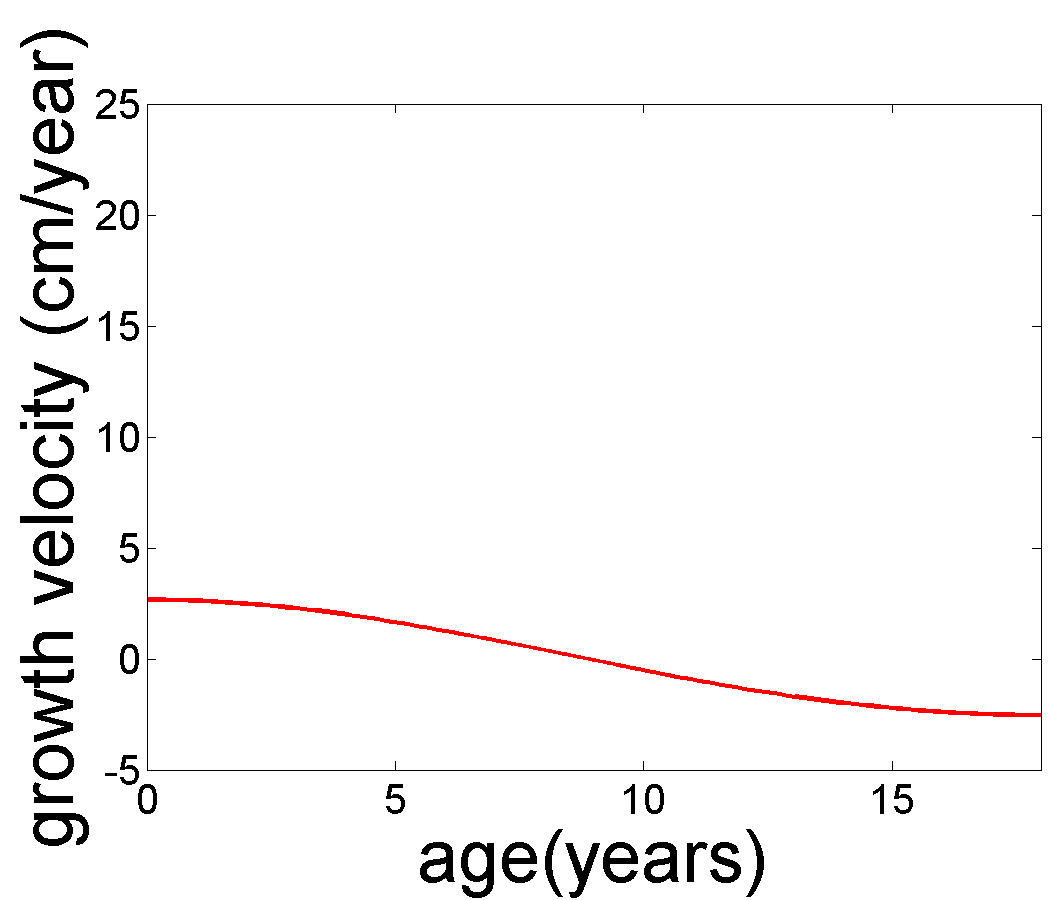}} 
   \subfloat[recovered seasonality]
	     {\includegraphics[scale=0.22]{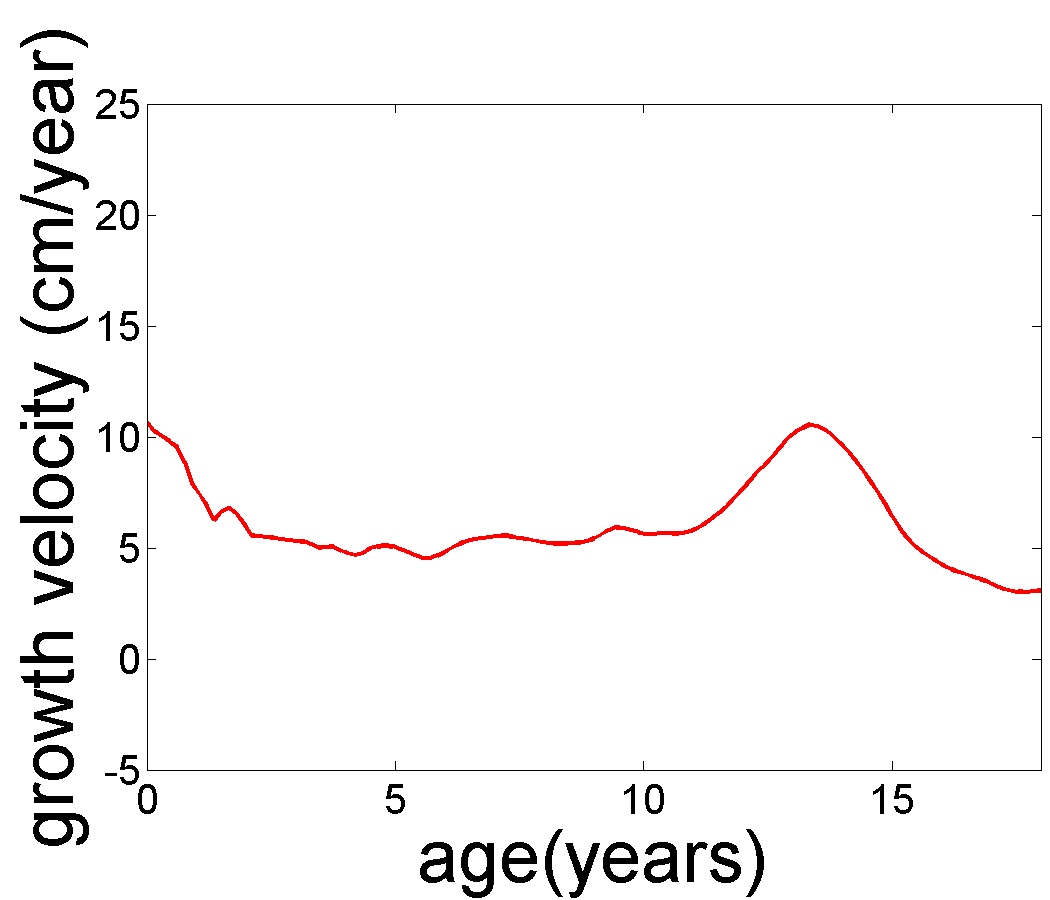}} 
   \subfloat[recovered warpings]
	     {\includegraphics[scale=0.22]{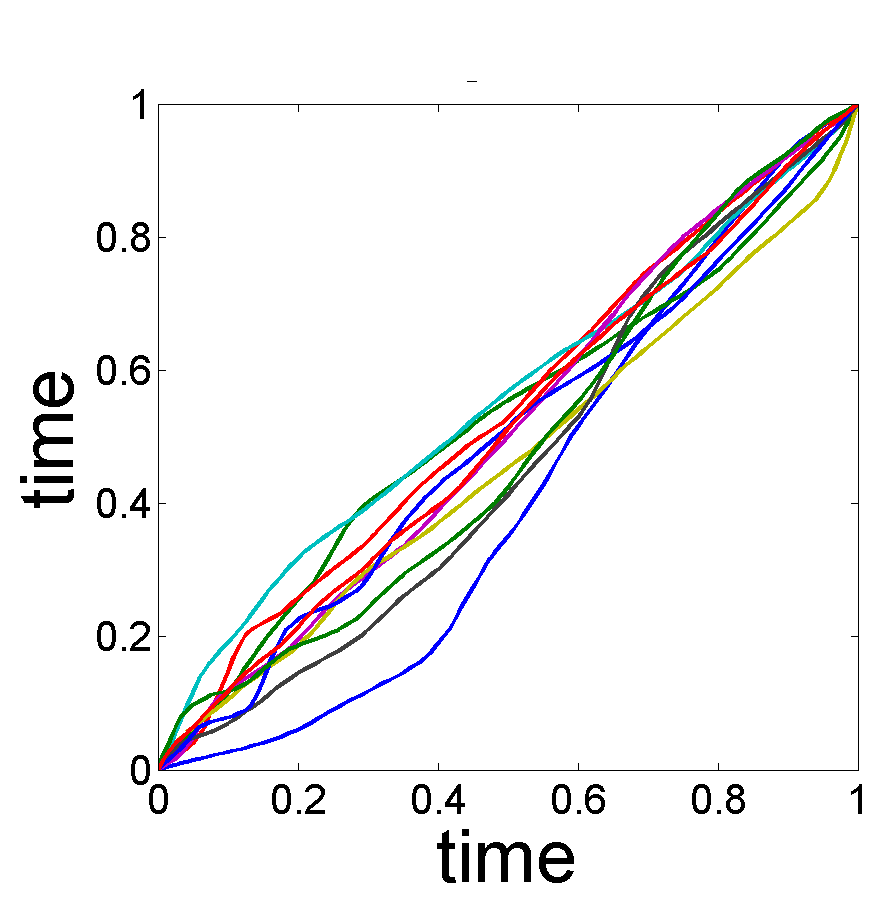} }
   \subfloat[negative log-likelihood]
	     {\includegraphics[scale=0.24]{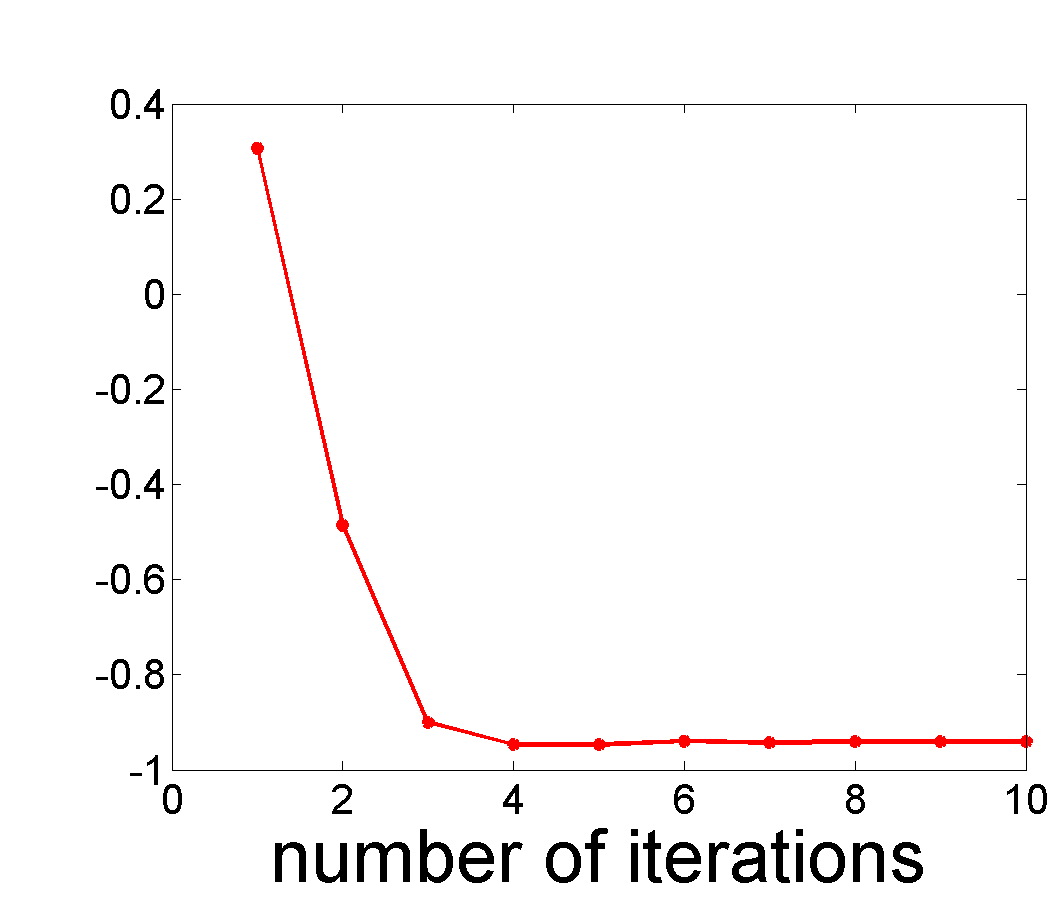}}
\caption{Estimated results of growth velocity data using Algorithm \ref{alg: CD for Signal Separation}.}
\label{fig:Male-Growth-Velocity-reconstruct}
\end{center}
\end{figure}

For this data, our subspace selection finds the optimal trend subspace to be ${\cal{H}}=span\{\cos(\pi t),\cos{(2 \pi t)}\}$. 
The estimated trend and seasonal components for this $\cal H$ are shown in Fig. \ref{fig:Male-Growth-Velocity-reconstruct}. 
As we can see, our algorithm detects a near-linear decay in growth velocity with age and a growth spurt around age 13. 

\begin{figure}[!htb]
\begin{center}
	 \subfloat[${\cal H}=span\{1\}$]
	      {\includegraphics[scale=0.17]{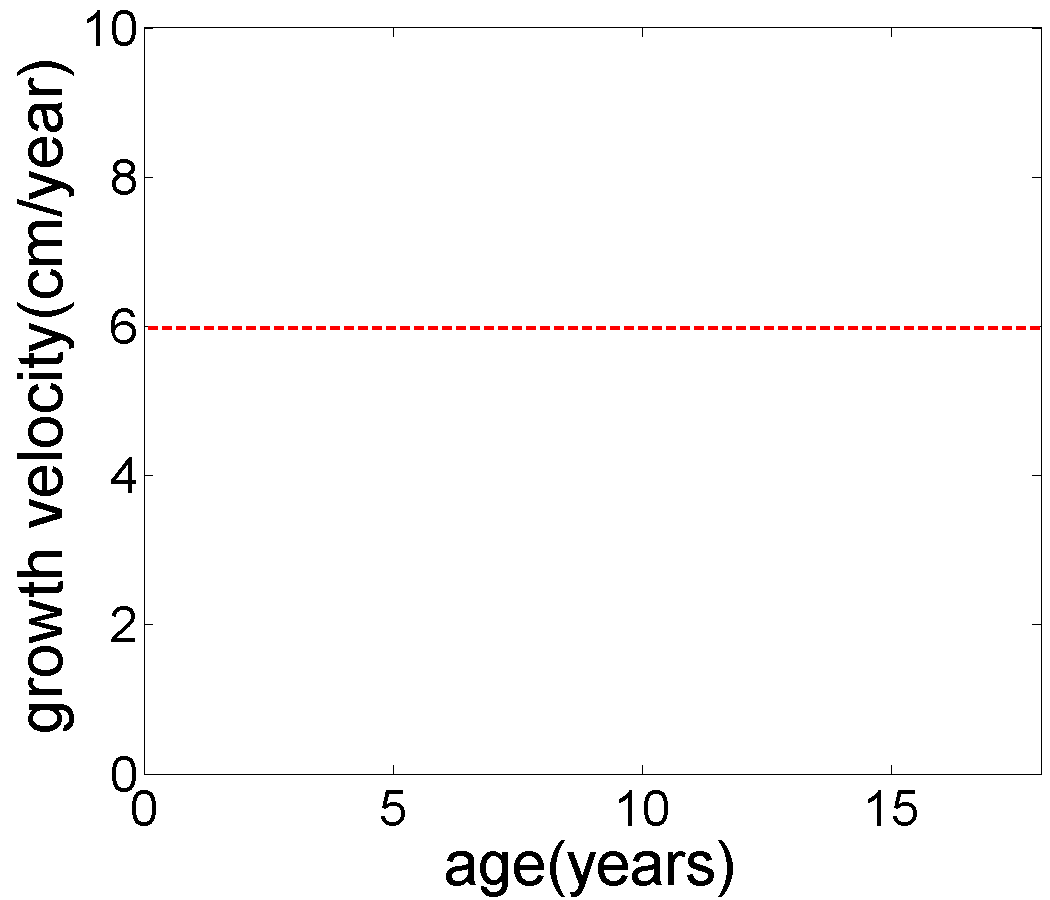}}
	 \subfloat[${\cal H}=span\{1,t\}$]
	      {\includegraphics[scale=0.17]{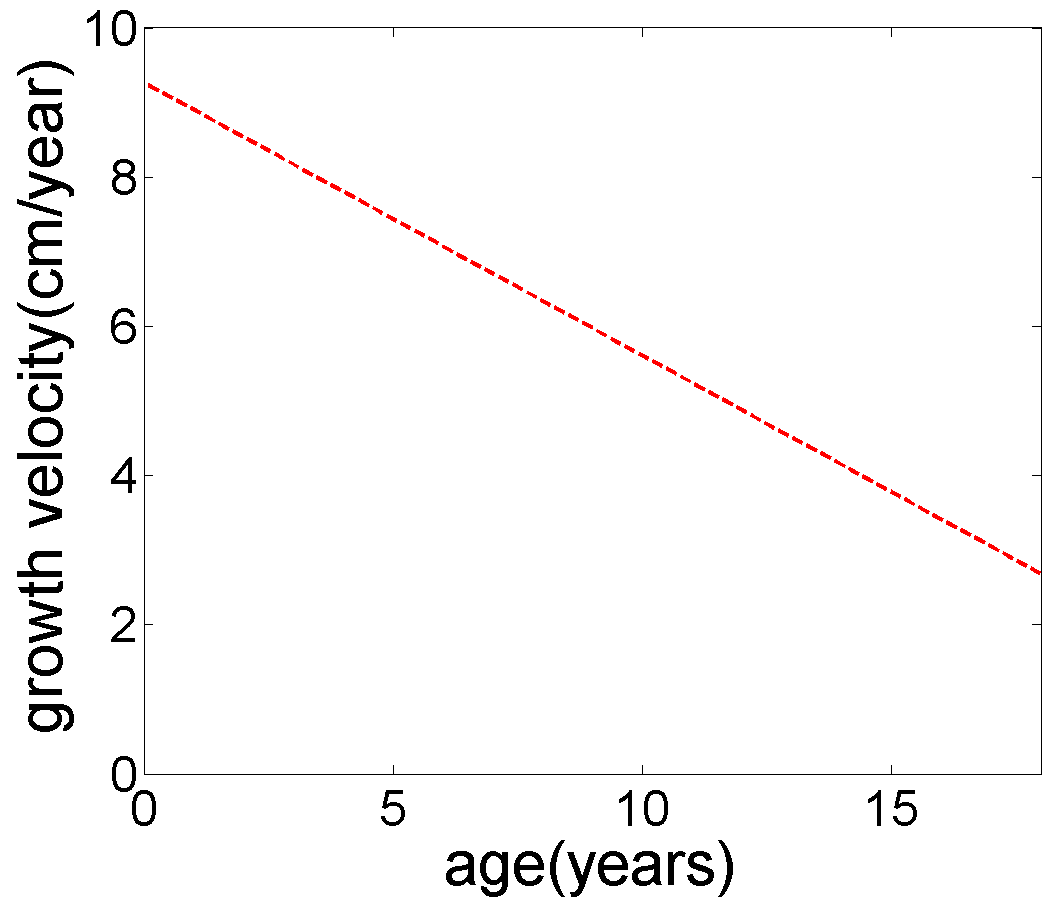}} 
	 \subfloat[Fourier, $\{\phi_1,\phi_2,\phi_3\}$]
 	      {\includegraphics[scale=0.17]{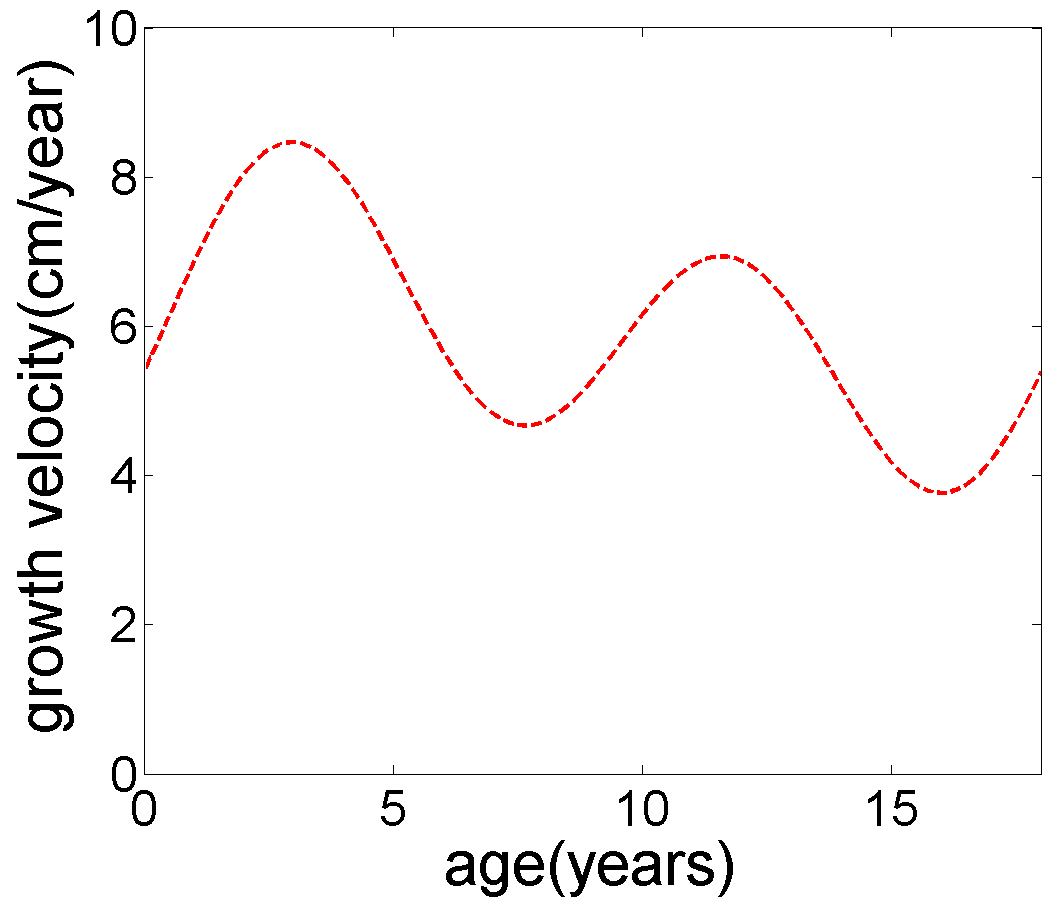}} 
	 \subfloat[${\cal H}=span\{1,\sin(\pi t)\}$]
	      {\includegraphics[scale=0.17]{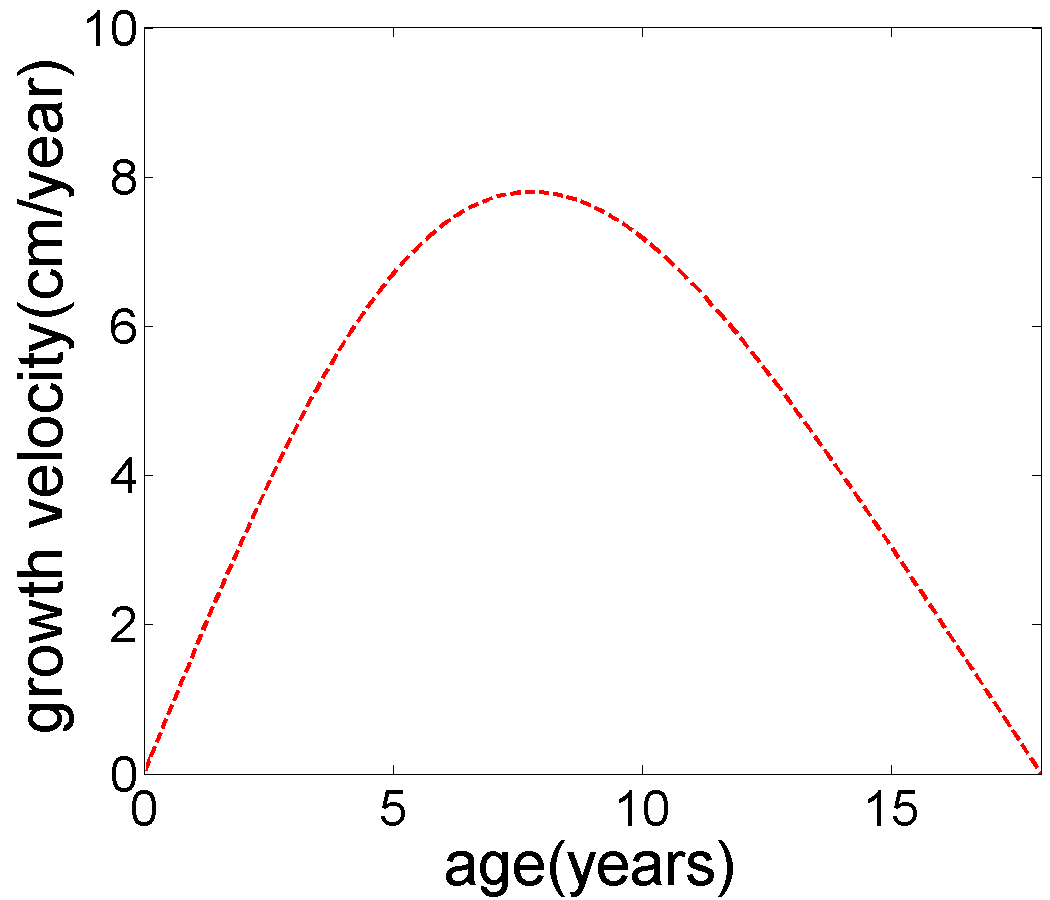}} 
	 \subfloat[${\cal H}=span\{1,\cos(\pi t)\}$]
	      {\includegraphics[scale=0.17]{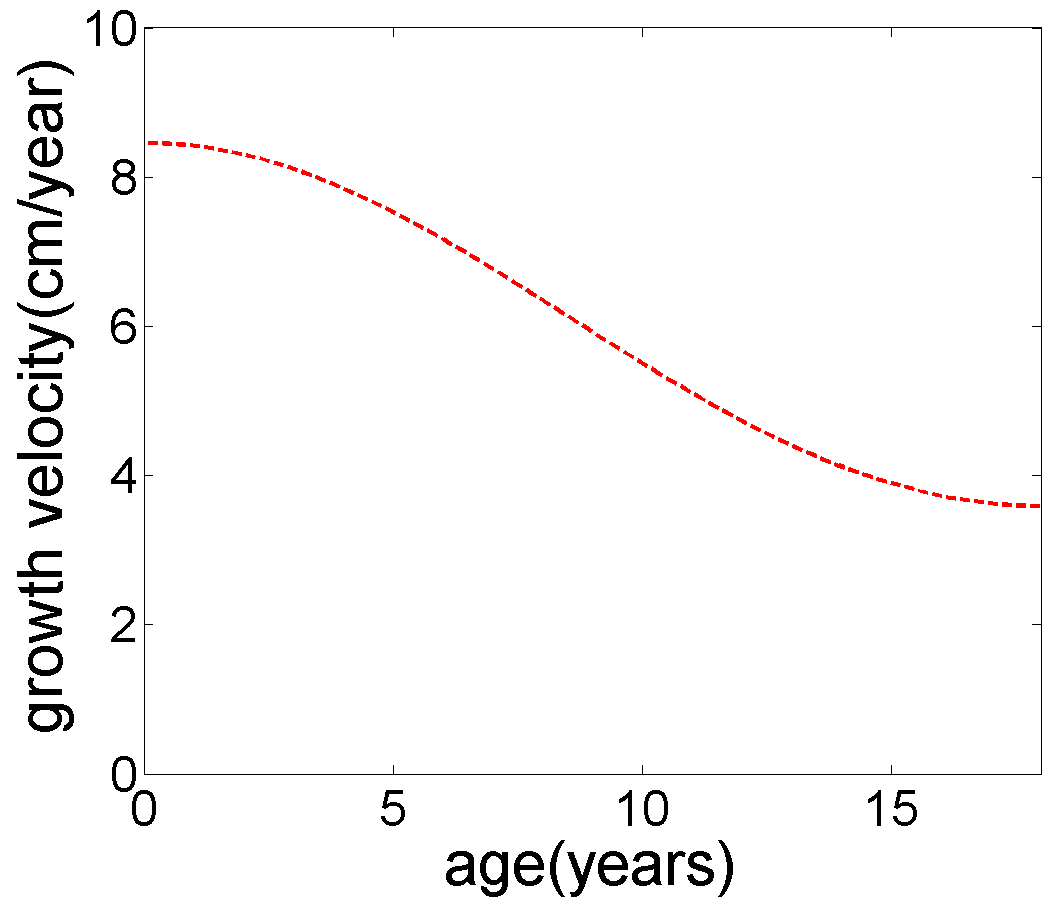}} 									
\caption{Trend estimation results of growth velocity data of the separation model using different subspace selections.}
\label{fig: simple separation for growth velocity}	
\end{center}
\end{figure}

For comparison purposes, the trend estimation results of applying the separation model $f_i(t)=h(t)+g(t)+\epsilon_i(t)$ 
 are shown in Fig. \ref{fig: simple separation for growth velocity}. 
This method requires a specification of ${\cal H}$, and the results are sensitive to this choice.  
Recall from Remark \ref{remark of separation model}, this approach cannot select ${\cal H}$, as is done in our MLE algorithm. 
Clearly the results do not capture the expected properties of the trend. 
Due to the lack of phase variability in the separation model, the estimates of $g$
do not capture the desired growth spurt and not presented in the paper.

The bootstrap technique yields the following results:
\begin{itemize}
   \item \textbf{Testing presence of a trend:} When testing for the null hypothesis $h=0$ 
	         using the bootstrap technique, we get $\rho_{h_0}=1.83$, $\hat{se}_B=0.44$, and 
           a $p$ value of $1.5\times 10^{-5}$. As a result, we reject the null hypothesis. 
	\item \textbf{Testing constant shape for a trend:} 
	          When testing for null hypothesis $h=c$, the test statistic is $\rho_{h_c}=1.83$ with $\hat{se}_B=0.44$.
	          Since the corresponding $p$ value is $1.5\times 10^{-5}$, we reject the null hypothesis: $h=c$.
	 \item \textbf{Testing linear shape for a trend:} When testing the linearity of a trend,
	              we obtain test statistic $\rho_{h_l}=2.52$ with $\hat{se}_B=8.08$, and a $p$ value of 0.37.
	              We fail to reject that the null hypothesis: the trend $h$ is a linear. 
								This conclusion is consistent with Fig. \ref{fig:Male-Growth-Velocity-reconstruct}(a). 
\end{itemize}

\begin{figure}[!htb]
\begin{center}
   \subfloat[confidence band of $\hat h$]
	      {\includegraphics[scale=0.25]{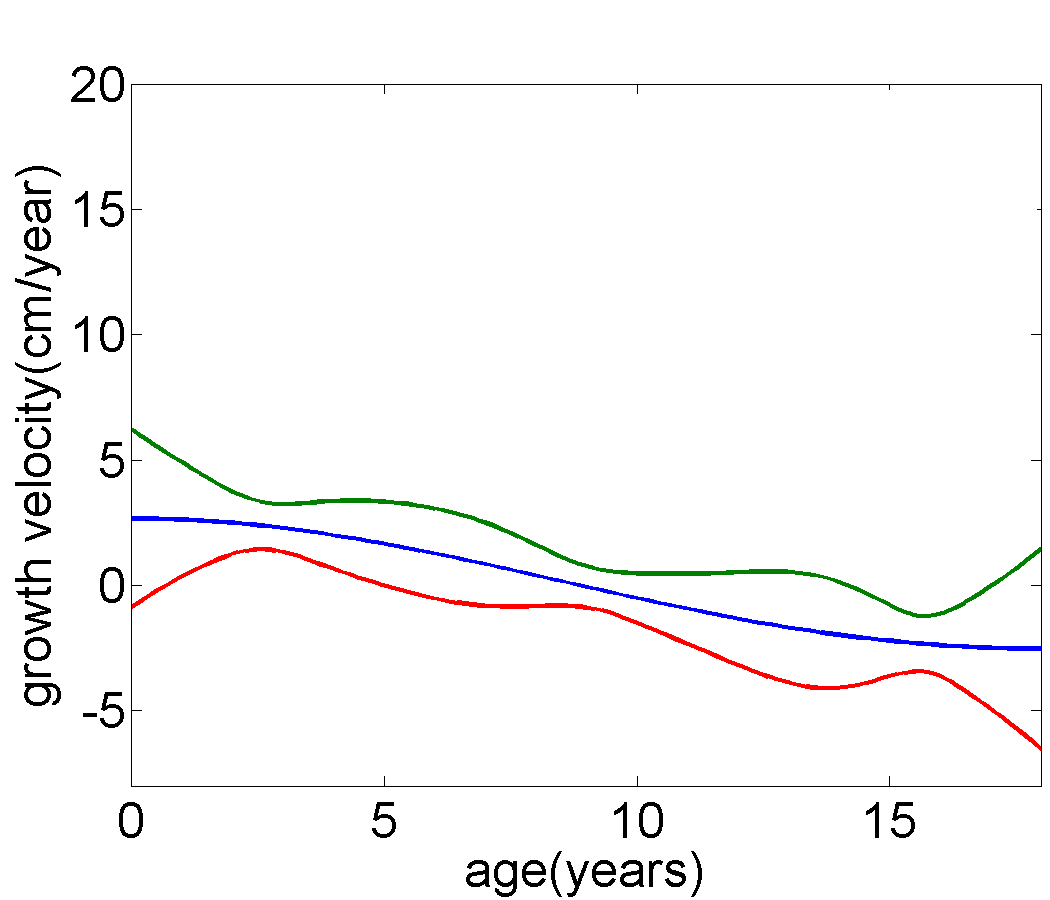}}
   \subfloat[confidence band of $\hat g$]
	      {\includegraphics[scale=0.25]{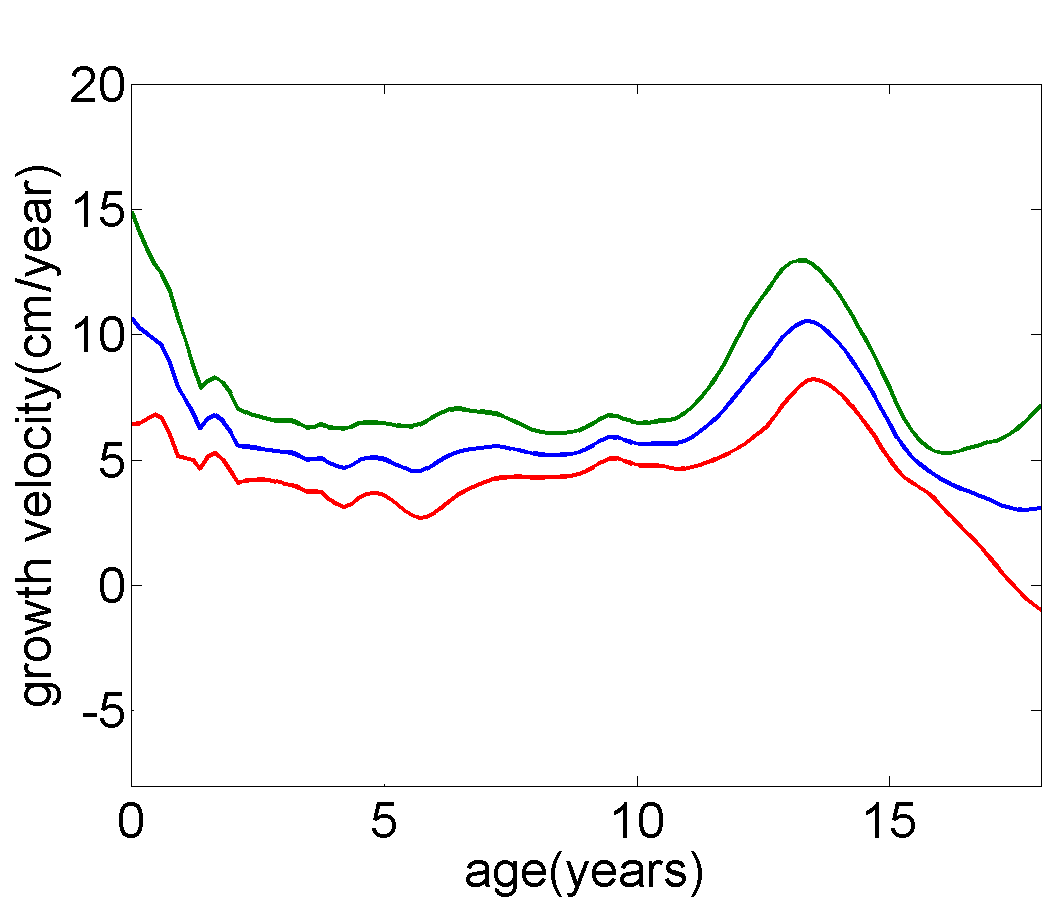}}
\caption{Cross-sectional confidence band of male Growth Velocity data.}
\label{fig:Male-Growth-Velocity-Bootstrap-replicate-g-h}
\end{center}
\end{figure}

Finally, Fig. \ref{fig:Male-Growth-Velocity-Bootstrap-replicate-g-h}, (a) and (b), shows cross-sectional confidence bands for
$\hat{h}$ and $\hat{g}$ at $95\%$ confidence level. 

%%%%%%%%%%%%%%%%%%%%%%%%%%%%%%%%%%%%%%%%%%%%%%%%%%%%%%%%%%%%%%%%%%%%%%%%%%%%%%%%%
\subsubsection{U.S. Electricity Price} \label{sec: U.S. Electricity Price}
\begin{figure}[!htb]
\begin{center}
\begin{tabular}{>{\raggedright}m{7.7cm} r}
   \includegraphics[scale=0.30]{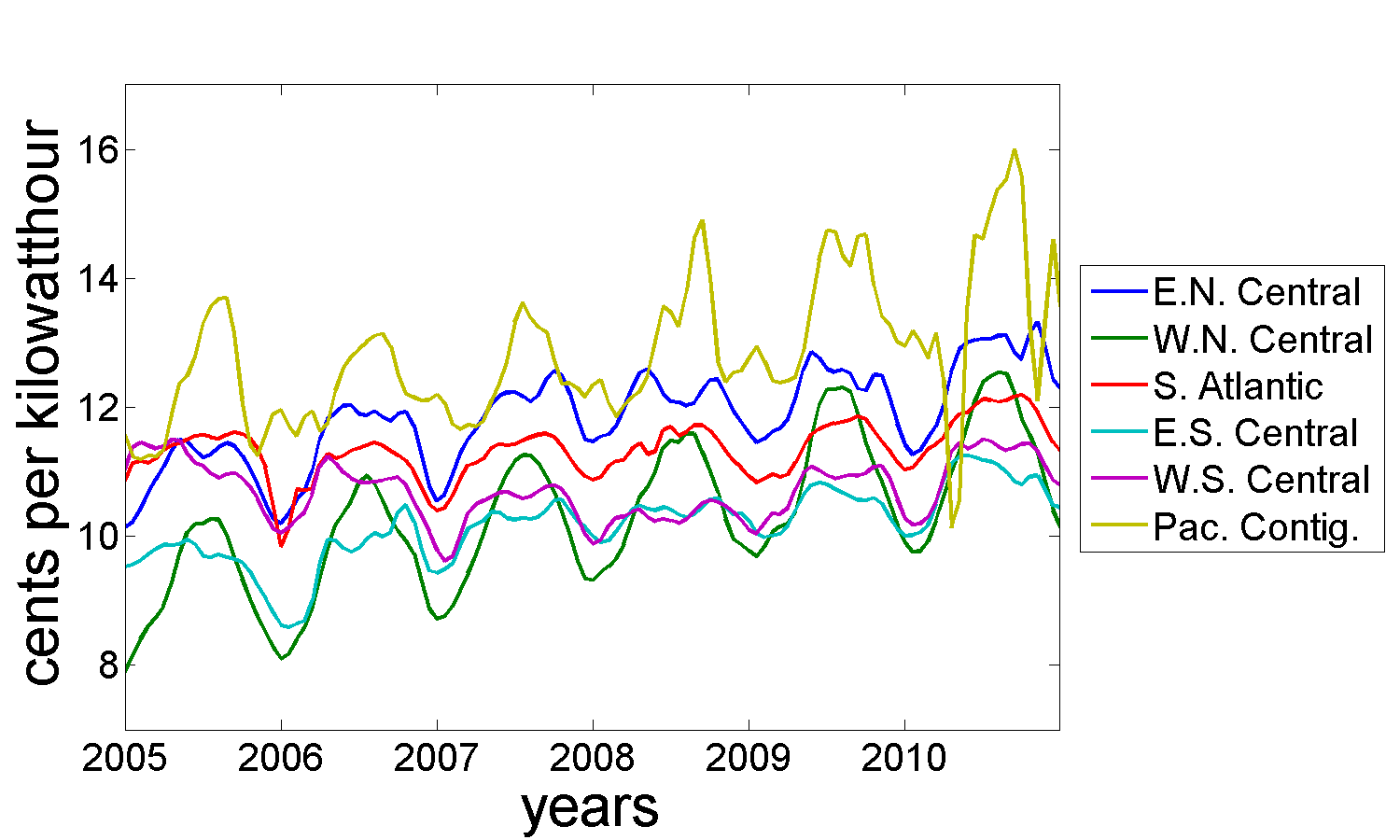} & 
\begin{tabular}{c}
   \subfloat[recovered $\hat h$]
	     {\includegraphics[scale=0.21]{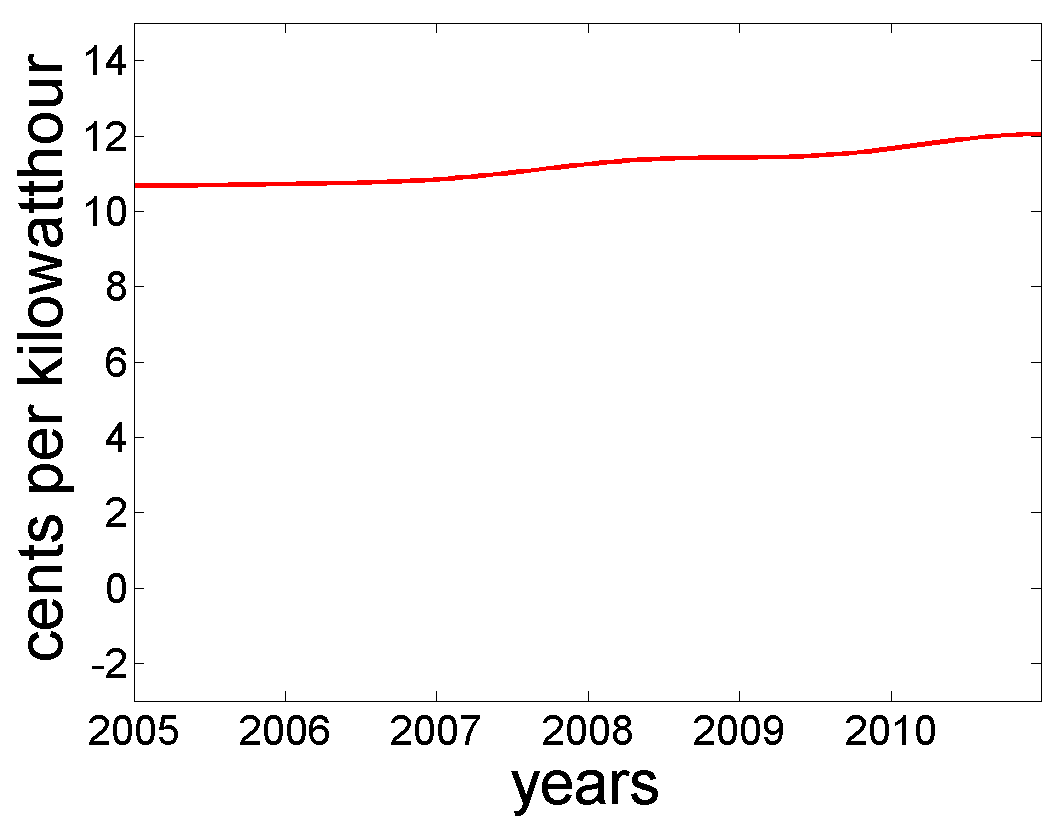}} 
   \subfloat[recovered $\hat g$]
	     {\includegraphics[scale=0.21]{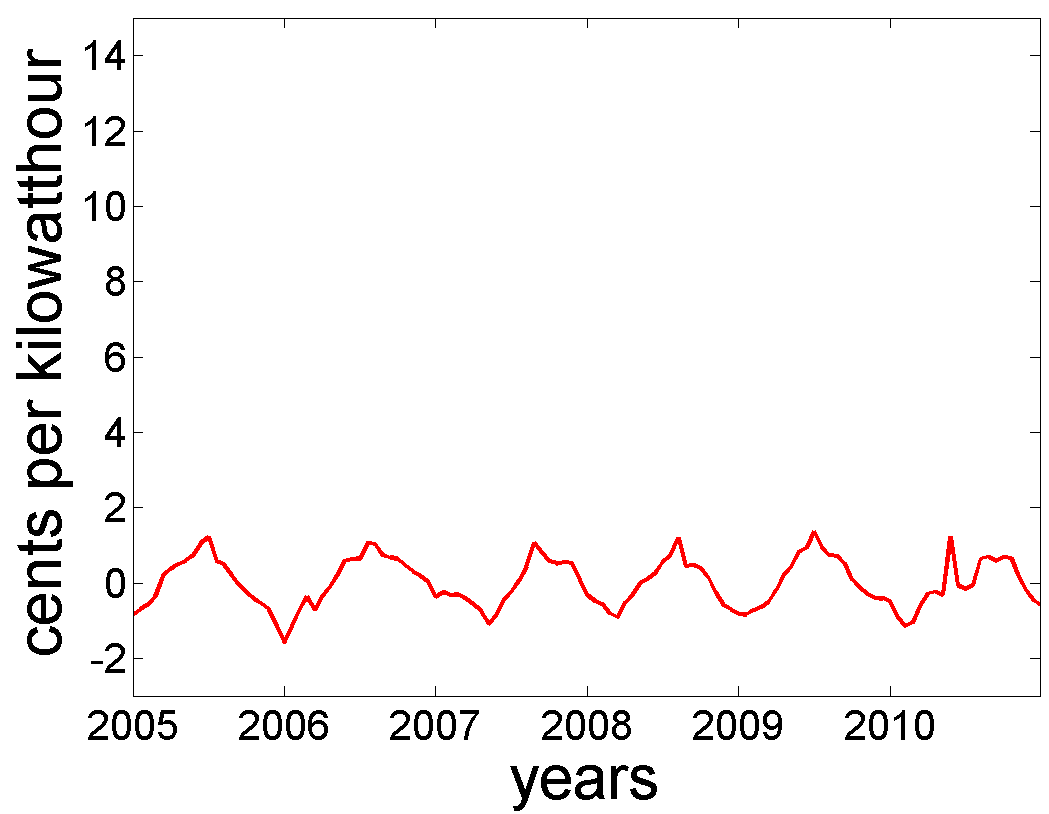}}\\
   \subfloat[recovered $\{\hat \gamma_i\}$]
	     {\includegraphics[scale=0.21]{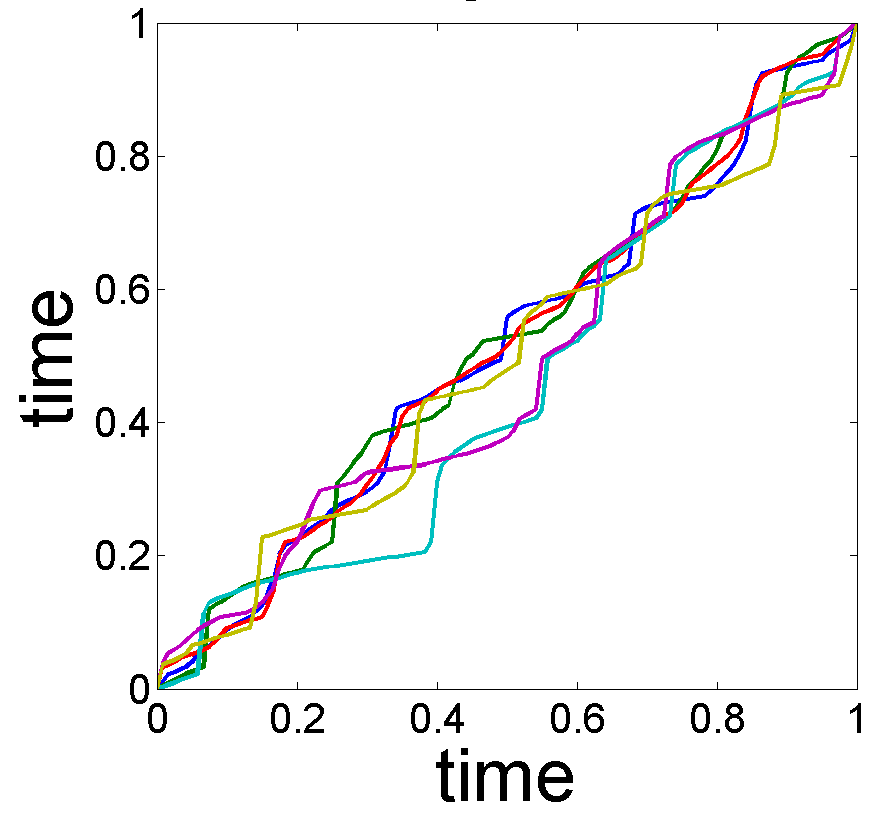}} 
   \subfloat[negative log-likelihood]
	     {\includegraphics[scale=0.235]{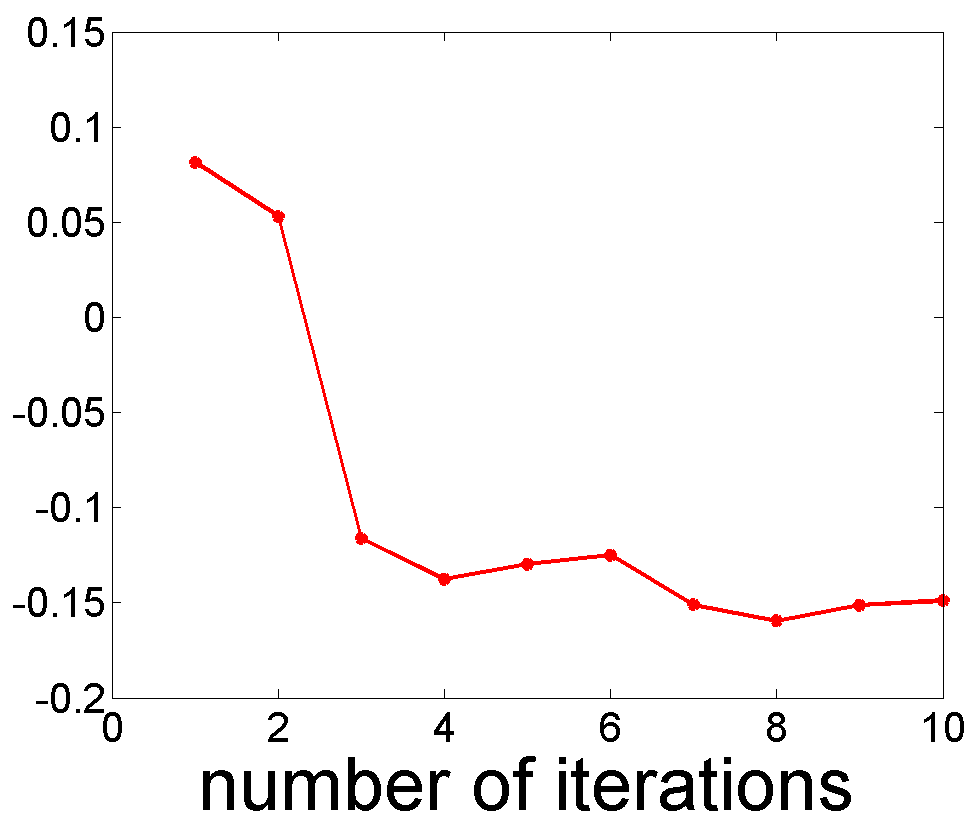}}
\end{tabular}
\end{tabular}
\caption{Left is U.S. electricity price data. Right depicts estimation results.}
\label{fig:US-Electricity-price-data}
\end{center}
\end{figure}

This example studies the monthly U.S. electricity prices for years 2005 to 2010, 
with data as shown in the left panel of Fig. \ref{fig:US-Electricity-price-data}. 
The $x$-axis represents time on a monthly scale from 2005.1 to 2010.12. The $y$-axis is unit price of electricity
in cents per kilowatt-hour. 
According to the data source, the U.S. Energy Information Administration \footnote{http://www.eia.gov/electricity/data.cfm}, 
divides U.S. into several regions: Alaska, Hawaii, New England, Middle Atlantic, 
East North Central, West North Central, South Atlantic, East South Central, 
West South Central, Mountain, Pacific Contiguous and Pacific Non-contiguous. 
We restrict to six regions in this analysis since these regions use the same electricity generation method. 
The raw data is shown in the left panel of Fig. \ref{fig:US-Electricity-price-data}. 
In this data,  we clearly see a seasonal effect on a yearly basis --
electricity prices increase during the summer and fall back during the winter. 
Since there are six annual cycles in the data, 
we expect six peaks in the estimated seasonality $\hat{g}$. 
Also, as the cycles are not quite synchronized, there is a potential for phase variability in the seasonal effect. 
Notice that there is a slowly increasing pattern from 2005 to 2010, pointing to the presence of a trend in the 
data. 

\begin{figure}[!htb]
\begin{center}
	 \subfloat[${\cal H}=span\{1\}$]
	      {\includegraphics[scale=0.17]{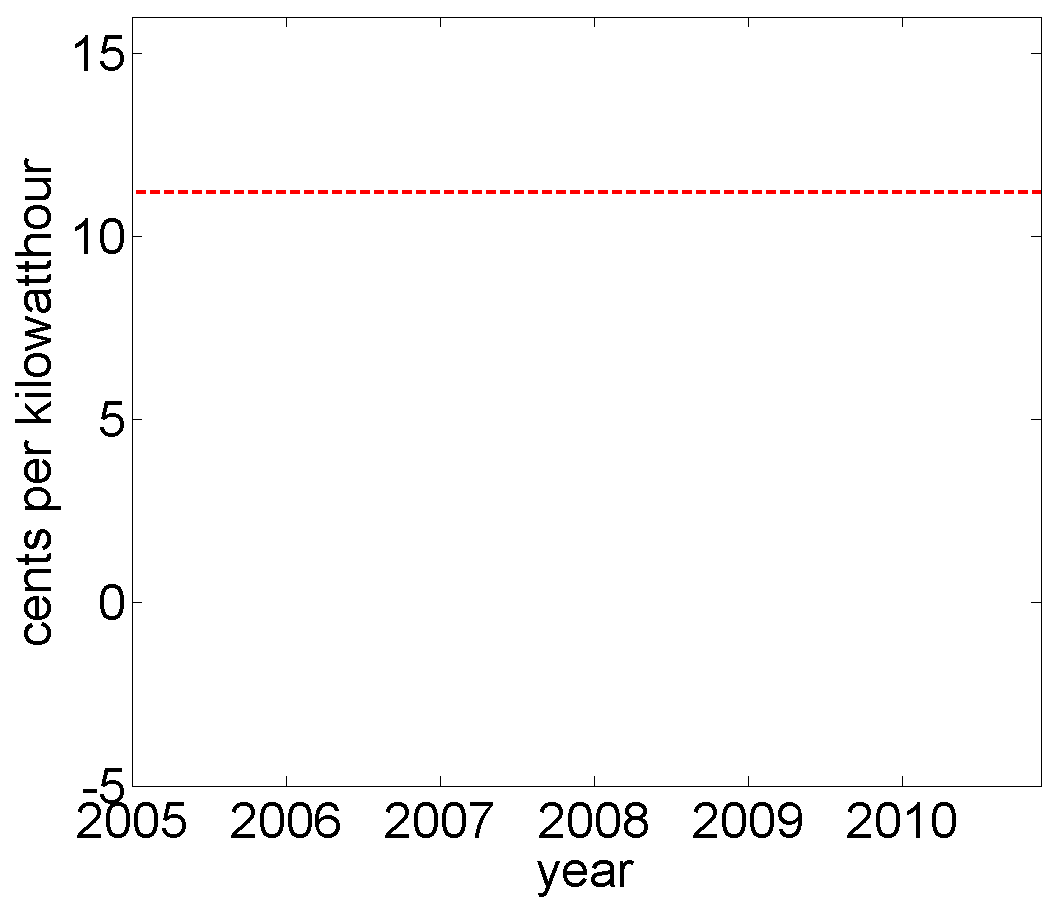}}
	 \subfloat[${\cal H}=span\{1,t\}$]
	      {\includegraphics[scale=0.17]{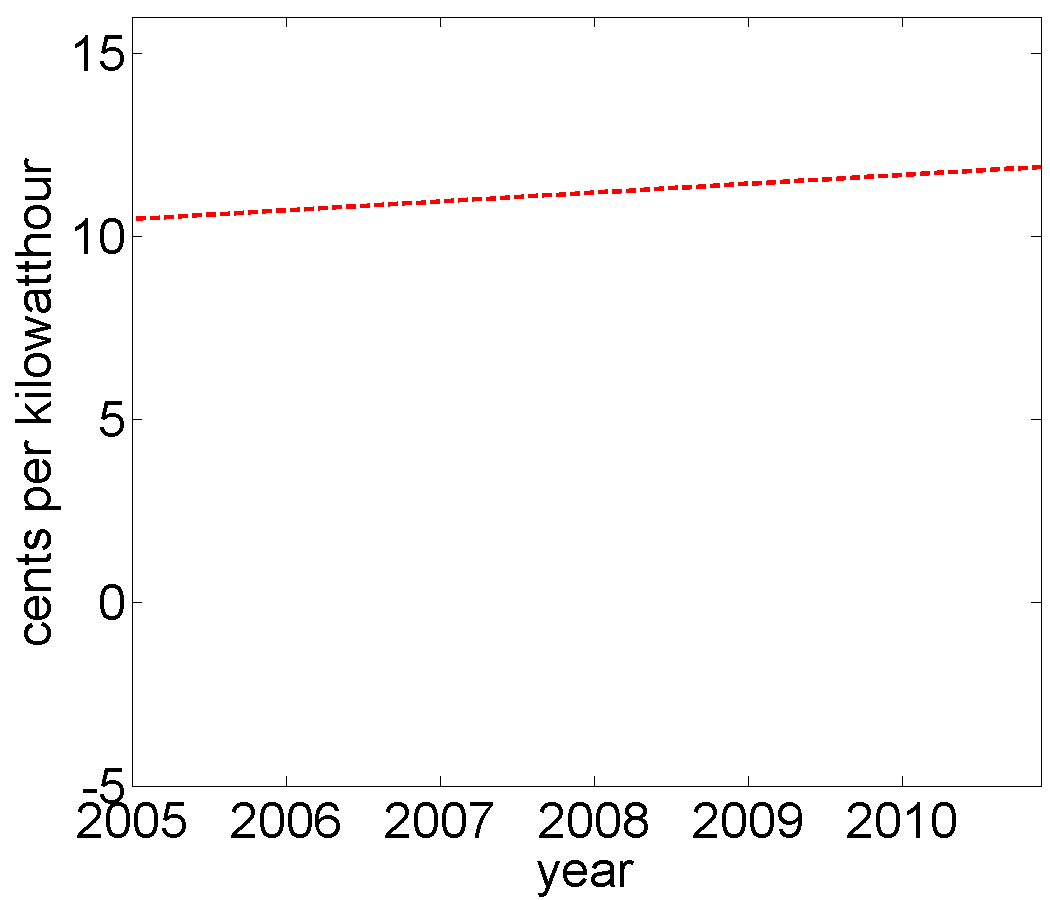}} 
	 \subfloat[Fourier, $\{\phi_1,\phi_2,\phi_3\}$]
 	      {\includegraphics[scale=0.17]{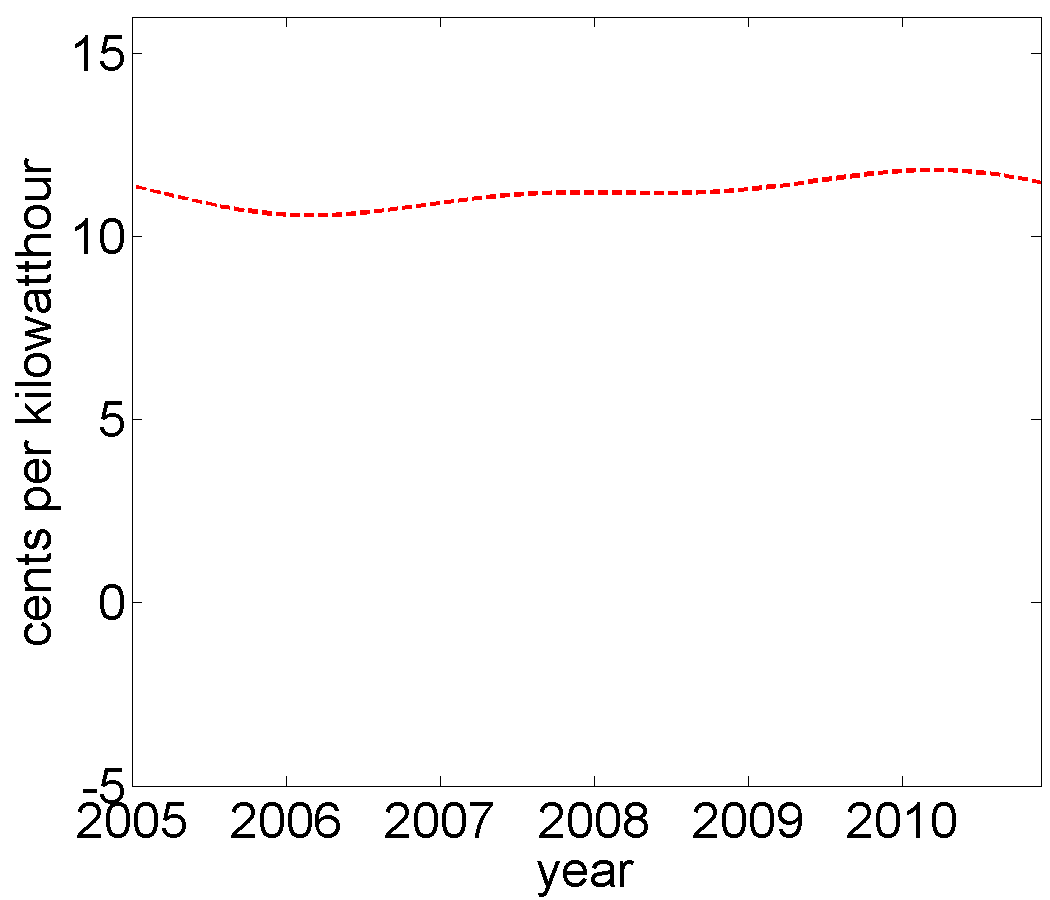}} 
	 \subfloat[${\cal H}=span\{1,\sin(\pi t)\}$]
	      {\includegraphics[scale=0.17]{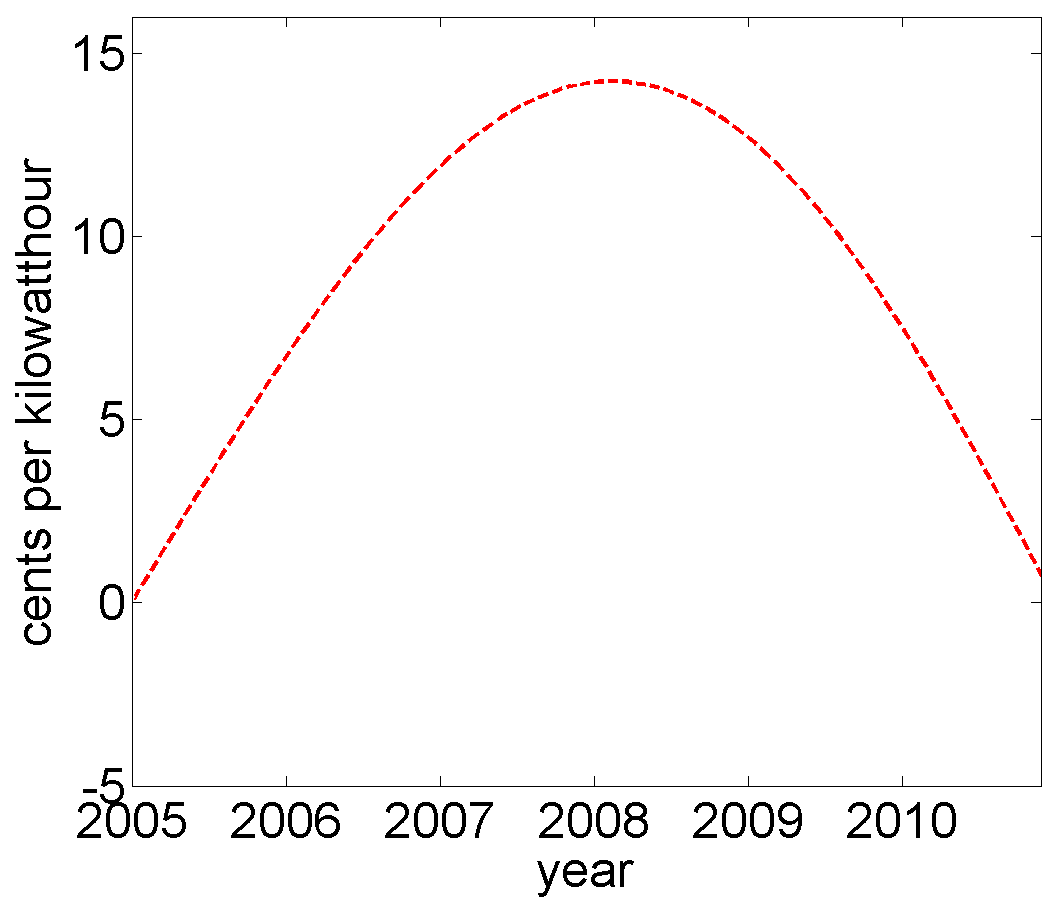}} 
	 \subfloat[Cosine $\{\phi_1,\phi_2,\phi_3\}$]
	      {\includegraphics[scale=0.17]{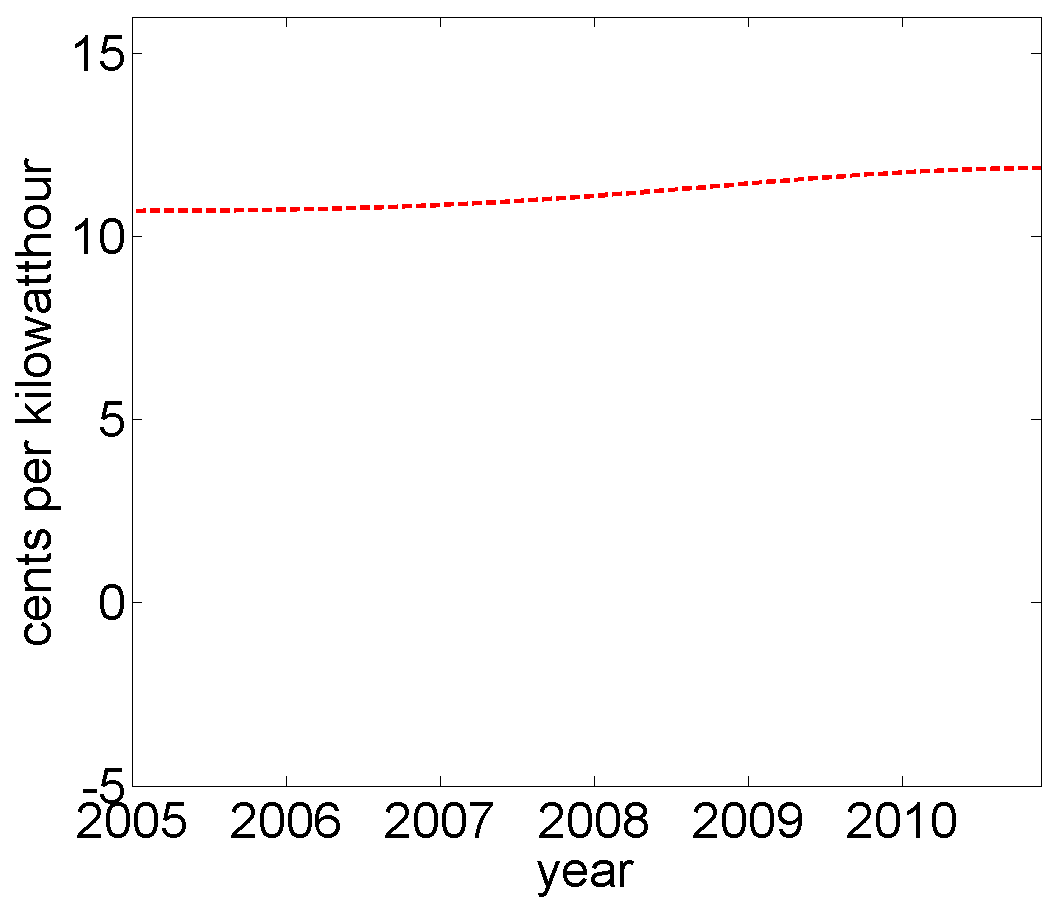}} 									
\caption{Trend estimation results of electricity price data of the separation model using different subspace selections.}
\label{fig: simple separation for electricity price}	
\end{center}
\end{figure}

We applied our subspace selection approach to this data, and found the 
optimal choice of trend subspace is ${\cal H}=span\{1,\cos(\pi t),...,\cos(5 \pi t)\}$. 
With this choice of ${\cal H}$, 
the MLE of the trend and seasonal effects are shown in Fig. \ref{fig:US-Electricity-price-data}, (a) and (b). 
The trend estimation results of the separation model $f_i(t)=h(t)+g(t)+\epsilon_i(t)$ with different choices of $\cal H$
are shown in Fig. \ref{fig: simple separation for electricity price}. 
While some results (Fig. \ref{fig: simple separation for electricity price}(c) and 
Fig. \ref{fig: simple separation for electricity price}(e)) seem decent, we do not have a way of 
selecting one over the other. 
Moreover, since the lack of phase variability in the separation model, the estimates of $g$
do not process the desired seasonal oscillation.

We now turn to bootstrap hypothesis testing of the results to the MLE algorithm.   

\begin{itemize}
   \item \textbf{Testing presence of a trend:} For testing if the trend is null, 
                  we obtain test statistic $\rho_{h_0}=11.21$ with $\hat{se}_B=0.39$.
	                The associated $p$ value is 0 and this trend $h$ is not null. 
									%Similarly for testing if the seasonality is null,
					        %we obtain $\rho_{g_0}=0.62$, bootstrap standard error $\hat{se}_B=0.42$, and a $p$ value of
									%$0.067$. This implies that the null hypothesis can be accepted for some confidence 
									%levels and rejected for other levels.
	 \item \textbf{Testing constant shape for a trend:} 
	            For testing null hypothesis $h=c$, we obtain test statistic $\rho_{h_c}=0.43$ and 
							its bootstrap standard error $\hat{se}_B=0.43$.  
							The $p$ value is $0.16$ so we fail to reject that the null hypothesis: trend $h$ is a constant. 
					    %For testing the seasonality $g$, we obtain $\rho_{g_c}=0.62$ with $\hat{se}_B=0.42$.
								% Hence, the null hypothesis, $g=c$, can be accepted for some confidence levels
								% and rejected for other levels because of the $p$ value 0.067. 
								 %we barely infer that this seasonality $g$ is not in a constant shape based on the $p$ value 0.067.
	 \item \textbf{Testing linear shape for a trend:} 
	               For testing the linearity of a trend $h$, we have $\rho_{h_l}=0.17$ and $\hat{se}_B=2.86$.
	               We fail to reject the null hypothesis, $h$ is linear, since a large $p$ value 0.47. 
							%	 For testing the seasonality $g$,
							%	 we get $\rho_{g_l}=7.64$ and its bootstrap standard error $\hat{se}_B=5.09$. 
							%	 Unlike the trend, the null hypothesis, seasonality $g$ is a linear, 
							%	 can be accepted for some confidence levels and rejected for other levels 
							%	 since the $p$ value is 0.068.
\end{itemize}

\begin{figure}[!htb]
\begin{center}
   \subfloat[confidence bands of $\hat h$]
	     {\includegraphics[scale=0.25]{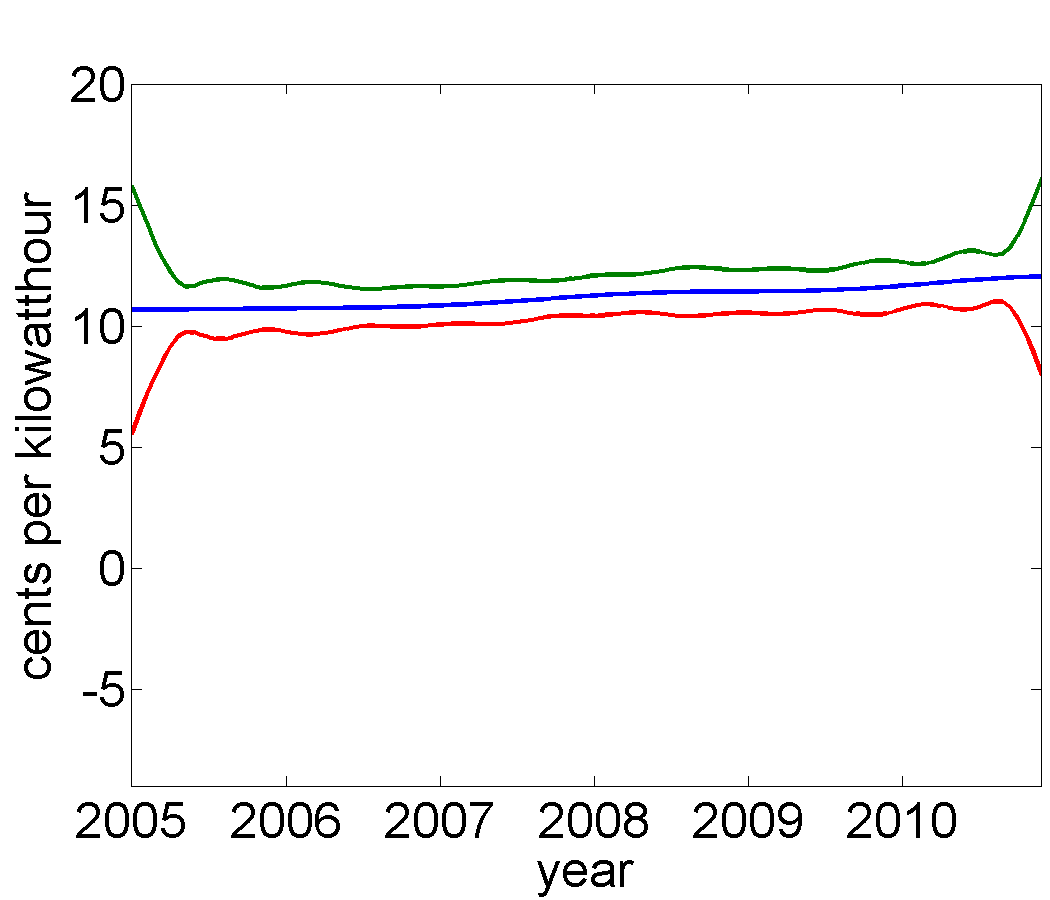}}
   \subfloat[confidence bands of $\hat g$]
	     {\includegraphics[scale=0.25]{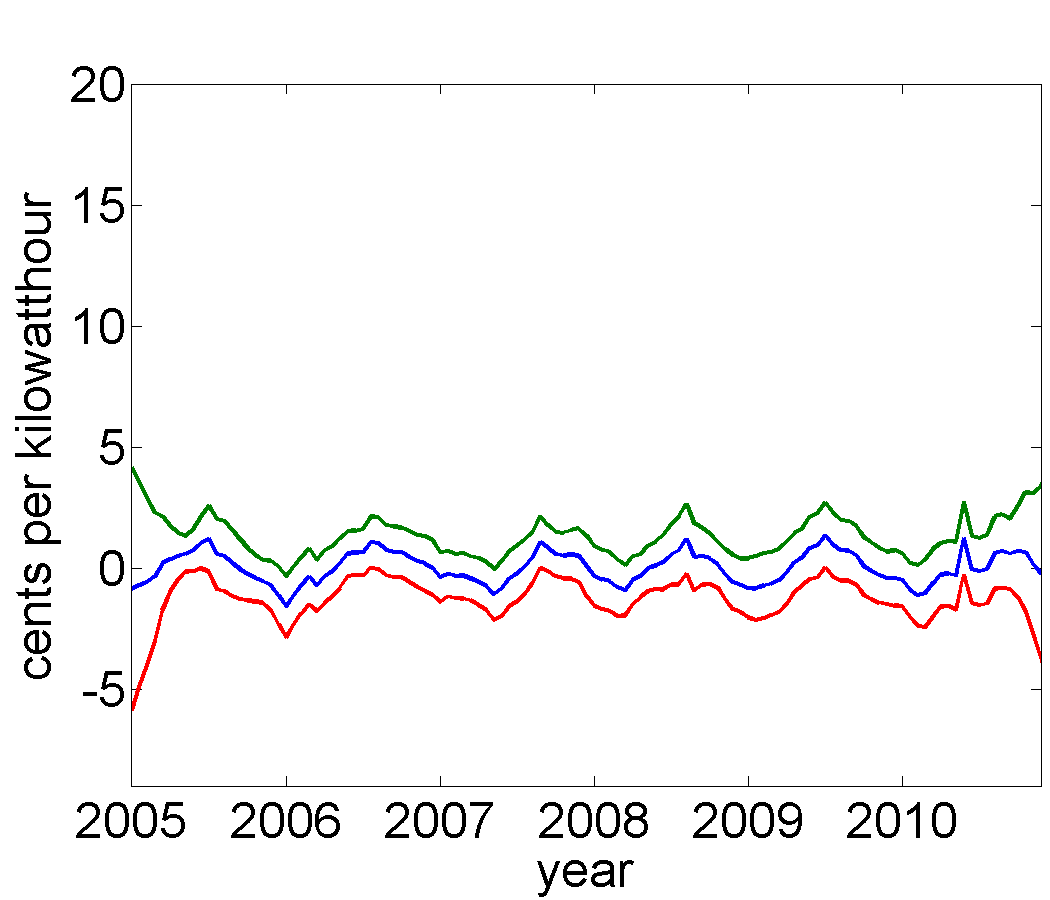}}
\caption{Cross-sectional confidence bands to U.S. electricity price data.}
\label{fig:Electricity-Price-Bootstrap-replicate-g-h}
\end{center}
\end{figure}

The cross-sectional confidence bands  with 95\% confidence level for $g$ and $h$,
are shown in Fig. \ref{fig:Electricity-Price-Bootstrap-replicate-g-h} for the electricity price data. 

%%%%%%%%%%%%%%%%%%%%%%%%%%%%%%%%%%%%%%%%%%%%%%%%%%%%
\subsubsection{U.S. Currency Exchange Fluctuation} \label{sec: U.S. Currency Exchange}
\begin{figure}[!tbh]
\begin{center}
   \subfloat[U.S. Dollar exchange rates]
	      {\includegraphics[scale=0.32]{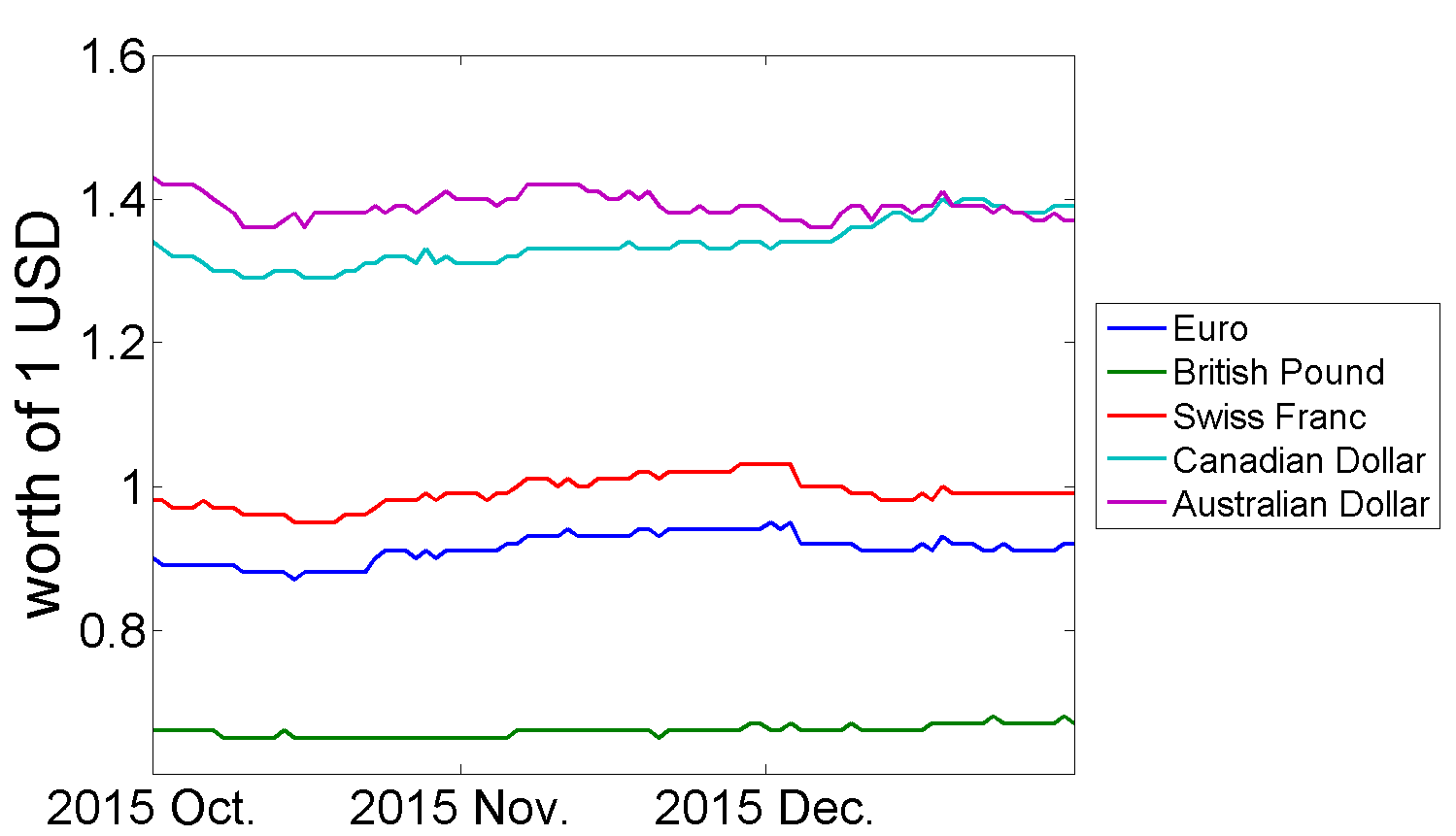}}
	 \subfloat[U.S. Dollar exchange fluctuations]
	      {\includegraphics[scale=0.32]{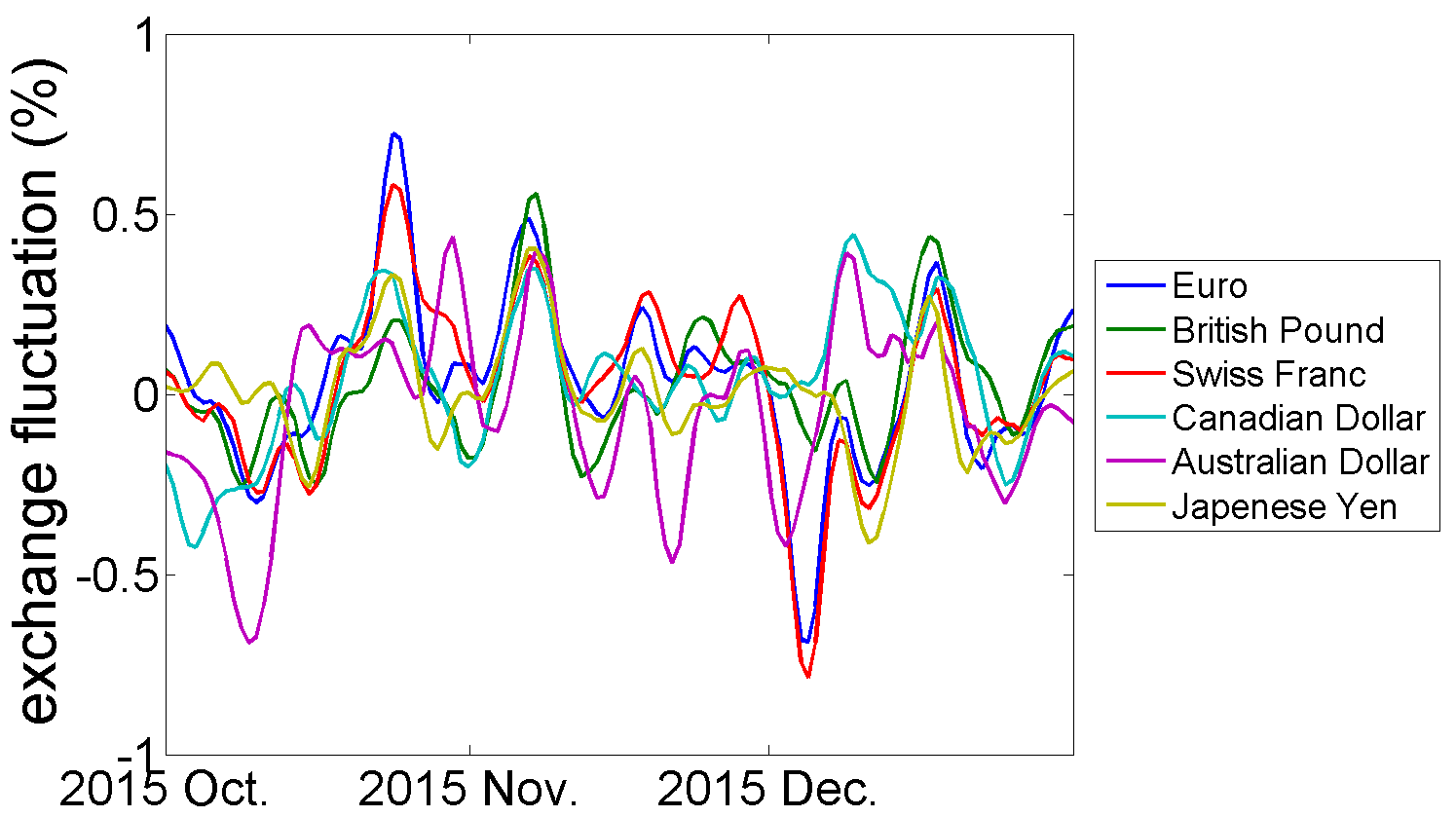}}
\caption{Three months of U.S. dollar exchange rates data and their fluctuations from October to December in 2015.}
\label{fig: US Dollar Exchange rate-data}
\end{center}
\end{figure}

In this experiment, we consider an application of financial data.
The US dollar foreign exchange rates from October 2015 to December 2015 are shown in Figure \ref{fig: US Dollar Exchange rate-data}(a). 
In finance, exchange fluctuation is studied rather than using exchange rates.
The exchange fluctuation is defined as
$ \tau=\frac{R_{1}-R_{0}}{R_{0}}\times100\%$
where $R_1$ is the current exchange rate and $R_0$ is the previous exchange rate. 
If the number $\tau$ is positive, the US dollar is undergoing revaluation 
and a negative $\tau$ value means devaluation. 
Figure \ref{fig: US Dollar Exchange rate-data}(b) displays the USD exchange fluctuation.
The analysis of exchange fluctuation is harder than Growth Velocity data 
and electricity price data since there is no physical interpretation. 
Moreover, the observed currency behavior may be related to less quantifiable 
considerations such as economic policies and governments. 

\begin{figure}[!tbh]
\begin{center}
   \subfloat[recovered $\hat h$]
	       {\includegraphics[scale=0.22]{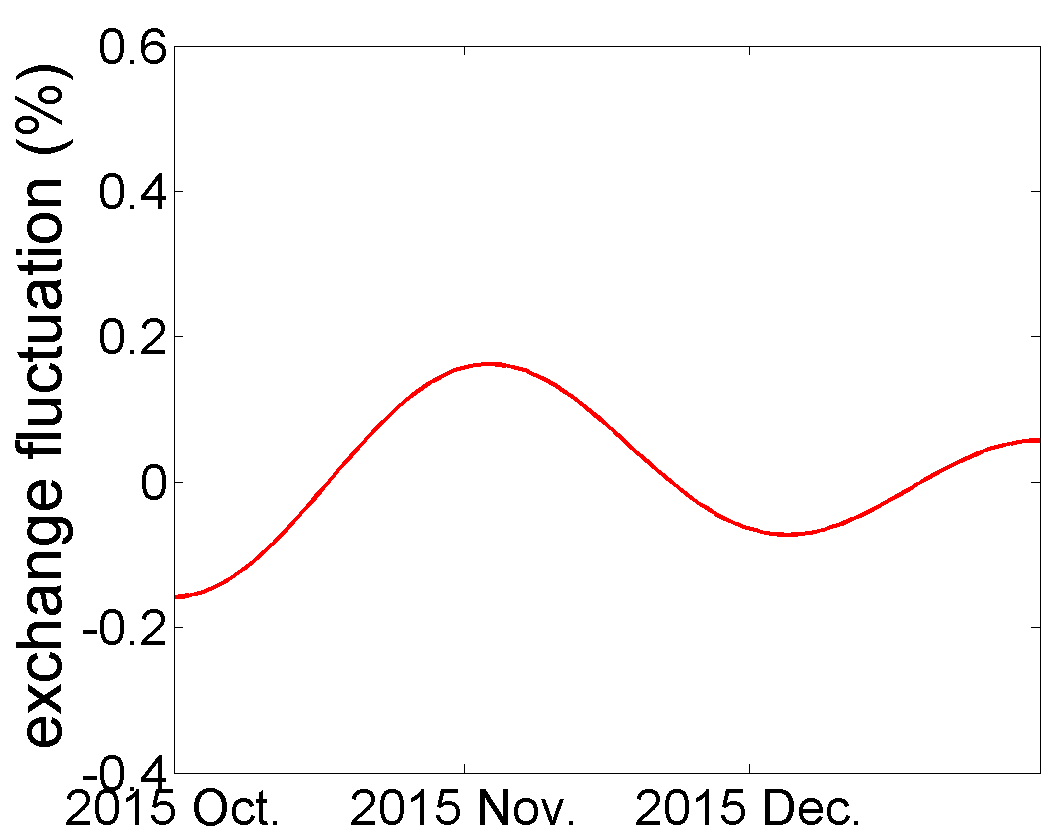}}
	 \subfloat[recovered $\hat g$]
	       {\includegraphics[scale=0.22]{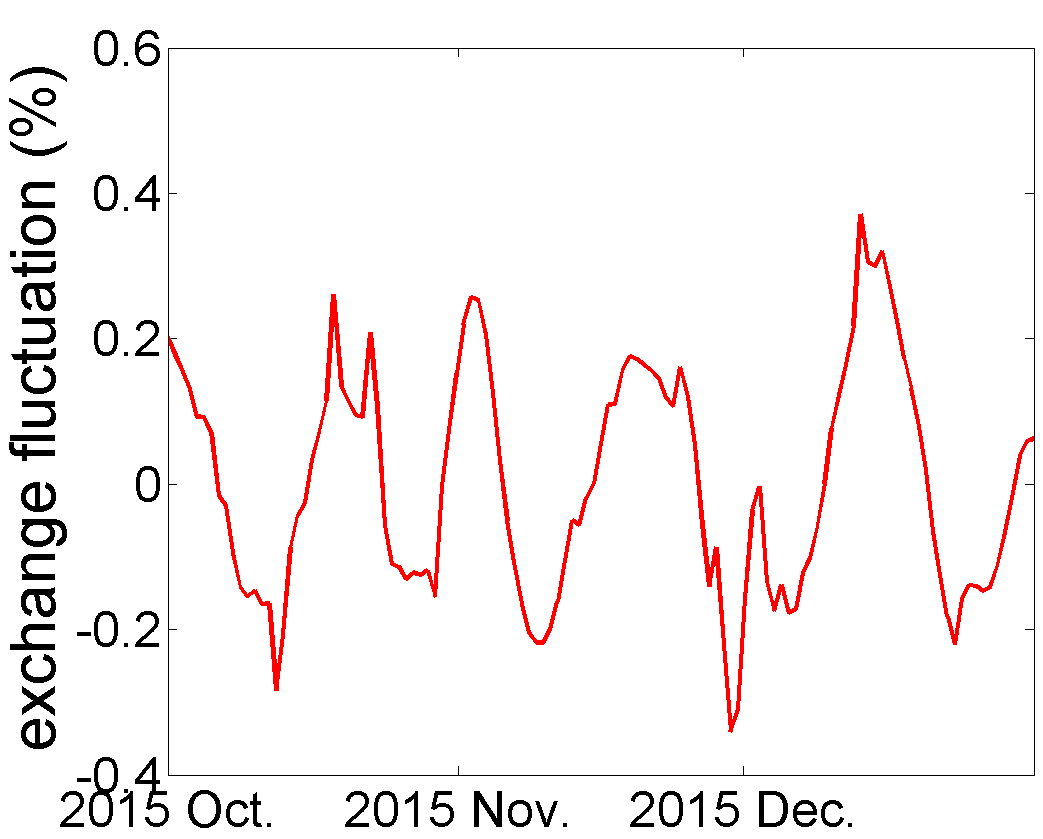}}
	\subfloat[recovered $\{\hat\gamma_i\}$]
	       {\includegraphics[scale=0.22]{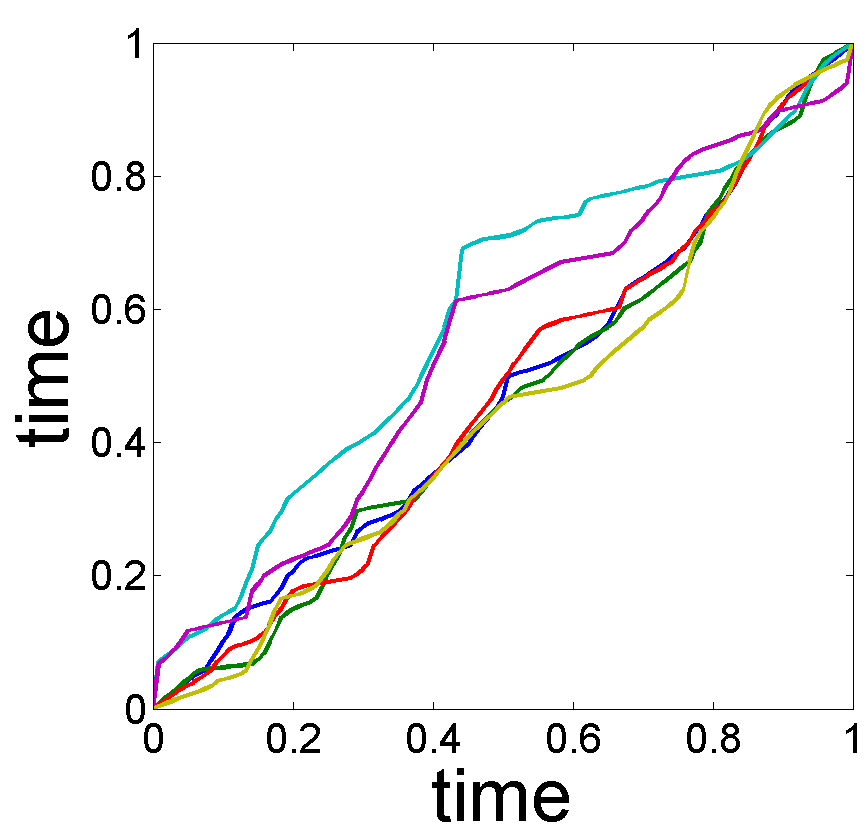}}
	\subfloat[negative log-likelihood]
	       {\includegraphics[scale=0.24]{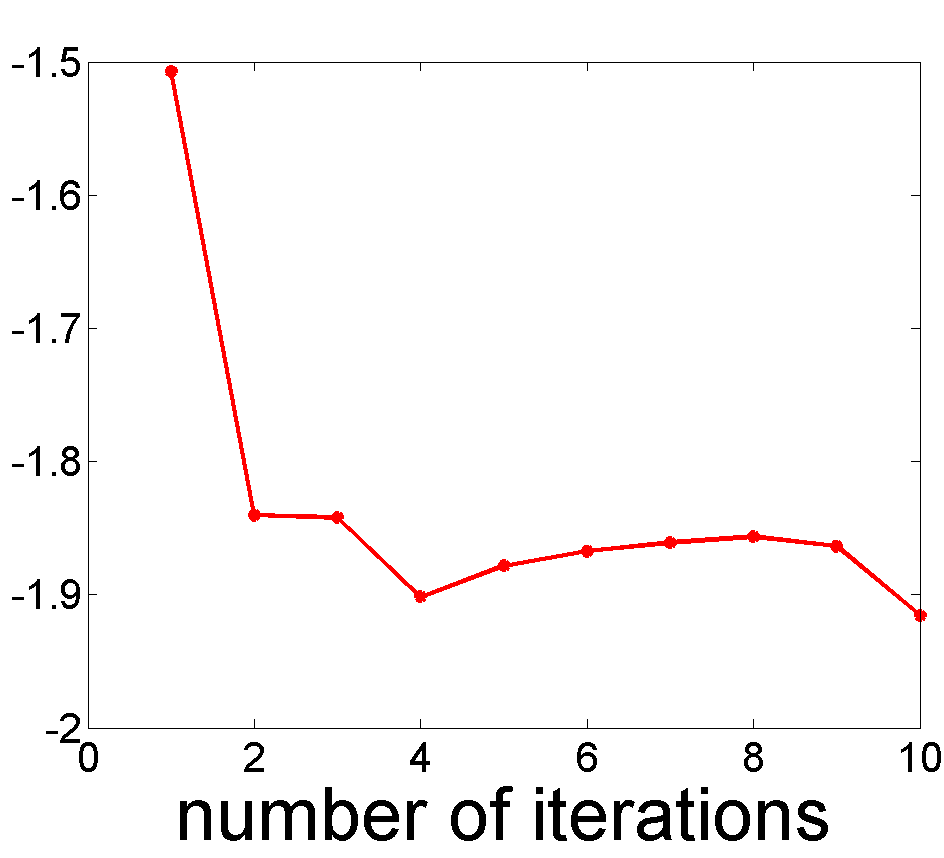}}
\caption{Estimation results of USD exchange fluctuation data.}
\label{fig: results of USD exchange fluctuation}
\end{center}
\end{figure}

\begin{figure}[!htb]
\begin{center}
	 \subfloat[${\cal H}=span\{1\}$]
	      {\includegraphics[scale=0.17]{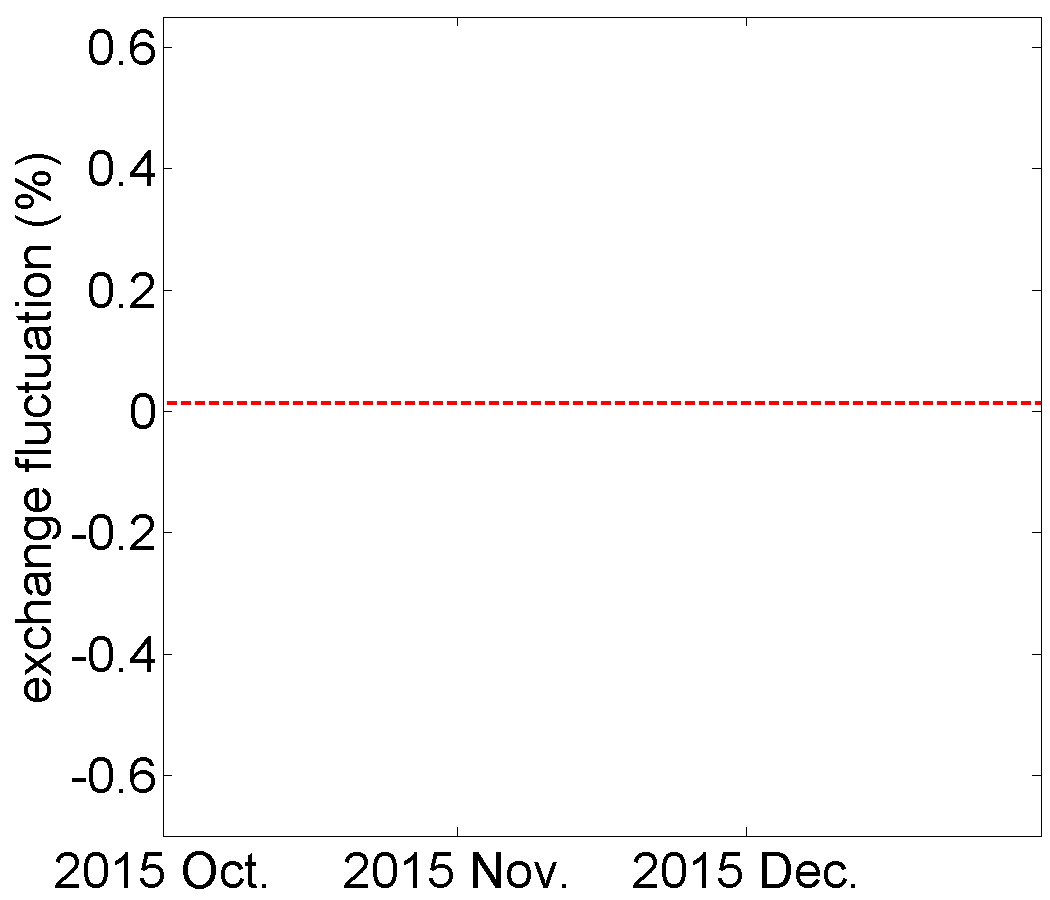}}
	 \subfloat[${\cal H}=span\{1,t\}$]
	      {\includegraphics[scale=0.17]{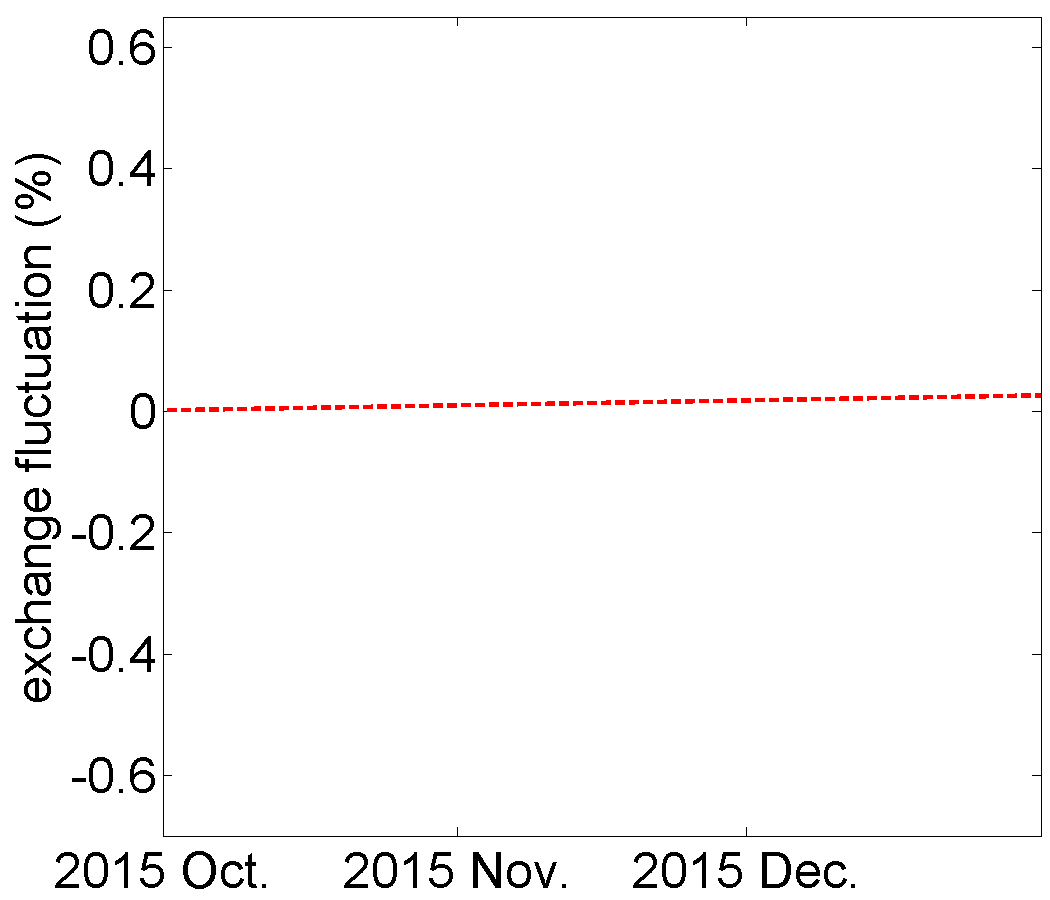}} 
	 \subfloat[Fourier, $\{\phi_1,\phi_2,\phi_3\}$]
 	      {\includegraphics[scale=0.17]{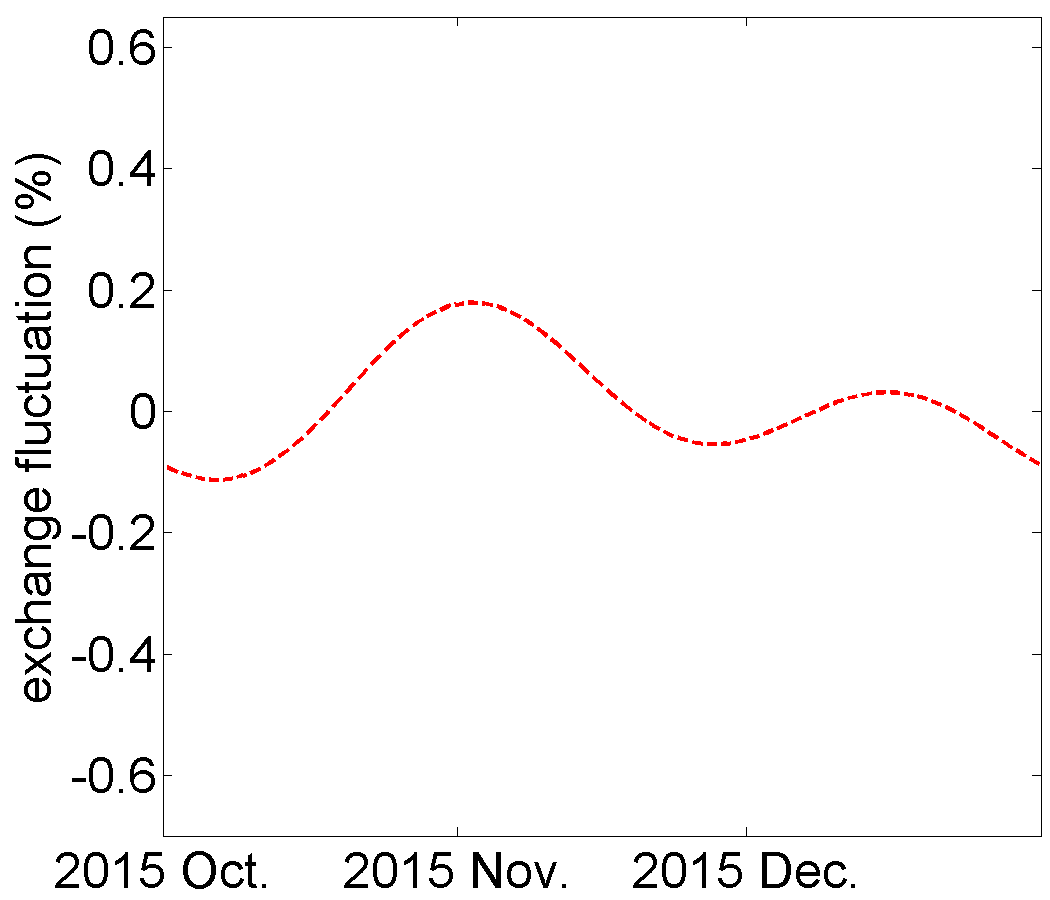}} 
	 \subfloat[${\cal H}=span\{1,\sin(\pi t)\}$]
	      {\includegraphics[scale=0.17]{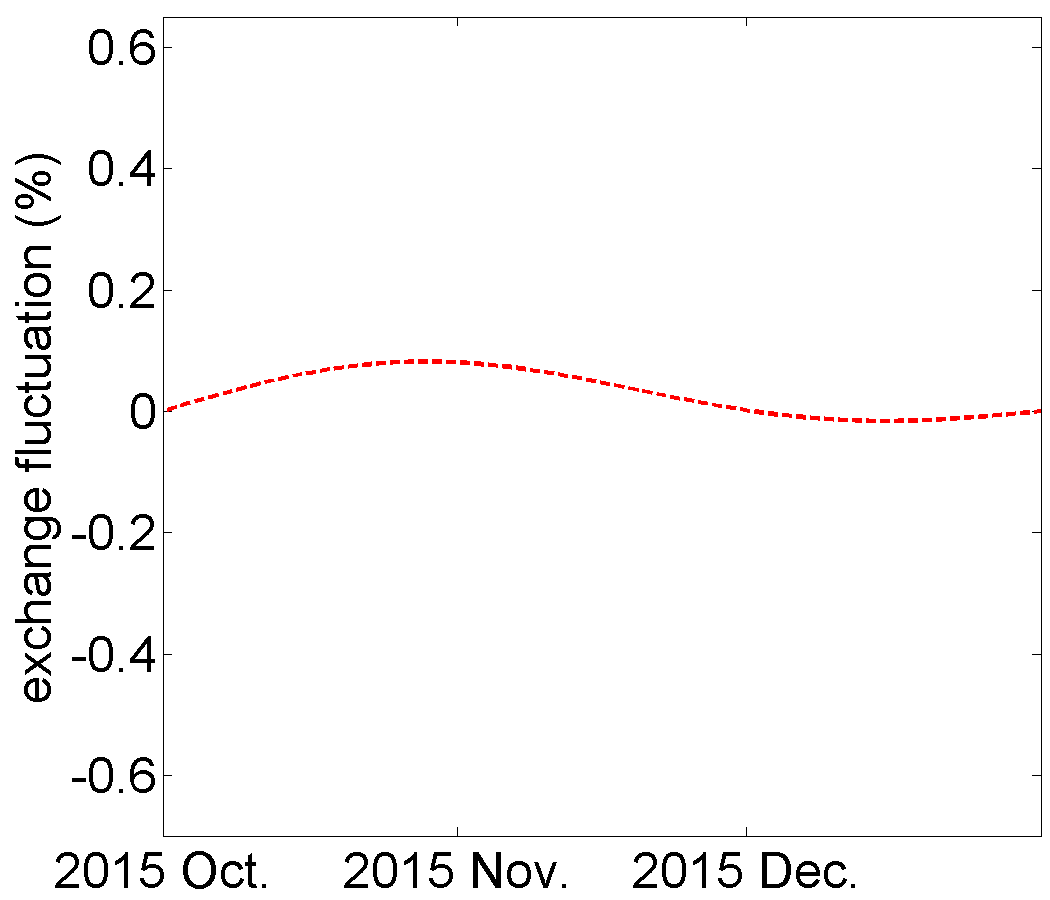}} 
	 \subfloat[Cosine $\{\phi_1,\phi_2,\phi_3\}$]
	      {\includegraphics[scale=0.17]{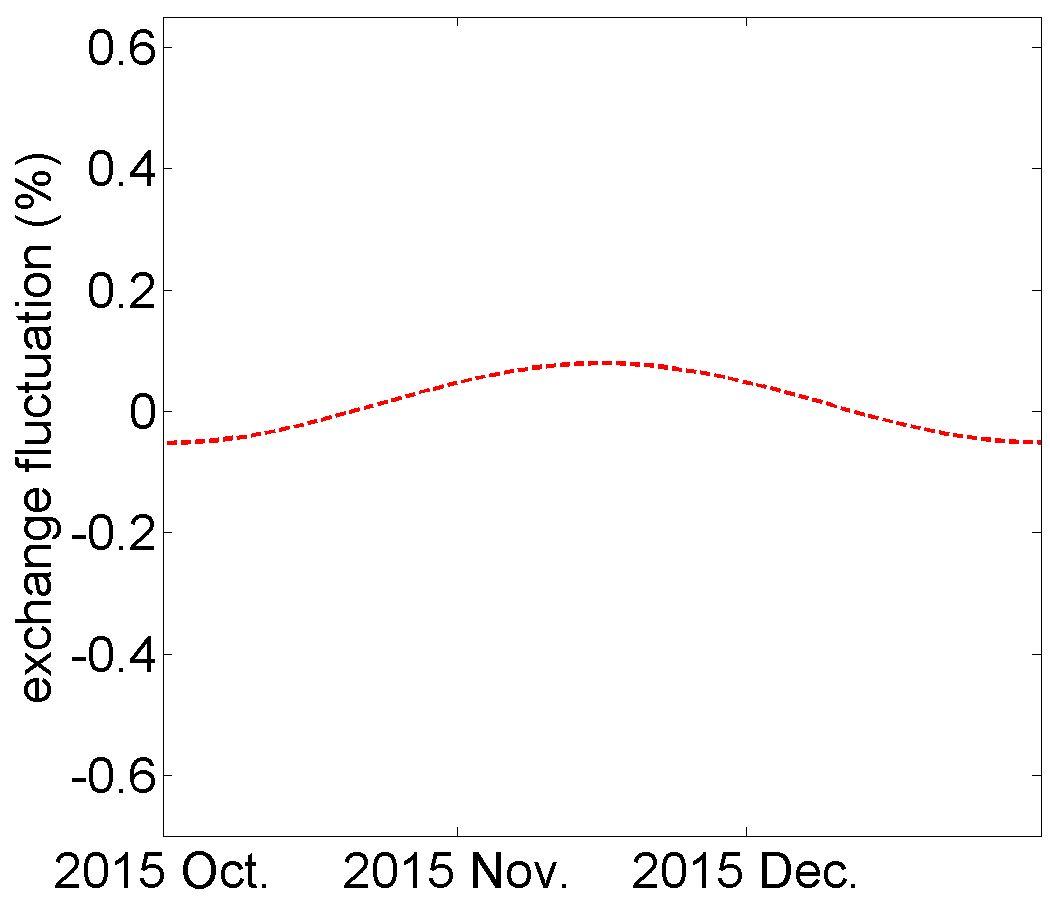}} 									
\caption{Trend estimation results of U.S. exchange fluctuation data of the separation model using different subspace selections.}
\label{fig: simple separation for exchange fluctuation}	
\end{center}
\end{figure}

Our MLE algorithm selects the trend space ${\cal{H}} = span\{1,\cos(\pi t), \cos(2\pi t), \cos(3\pi t)\}$ and 
produces the estimated trend and seasonal effects given in Fig. \ref{fig: results of USD exchange fluctuation}.
The estimated trend has a slow oscillation, while the estimated seasonal effect oscillates with a significantly higher frequency. 
Applying the separation model $f_i(t)=h(t)+g(t)+\epsilon_i(t)$ to the data yields the estimated trends 
shown in Fig. \ref{fig: simple separation for exchange fluctuation}.
While the trends in Fig. \ref{fig: simple separation for exchange fluctuation}, (c)-(e), seems
reasonable, the lack of physical interpretation of the data and the inability to select the subspace $\cal H$ due
to the invariance of the negative log-likelihood function, it is not possible to quantify the selection
of these trends.

Bootstrapping hypothesis testing for the currency fluctuation yields the following:
\begin{itemize}
   \item \textbf{Testing presence of a trend}: 
	        For testing the null hypothesis $h=0$, we have $\rho_{h_0}=0.092$, $\hat{se}_B=0.03$, and
	        a $p$ value of $1.08\times 10^{-3}$. Therefore we reject the null hypothesis. 
					%Similarly, we have $\rho_{g_0}=0.149$ with $\hat{se}_B=0.016$ when testing the null hypothesis $g=0$.
					%The resulting $p$ value is 0 and the null hypothesis, $g=0$, is rejected. 
					%We conclude that the exchange fluctuation data has a non-zero trend and a non-zero seasonality.
	 \item \textbf{Testing constant shape for a trend:} 
	        For testing the null hypothesis, $h=c$, the test statistic using bootstrap is $\rho_{h_c}=0.091$ with $\hat{se}_B=0.034$. 
					The corresponding $p$ value is $3.9\times 10^{-3}$ and therefore, the null hypothesis is rejected.	
					%When testing the null hypothesis, $g=c$, we get $\rho_{g_c}=0.149$ and $\hat{se}_B=0.016$. 
					%Based on the $p$ value $0$, we reject the null hypothesis. 
					%We conclude that both the trend $h$ and the seasonality $g$ are not constant. 
	 \item \textbf{Testing linear shape for a trend:} 
	       For testing the linearity of a trend, the test statistic using bootstrap is $\rho_{h_l}=0.76$ with $\hat{se}_B=0.807$. 
	        Since the  $p$ value is $0.17$, we fail to reject the null hypothesis. 
				 %Similarly, the test statistic for the seasonality $g$ is $\rho_{g_l}=7.63$ with $\hat{se}_B=2.14$. 
				 %Because the associated $p$ value is $1.8\times 10^{-4}$, the null hypothesis, the seasonality $g$ is a linear function,
				 %is rejected.
\end{itemize}

\begin{figure}[!tbh]
\begin{center}
    \subfloat[confidence band of $\hat h$]
	      {\includegraphics[scale=0.225]{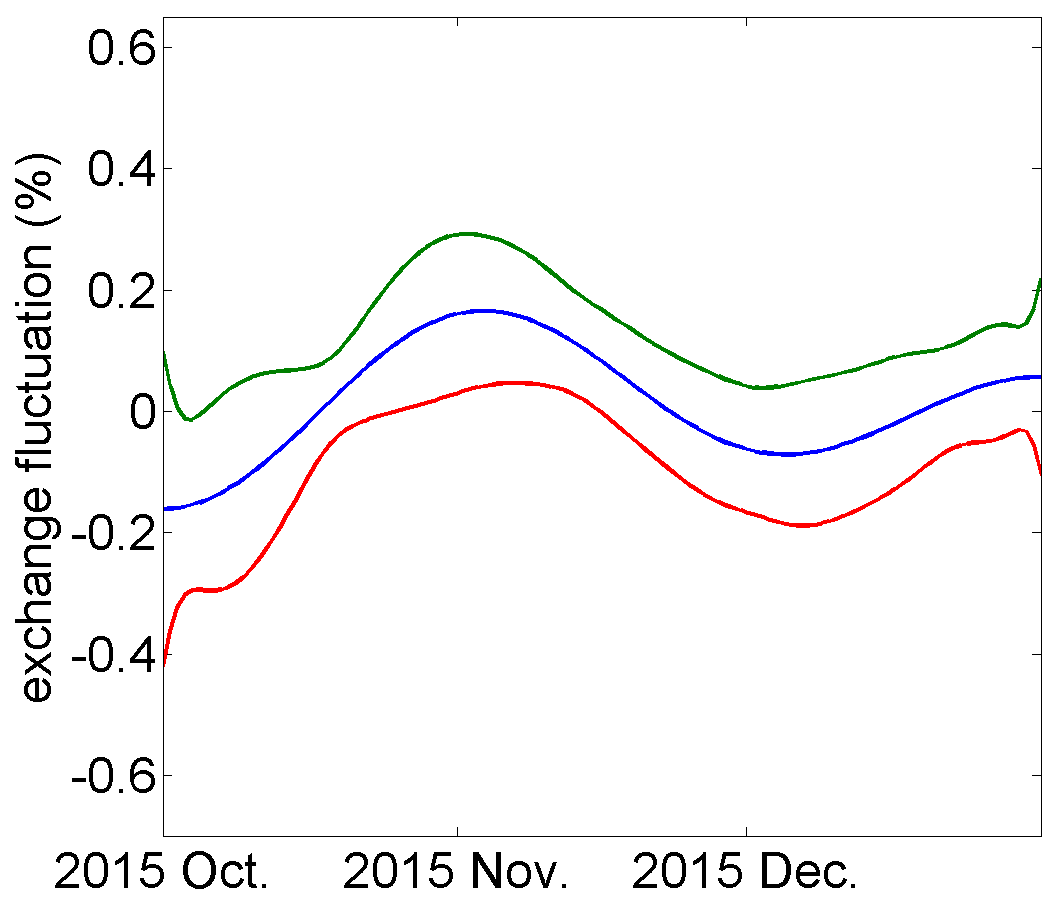}}
	  \subfloat[confidence band of $\hat g$]
	      {\includegraphics[scale=0.225]{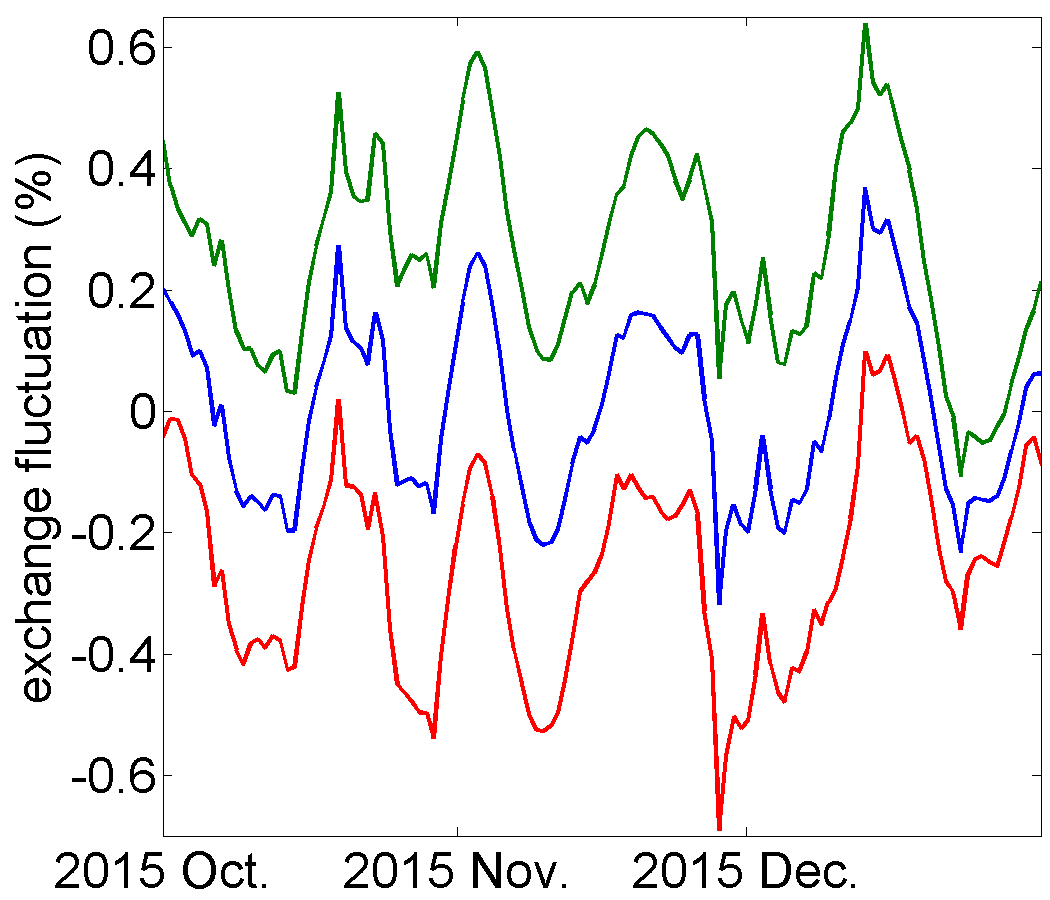}}
\caption{Cross-sectional confidence band of USD exchange fluctuation data.} 
\label{fig: Bootstrap replicates of USD exchange fluctuation}
\end{center}
\end{figure}

Finally, Fig. \ref{fig: Bootstrap replicates of USD exchange fluctuation} shows cross-sectional confidence bands 
for the trend and the seasonality at the 95$\%$ confidence level. 

%%%%%%%%%%%%%%%%%%%%%%%%%%%%%%%%%%%%%%%%%%%%%%%%%%%%%
\section{Conclusion} \label{sec: Conclusion}
We have developed a novel, model-based framework to solve the trend and variable phase seasonality estimation 
problem by  estimating the trend $h$ and seasonality $g$
components from time series data, in situations where the seasonal component exhibits random time warpings.   
The model subsumes those used by related approaches in the literature. 
We assume that the subspaces associated with these two components -- trend and seasonality -- are orthogonal, 
and the Karcher mean of warping functions is identity, to ensure that the components are identifiable.  
Under these conditions we seek MLE of $h$ and $g$, using a coordinate-descent algorithm that 
iteratively updates one component at a time, while fixing the others. 
We also use maximized likelihood to 
select an appropriate subspace ${\cal H}$ for the trend component using a increasing sequence of nested subspaces. 
The estimated quantities -- trend and seasonality -- are tested, using bootstrap replication, for being null or having a specific simple shape, 
such as constant and linear. 

Both synthetic data and real data have been used to demonstrate the effectiveness of this method. 
Using synthetic data, where the ground truth is known, we have demonstrated the robustness of  
our method in the presence of noise. We demonstrate this framework's ability to extract trend and seasonality 
using three real application datasets: the Berkeley Growth data, U.S. electricity price data, and USD exchange fluctuation. 
Bootstrap hypothesis testing supported the presence of a trend in all of the datasets. 
For the USD exchange fluctuation data, a low frequency oscillation between revaluation and devaluation was extracted,
along with a higher frequency seasonal oscillation.
For the Berkeley Growth data and  electricity price data, 
we obtain estimates of the trends and seasonal effects that are expected from the nature of the applications.

\section*{References}
\bibliography{myrefs}
\end{document}